\documentclass[preprint, 12pt, numafflabel]{elsarticle}

\usepackage{array}
\usepackage[title, titletoc]{appendix}
\setcitestyle{square}
\usepackage{adjustbox}
\usepackage{tabularx}
\usepackage{afterpage}

\makeatletter
\g@addto@macro\@floatboxreset\centering

\makeatother
\usepackage{fancyhdr}
\usepackage[utf8]{inputenc}
\usepackage{graphicx} 
\usepackage{amsmath}
\usepackage{subcaption}
\usepackage{upgreek}
\usepackage{booktabs}
\usepackage{multirow}

\usepackage{pdflscape}
\usepackage{rotating}

\usepackage{xr-hyper} 
\usepackage{hyperref} 
\usepackage{catchfilebetweentags} 
\hypersetup{colorlinks,allcolors=black}
\usepackage[top=2cm, bottom=2cm, left=2cm, right=2cm]{geometry}

\newcommand{\exvivo}{\textit{ex vivo}}

\usepackage{amssymb}

\usepackage{lineno}
\usepackage{soul} 
\usepackage{empheq} 
\usepackage{lineno}
\makeatletter
\soulregister\cite7 
\soulregister\ref7  
\soulregister\caption7
\soulregister\equation7
\soulregister\label7
\soulregister\exvivo7
\soulregister\~7    
\soulregister\paragraph7
\date{July 2025}
\renewcommand{\hl}[1]{#1}
\begin{document}
    \journal{Ultrasonics}
    \begin{frontmatter}
    

        \title{Imaging of ultrasound specularity at cortical bone interfaces \\to detect unbalanced remodeling: a preliminary study}
        \affiliation[lib]{{organization}={Sorbonne Université, INSERM, CNRS, Laboratoire d’Imagerie Biomédicale, LIB, F-75006},
                    city={Paris},
                    country={France}}
        \affiliation[delft]{organization={Department of Imaging Physics, Delft University of Technology},
                    country={The Netherlands}}
        
        \author[lib]{Amadou S. DIA \texorpdfstring{\footnote{Corrresponding author: amadou.dia@sorbonne-universite.fr}{}}}
        
        \author[lib]{Salomé VIGNAT}
        \author[lib,delft]{Guillaume RENAUD}
        \author[lib]{Quentin GRIMAL}
    \end{frontmatter}
\section*{Abstract}
 The region near the interface between cortical bone and the medullary cavity—the endosteal surface—is of particular interest for the early detection of osteoporosis. This region is typically the first to exhibit signs of unbalanced remodeling (enlarged pores and increased surface roughness). In this study, we analyse ultrasound reflection at the endosteal surface to introduce a novel potential biomarker of cortical bone health based on specularity. Our hypothesis is that increased pore size and surface roughness enhance ultrasound scattering, thereby reducing specular reflection. We reconstruct maps of specularity by combining a signal processing technique initially designed to enhance specular reflectors in soft tissues with a beamforming technique that accounts for refraction at bone–soft tissue interfaces. Specularity values (between 0 and 1) quantify the similarity between received signals and signals from an ideal specular reflector. Using numerical simulations and \exvivo~measurements with a 2.5 MHz phased array we highlight a strong relationship between specularity and bone microstructure, as assessed by high-resolution micro-computed tomography. Among 12 regions of interest (ROI) in the femoral bone of three donors, 8 ROIs without large pores showed high specularity ($>$0.5) in over 50\% of pixels, in contrast to 4 ROIs with signs of extensive remodeling. Both pore volume fraction and pore size were strongly associated with specularity. 
 In simple linear regression analyses, each parameter individually explained 84\% of the variability in specularity.
 These findings suggest that specularity reflects bone microstructure and may \hl{potentially} serve as a sensitive marker for identifying cortical bone degradation near the endosteal surface.
\section{Introduction}
Each year, approximately nine million fragility fractures related to osteoporosis occur worldwide \cite{cooper_iof_2017}. Fracture risk assessment traditionally relies on clinical risk factors and areal bone mineral density (aBMD) measurements obtained using dual-energy X-ray absorptiometry (DXA). However, many individuals at high risk for fractures are not identified through aBMD alone \cite{siris_bone_2004, briot_frax_2013}. Bone strength is also influenced by factors beyond those captured by DXA, including microarchitectural features and cortical bone properties. Cortical bone is particularly important, as it plays a key role in mechanical stability, and approximately 80\% of fragility fractures occur in regions comprising large amounts of cortical bone within the appendicular skeleton \cite{bala_role_2015}. Age-related bone loss in the cortex is characterized by reduced cortical thickness and increased porosity \cite{zebaze_intracortical_2010}. High-resolution peripheral quantitative computed tomography (HR-pQCT) can assess thickness and, to some extent, porosity, however, this modality remains largely restricted to research settings.

Ultrasound (US)-based diagnostic approaches offer several advantages: they are radiation-free and cost-effective compared to other imaging modalities, and they can be used at point-of-care \cite{hans_quantitative_2022}. These approaches provide parameters reflecting cortical thickness and bone material properties indicative of mechanical competence \cite{karjalainen_ultrasonic_2008,grimal_quantitative_2019,minonzio_ultrasoundbased_2019}. Existing ultrasound methods, however, provide estimates of material properties averaged over a large region of interest encompassing the entire thickness of the cortex\cite{karjalainen_ultrasonic_2008,schneider_ex_2019}. They do not resolve the spatial heterogeneity of material properties, which may prevent the detection of localized degradations that could reveal a pathological remodeling process. Indeed, intracortical remodeling typically initiates at bone surfaces, particularly near the interface with the medullary cavity (endosteal surface). In aged and osteoporotic individuals, the unbalanced remodeling process leads to a heterogeneous degradation of the microstructure: enlarged and clustered pores \cite{bell_regional_1999, andreasen_understanding_2018}, a radial porosity gradient--increasing porosity from the inner (endosteal) to the outer (periosteal) surface--and increased roughness of the endosteal surface resulting from remodeling cavities opening into the medullary cavity \cite{zebaze_intracortical_2010}. This region where dense cortical bone gradually merges with more porous trabecular-like tissue, referred to as the transition zone \cite{zebaze_new_2013}, is an interesting target for early detection of osteoporotic changes.

Some authors have proposed to assess the gradient of material properties in the cortex using guided waves \cite{haiat_influence_2009} but such method has not been implemented in vivo as far as we know. Another group has proposed assessing cortical pore-size distribution using ultrasound backscatter measurements assisted with imaging \cite{iori_estimation_2020}. While this method shows promise for identifying individuals with imbalanced remodeling\cite{dehnen_quantitative_2024}—particularly those characterized by the presence of very large pores—it does not specifically capture microstructural changes near the endosteal interface.

The method proposed by Renaud et al. \cite{renaud_vivo_2018} extends B-mode imaging into the bone by employing a delay-and-sum image reconstruction algorithm including correction of refraction at the bone-soft tissue interfaces. This technique provides anatomical images of the cortex and of the medullary cavity. In the present study, we build on this approach to perform a focal assessment of bone tissue near the endosteal surface and in the transition zone.


The objective of this study is to test the hypothesis that microarchitectural changes near the endosteal interface—within a region on the order of the ultrasound wavelength—lead to a measurable decrease in the specularity of reflected ultrasound signals. Indeed, the large intracortical pores in the transition zone and the increased roughness of the endosteal interface likely lead to an enhanced ultrasound scattering and a decreased amplitude of the specular reflections \cite{sukmana_surface_2006, sjoj_ultrasound_2012,karbalaeisadegh_ultrasound_2022,dia_influence_2023}. A further aim is to identify which structural parameters (e.g., pore size, porosity, surface roughness) most significantly correlate with reduced specularity. To this end, we adapted a method developed by Rodriguez et al. \cite{rodriguez-molares_specular_2017} to enhance specular reflections in a beamforming algorithm. 
We conducted numerical simulations to validate the implementation of the technique
and \exvivo~measurements in human bones to investigate how microstructure near the endosteal surface relates to \hl{specularity} metrics.

\section{Materials and methods}
\subsection{Samples}
\begin{figure}[htb!]
    \centering
    \includegraphics[width=.6\textwidth]{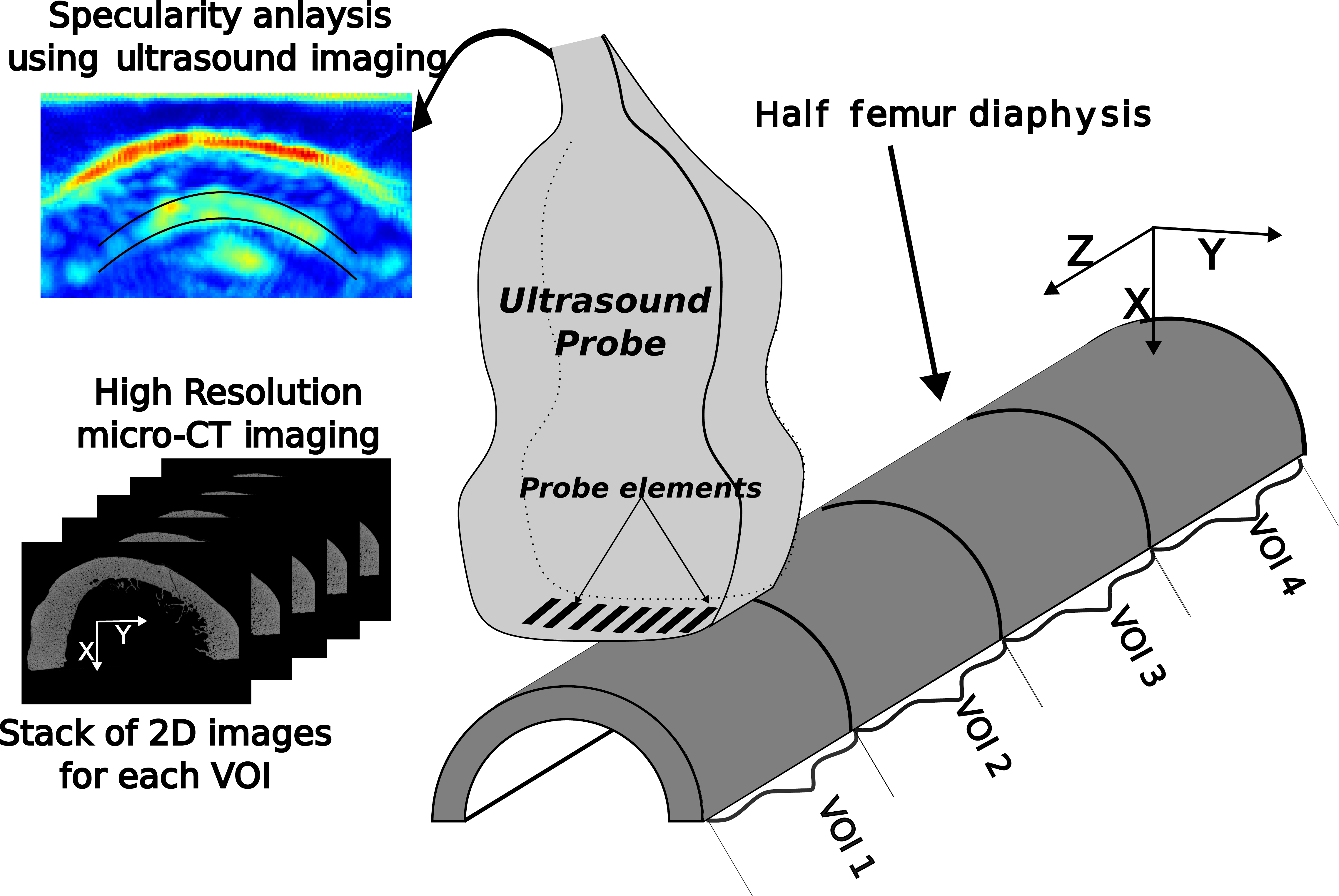 }
    \caption{Experimental protocol. A half-cylinder femur sample was immersed in physiological saline solution and imaged in a plane perpendicular to the bone axis using an ultrasound array operating at a central frequency of 2.5 MHz. Four measurement zones were defined in each sample. Ten US acquisitions were performed for each zone, with slight lateral ($y$-direction) or rotational adjustments of the probe between repetitions to vary the imaging area.
    Samples were scanned with X-ray micro-computed tomography (voxel size: 8.8~\textmu m) and four Volumes of Interest (VOI) were defined as stacks of 2D slices corresponding to each measurement zone.}
    \label{fig:method:protocol}
\end{figure}

Bone samples were obtained from the femurs of three female donors. The age of the donors were 78, 78 and 80 years old for samples 1 to 3, respectively. Ethical approval for sample collection was granted by the Human Ethics Committee of the Centre du don des Corps at the University Paris Descartes (Paris, France).
Informed written consent was provided by the tissue donors or their legal guardians, in accordance with the legal requirements outlined in the French Code of Public Health.
The fresh bone material, stored at -20°C, was sawed, and soft tissues were totally removed. For each femur, a half-cylinder measuring 7 cm in length was extracted from the region between the proximal third and the mid-diaphysis. The samples were subsequently fixed in alcoholic formalin-free fixative F13 (Morphisto, Germany) for 48 hours and rinsed in distilled water for 6 hours using an ultrasonic bath. Four regions of interest, each approximately 1.5 cm in length, were delineated in each sample, resulting in 12 measurement zones (Figure~\ref{fig:method:protocol}).

\subsection{High-resolution X-ray micro-computed tomography (HR-\textmu CT) imaging and definition of volumes of interest} 
The samples were scanned using HR-\textmu CT using a Skyscan 1176 (SkyScan-Brücker, Kontich, Belgium) as described in details in a previous work \cite{dia_ultrasound_2026}. Briefly, 
Image reconstruction with a voxel size of 8.8~\textmu m was performed with the Feldkamp filtered back-projection algorithm (NRecon, version 1.7.4.6, SkyScan-Brücker, Kontich, Belgium).
Four volumes of interest (VOI) were defined within the three-dimensional (3D) HR-\textmu CT image of each sample. The thickness (direction of the bone axis) and width of the VOIs were 15 and 20~mm, respectively. The height of the VOIs varied between 8 and 10~mm, depending on the cortical thickness and diameter of each sample. The VOI's dimensions approximately correspond to the thickness (probe elevation direction), lateral dimension (probe aperture) and depth of the US image.

\subsection{Microstructure and endosteal roughness from HR-\textmu CT images} 
In this section, we present the parameters related to the pore microstructure and the roughness of the endosteal surface, which were extracted from HR-\textmu CT images. These parameters are considered potential contributors to the observed variation in specularity among the VOI of each samples.

All processing steps were performed using CTAn software (CTAn; SkyScan-Brücker, Kontich, Belgium) except where otherwise indicated. These steps were described in details in a previous study \cite{dia_ultrasound_2026}. 
The initial step was to apply a Gaussian filter (round kernel of radius 2 or 3, depending on the sample) on 3D reconstructed volumes to remove acquisition noise following a standardised procedure for the analysis of bone microstrure \cite{campbell_quantitative_2014}.
Then, the VOI were binarized using 3D Otsu method \cite{behrooz_automated_2017}. 

\begin{figure}[ht!]        \includegraphics[width=.45\linewidth]{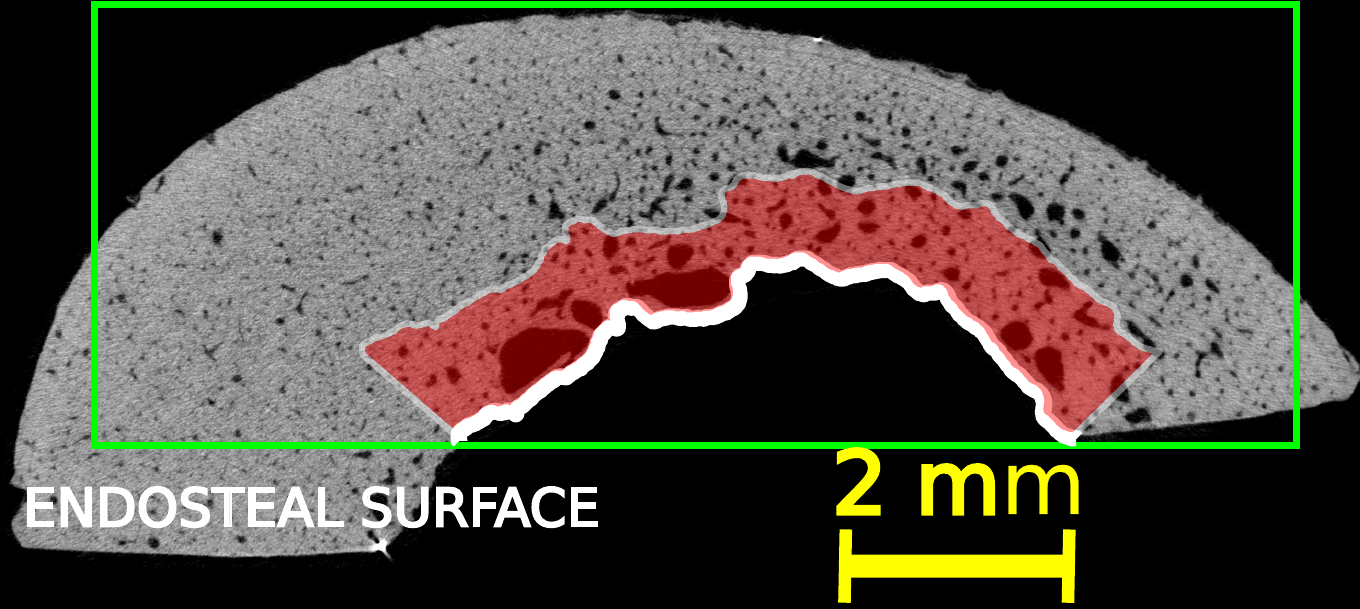}
    \caption{Representative 2D slice of a HR-\textmu CT VOI illustrating the segmented endosteal surface (thick white line) and the definition of the endosteal region (in red). Note that the pores in this region are larger compared to those in other areas of the cortex.}
    \label{meth:e_microstruct}
\end{figure}

In each 2D slice of a VOI, we performed the segmentation of the endosteal surface using MATLAB 2023b (MathWorks Inc., Natick, MA, USA, \textit{bwboundaries} function). We then defined the endosteal region as a region of thickness 1.4~mm bounded by the endosteal surface (Figure~\ref{meth:e_microstruct}). The 1.4~mm-thickness corresponds to one US wavelength in bone at 2.5~MHz assuming a speed of sound of 3500~m/s. The endosteal volume of interest (EVOI) was defined as the stack of segmented 2D endosteal regions.
The microstructural parameters characterizing the pores within each EVOI were computed using MATLAB from each 2D slice: (i)~ the endosteal porosity representing the percentage of pore surface relative to the total surface of the endosteal region; (ii)~the distribution of endosteal pore diameters (the diameter of each pore was defined as that of a disk with an equivalent area). For each EVOI, we then computed the median endosteal porosity (E.Por) and the median endosteal pore diameter (E.Po.Dm) across the EVOI 2D slices. The variations along the bone axis of the endosteal porosity and pore diameter were quantified from the interquartile ranges (IQR) of these quantities across the EVOI 2D slices, yielding IQR.E.Por and IQR.E.Po.Dm, respectively.

\begin{table}[ht!]
    \centering
    \begin{tabular}{lll}
        \toprule
        Variable & Unit & Definition\\
\hline
        E.Por & \% & Median porosity in the endosteal region (EVOI) \\
        E.Po.Dm & $\upmu$m & Median diameter of pores in the endosteal region (EVOI)\\
        IQR.E.Por & $\upmu$m & Interquartile range of porosity across 2D slices in the EVOI\\&& (to quantify microstructural heterogeneity in the probe elevation direction) \\
        IQR.E.Po.Dm & $\upmu$m & Interquartile range of pore diameter across 2D slices in the EVOI\\
        && (to quantify microstructural heterogeneity in the probe elevation direction)\\
        $N_{40}$ or $N_{50}$ & \% & Number of pixels with specularity $>$ 0.4 or $>$ 0.5 expressed as \\&&a percentage of the total number of pixels in the endosteal region \\
        MSI & \% & Mean Specularity Index, expressed as a percentage \\
        \hline
        $R_q$ & $\mu$m & Root Mean Square height relative to a smooth endosteal surface \\
        $\rho_{\text{length}}$ & mm & Correlation length \\        
        \bottomrule
    \end{tabular}
    \caption{Quantitative variables used for the analysis.}
    \label{tab:definition_variables}
\end{table}

The roughness of the endosteal boundary was quantified after removing its curvature by applying a high-pass Butterworth filter to the line profiles. The cut-off frequency of the filter was adapted individually for each sample to account for variations in gross geometry. Its value was ranging between 0.045~mm$^{-1}$ and 0.057~mm$^{-1}$. We used two parameters: the root mean square (rms) height ($R_q$) and the correlation length of the height variations ($\rho_{\text{length}}$) \cite{wilhjelm_influence_2001}. 
The correlation length $\rho_{\text{length}}$ was defined as the spatial lag at which the profile autocorrelation decreases by a factor of $1/e$ of its maximum value. This parameter characterises the lateral distance over which surface changes occur. A longer $\rho_{\text{length}}$ indicates smoother transitions, whereas a shorter one indicates more abrupt changes.

\subsection{Ultrasound data acquisition}
Experiments were conducted as described in \cite{dia_ultrasound_2026}. Femur samples were immersed in physiological saline solution and imaged in a plane perpendicular to the bone axis. We used a fully programmable US system (Vantage, Verasonics Inc., Redmond, WA, USA) with a phased array US transducer with 96 elements, operating at a central frequency of 2.5 MHz (P4-1 ATL/Philips, Bothell, WA, USA) with a pitch of 0.295 mm and an elevation aperture of 14 mm (US image thickness). The emitted pulse had a 80\% relative frequency bandwidth at -6 dB. The US probe was positioned approximately 4 mm above the outer bone surface at the center of each region of interest, each corresponding to a VOI (Figure \ref{fig:method:protocol}). Ten acquisitions were performed for each region, with slight lateral (y-direction) or rotational adjustments of the probe between repetitions to vary the imaging area. After each repositioning, probe placement was fine-tuned relying on the real-time visualization of intra-cortical bone images, as in in vivo acquisitions \cite{renaud_vivo_2018}, to ensure clear visualization of the periosteal and endosteal surfaces. We used a synthetic aperture imaging sequence with single-element transmissions \cite{jensen_synthetic_2006}. 
The signals recorded by the elements were stored in a 3D-matrix $\textbf{S}$ with dimensions [$N\times N_R \times N_T$], where $N$ is the number of time samples, $N_R$ is the number of receiving elements and $N_T$ is the number of transmissions.

\subsection{Simulations}
We performed numerical simulations of the imaging sequence to validate the implementation of our method  and assess \hl{the theoretical impact of endosteal-surface microtopography on specularity; however, we did not model porosity.}
%
The nominal model is a three-layer medium: an elastic slab representing the bone layer (cortex) surrounded by two perfect fluid layers representing soft tissues between the probe and bone and marrow. 
We designed 40 models with different rms height ($R_q$) and correlation length ($\rho_{\text{length}}$), ranging from 30 to 480 $\upmu$m and 0.5 to 4~mm, respectively (Figure~\ref{fig:config_simull}). 
In practice, the endosteal surface profile was generated from a normal distribution of height values with a standard deviation equal to the desired $R_q$. This profile was then smoothed using a centred Gaussian filter with a standard deviation equal to $\rho_{\text{length}}$, to enforce the desired spatial correlation.
The cortical bone layer was modeled as an elastic slab with a mass density of 2 kg/dm$^3$ and a compressional and shear wave speeds of 3500~m/s and 1800~m/s, respectively. The soft tissue-mimicking layers were modeled as a fluid with compressionnal wave-speed of 1540 m/s and a mass density of 1 kg/dm$^3$
The parameters of the imaging sequence and the properties of the probe matched those of the experiment described above. 
Elastic wave propagation was simulated using the open-source Finite Difference Time-Domain (FDTD) code SimSonic \cite{bossy_simsonic_nodate, bossy_three-dimensional_2005}. To prevent reflections at the boundaries of the simulation domain, a Perfectly Matched Layer (PML) boundary condition was applied. 
The simulations were performed with a grid step of 10 $\upmu$m. The time step was chosen to satisfy the Courant–Friedrichs–Lewy (CFL) stability condition for 2D simulations, with a constant CFL value of 0.99.

\begin{landscape}{
\begin{figure}[ht!]
    \centering
    \includegraphics[width=1\linewidth]{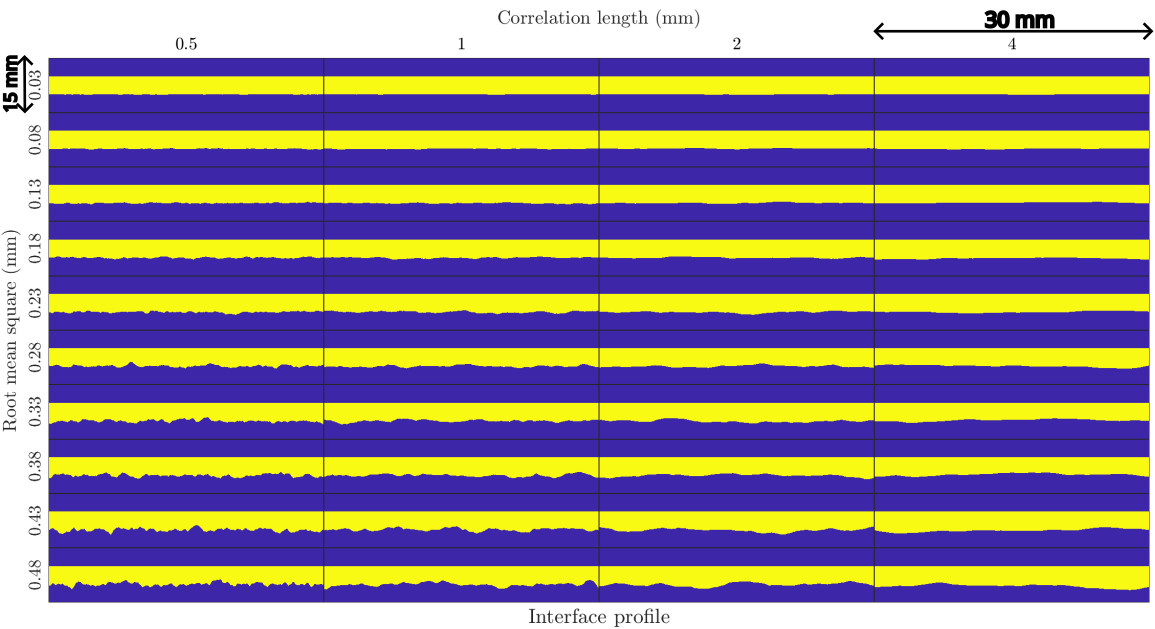}
    \caption{The 40 simulation models. Each thumbnail delimited by black lines is a three-layer medium (bone cortex in yellow) with specific values of rms height ($R_q$, different lines) and correlation length ($\rho_{\text{length}}$, different columns).} 
    \label{fig:config_simull}
    \end{figure}
}
\end{landscape}

\subsection{Ultrasound image reconstruction and map of specularity}

\begin{figure}[ht!]
    \centering
    \includegraphics[width=.4\linewidth]{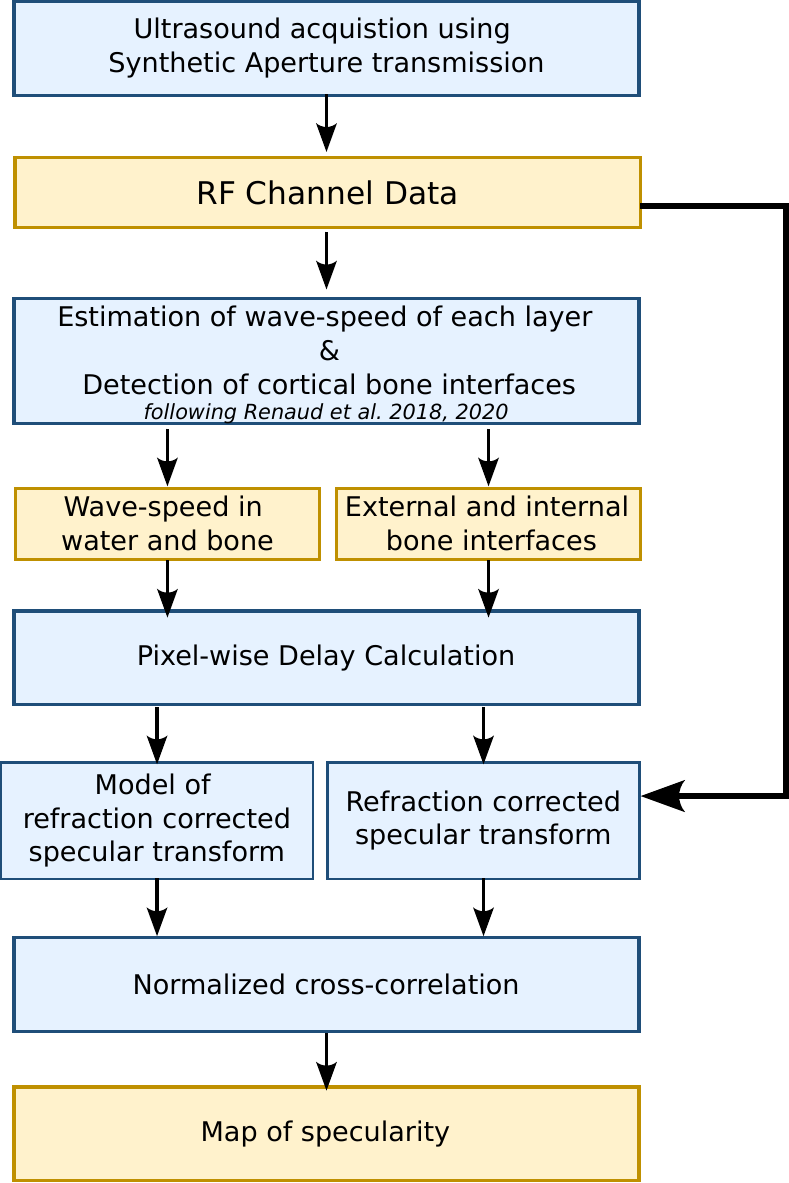}
\caption{Ultrasound processing pipeline to obtain the ultrasound specularity map.}
\label{fig:meth-flowchart}
\end{figure}

Large intracortical pores within the transition zone, together with the increased roughness of the endosteal interface, are expected to enhance ultrasound scattering and thereby reduce the amplitude of specular reflections. Therefore, the raw ultrasound radio-frequency data were processed to generate an image representing the specularity of the medium. An overview of the complete processing pipeline is provided in Figure~\ref{fig:meth-flowchart}.

We employed a specular transform algorithm similar to that described in \cite{rodriguez-molares_specular_2017}, modified to account for refraction at the soft tissue-bone interface.

\begin{figure}[ht!]
    \centering
    \includegraphics[width=.6\textwidth]{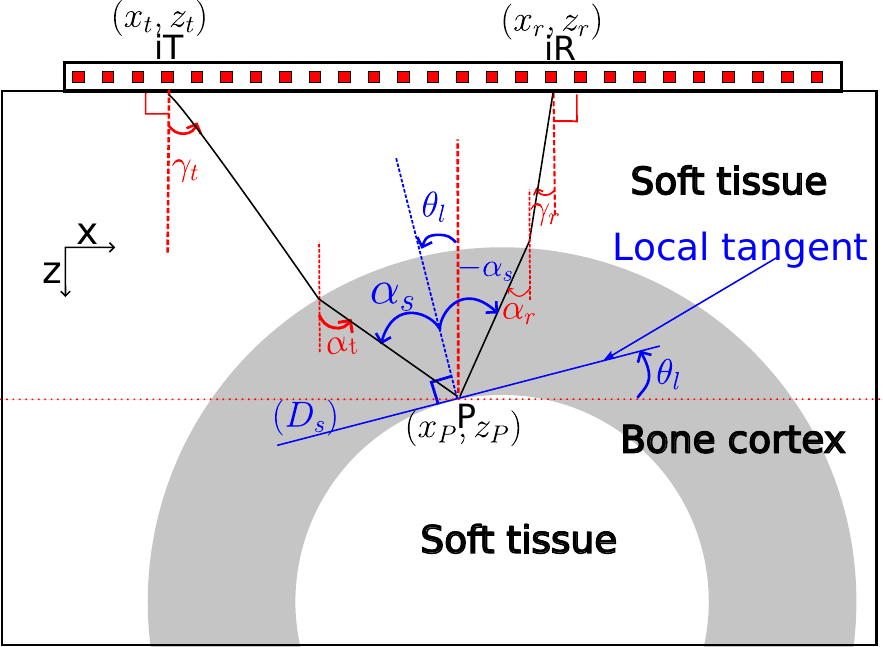}
    \caption{Illustration of ray tracing for the calculation of specular wave travel time for a point located at the internal (endosteal) bone surface. The figure shows the path of an ultrasound wave propagating from a probe array element, refracting at tissue interfaces according to Snell’s law, and undergoing specular reflection. $\gamma_r$ and $\gamma_t$ are the receive and emit angles at the elements of the probe, $\alpha_r $ and $\alpha_t $ are the receive and transmit angles at the pixel (after refraction). The angle of specular reflection for an hypothetical specular reflector oriented at angle $\theta_l$ is denoted $\alpha_s$ where $\alpha_s=\alpha_t-\theta_l$ and $\alpha_s=\theta_l-\alpha_r$. For this transmitter-receiver pair, the transmit-receive mid-angle defined as $\beta=\frac{\alpha_t+\alpha_r}{2}$ is equal to the local orientation of the reflector e.g. $\beta=\theta_l$.}
    \label{meth:refraction}
\end{figure}

\paragraph{Calculation of refraction-corrected delays and angles}
The first step involved calculating the round-trip travel time for each image point $P$, for each transmit beam, and for each receive element of the probe array. This calculation requires prior knowledge i)~of the geometry of the silicone front layer of the ultrasound probe and of the position of the bone external surface; ii)~the wave speed in the silicone, water and bone. These wave speeds, along with the thickness of the silicone front layer of the ultrasound probe, were determined in a separate experiment, as described in \cite{dia_ultrasound_2026}.
The travel times accounting for refraction at all interfaces were computed using a bending ray tracing technique based on Fermat’s principle of least travel time. This method was previously applied in vivo \cite{renaud_vivo_2018, renaud_measuring_2020}. 
A schematic of this process, illustrating the ray path and refraction at boundaries, is shown in Figure~\ref{meth:refraction}.

\paragraph{Specular transform $f(\beta,P)$}
The angles of incidence $\alpha_t$ and reflection $\alpha_r$ at each pixel $P$ were calculated for every transmitter–receiver pair (see Figure ~\ref{meth:refraction}). Then, following the approach outlined in \cite{rodriguez-molares_specular_2017}, the specular transform $f(\beta,P)$ at each pixel was obtained by coherently summing all signals originating from ray paths corresponding to a reflection on a same hypothetical specular reflector passing through $P$ but of unknown orientation. 
This is equivalent to summing all time-delayed signals that share the same transmit–receive mid-angle, denoted by $\beta=\frac{\alpha_t+\alpha_r}{2}$:

%
\begin{equation}
\label{chap5:eq:specular_transform}
f(\beta;P) = \sum_{i_T=1}^{N_T} 
S\big(\tau_{i_T,i_R}(P), \alpha_r, \alpha_t\big)
\quad \text{with $\frac{\alpha_r + \alpha_t}{2} = \beta$} .
\end{equation}

where $\tau_{iT,iR} $ is the round-trip time of flight from transmitter $iT $ to receiver $iR$.

\paragraph{Map of specularity $\Psi(P)$} 
Similarly to \cite{rodriguez-molares_specular_2017}, we 
computed a specularity index $\Psi(P)$ by cross-correlating the specular transform $f(\beta,P)$ with a model of perfect specular reflection. We extended the original one-layer model proposed in \cite{rodriguez-molares_specular_2017} into a multi-layer version that considers refraction and curvature of the interface. The model used the experimentally emitted tone burst and time delays corresponding each ray path. Further details are given in \cite{dia_refraction_2025}. Normalizing the cross-correlation by the maximum correlation of a perfect match yields values of $\Psi(P)$ between 0 and 1. Values close to 1 indicate that the signal is characteristic of a specular reflection, whereas values close to 0 indicate that the signal originates from highly scattered waves.

\paragraph{Endosteal specularity metrics}
The endosteal surface was segmented from the specularity $\Psi(P)$ maps using a Dijkstra-based shortest-path approach \cite{hong_medical_2012}. The graph search was restricted to the cortical layer mask and further constrained to pixels exhibiting high specularity values, ensuring that the extracted path followed the strong specular reflection associated with the endosteal interface. Edge weights were designed to promote spatial smoothness by penalizing abrupt depth variations. The resulting segmentation of the endosteal surface was subsequently fitted with a parabola.

An endosteal region of interest (ROI) was defined as follows around the fitted endosteal interface. The ROI was between two parabolas (translations of the fitted interface) separated by a certain ROI thickness. The ROI thickness was determined based on simulations results (without intracortical pores) by analyzing the average value of $N_{50}$ at each depth within cortical bone for all combinations of $R_q$ and $\rho_{length}$ (results not shown). The full width at half maximum (FWHM) of the resulting curve was on the order of one wavelength in bone, i.e. 1.4~mm at 2.5~MHz for a speed of sound of 3500~m/s, consistent with the resolution of the imaging system. This value was then used as the ROI thickness to process both simulation and experimental data.

Within this region, we used two metrics to quantify endosteal specularity: (i) the average value of $\Psi(P)$ (mean specularity index, $MSI$) and (ii) the \hl{proportion} of pixels with $\Psi(P)$ greater than a given threshold, $N_{th}$, where $th$ is the chosen threshold value in \%. 
For \exvivo~ experiments, the threshold was set at 40\% ($\Psi(P)>0.4$).
In simulations, 50\% ($\Psi(P)>0.5$) was used. The lower threshold in \exvivo~measurements was chosen to account for increased noise, which can reduce the observed specularity compared to idealized 2D simulations.

The metrics values presented in the rest of this article for the \exvivo~ data are the average values calculated from the 10 measurement repetitions in each measurement zone. 

\begin{figure}[ht!]
    \begin{subfigure}{.45\linewidth}
        \includegraphics[height=.55\linewidth,trim={0 0 0 40},clip]{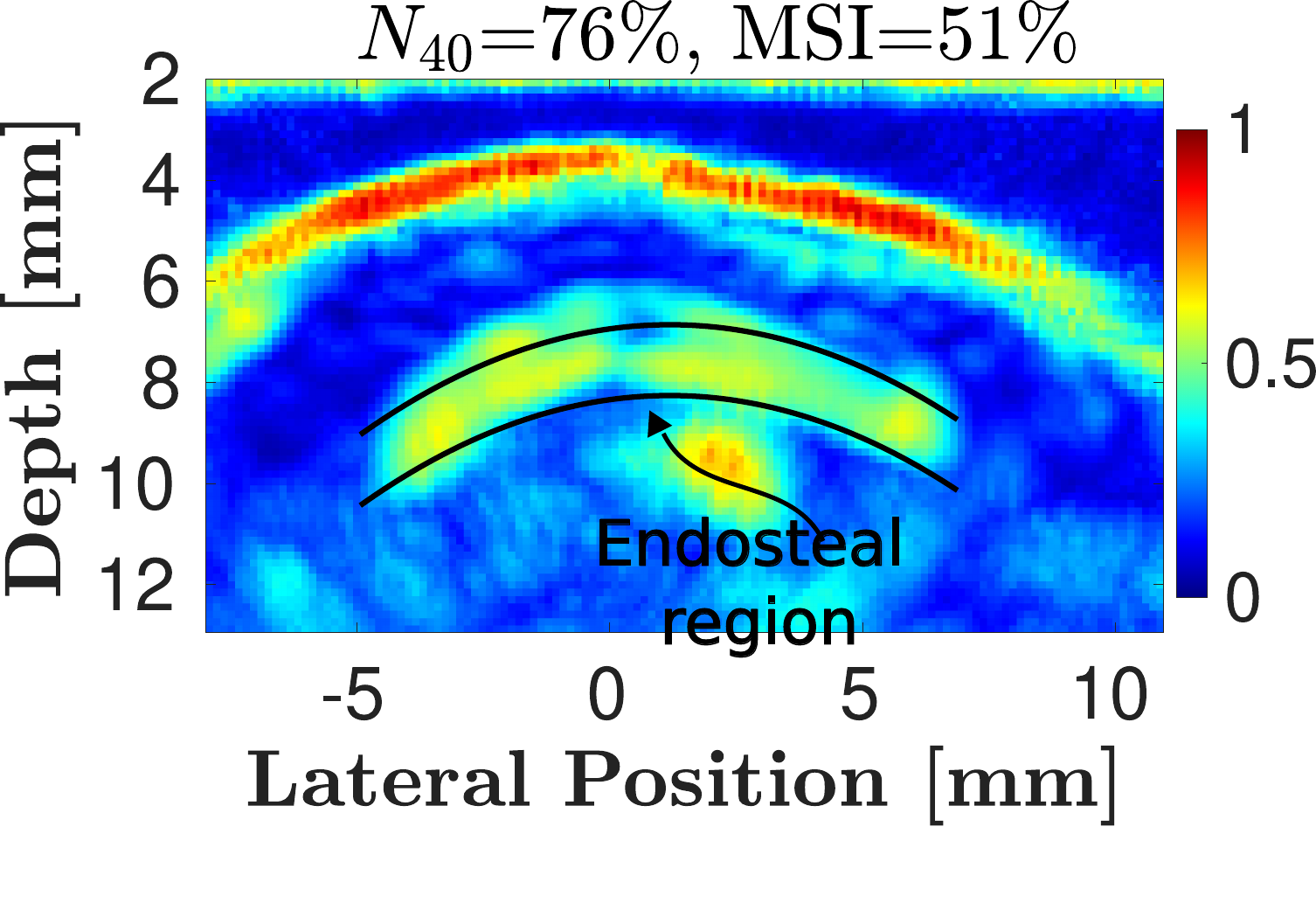}
        \caption{}
    \end{subfigure}
    \begin{subfigure}{.45\linewidth}
        \includegraphics[height=.55\linewidth,trim={0 0 0 28},clip]{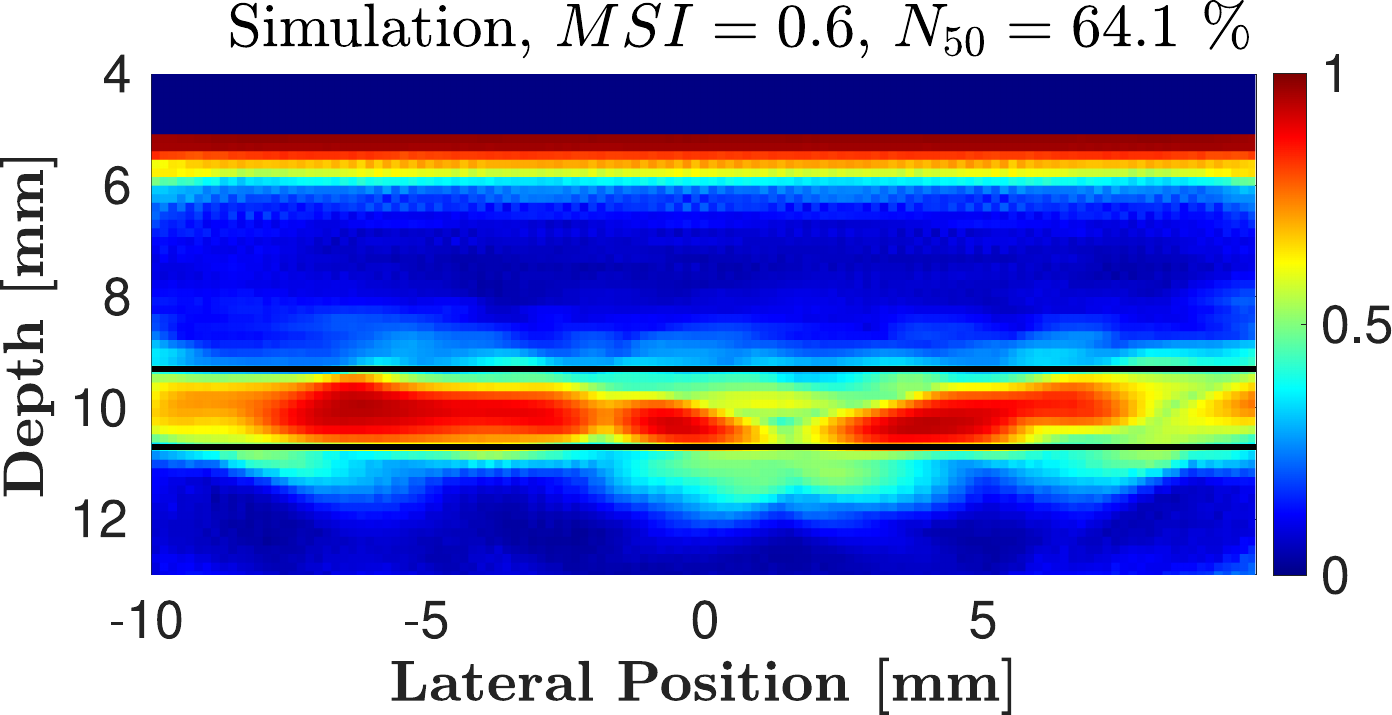}
        \caption{}
    \end{subfigure}
     
    \caption{Maps of specularity $\Psi(P)$ for (a) one representative \exvivo~ measurement and (b) one simulation result.  The ROI for the calculation of specularity metrics is defined in between the black lines.}
    \label{meth:enndosteal_region_specularity}
\end{figure}

\subsection{Data analysis}
The aim of this study was to assess the impact of endosteal microstructure (rms height $R_q$, correlation length $\rho_{\text{length}}$, cortical porosity $E.Por$, and endosteal pore diameter $E.Po.Dm$) on specularity metrics ($MSI$, $N_{th}$).
To this end, we first conducted univariate analyses by computing Spearman rank correlation coefficients between each microstructure parameter and specularity metrics. Correlations were considered statistically significant at $p < 0.05$.
We then fitted simple linear regression models separately for each microstructure parameter to evaluate their individual ability to predict specularity metrics.
Model performance was assessed using the adjusted coefficient of determination ($\text{Adj}-R^2$) and the root mean square error (RMSE). Parameters with the largest adjusted coefficients in the final models were identified as having the dominant influence on the specularity metrics.

\section{Results}
\subsection{Simulation results}
\begin{figure}[ht!]
    \centering
    \setlength\tabcolsep{0pt}

        \begin{tabularx}{\linewidth}{m{.09\linewidth}|m{.3\linewidth}|m{.3\linewidth}|m{.3\linewidth}}
        $\rho_{\text{length}}$ & \centering\arraybackslash $Rq \sim 30~$\textmu m & \centering\arraybackslash $Rq \sim 300~$\textmu m & \centering\arraybackslash $Rq \sim 500~$\textmu m \\
        \toprule
    \multirow{2}{*}{$0.5~$mm}
 &\begin{subfigure}{.3\textwidth}
        \includegraphics[width=\linewidth]{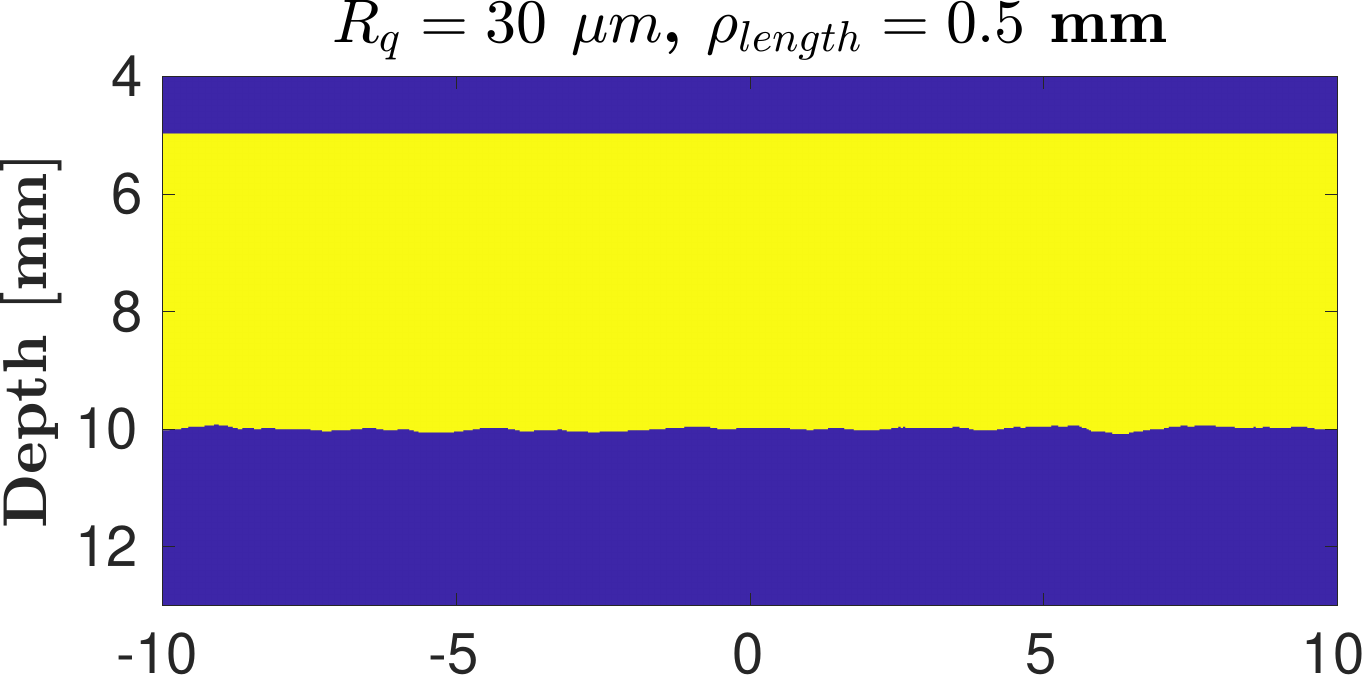}
        \caption{}
    \end{subfigure}
 &\begin{subfigure}{.3\textwidth}
        \includegraphics[width=\linewidth]{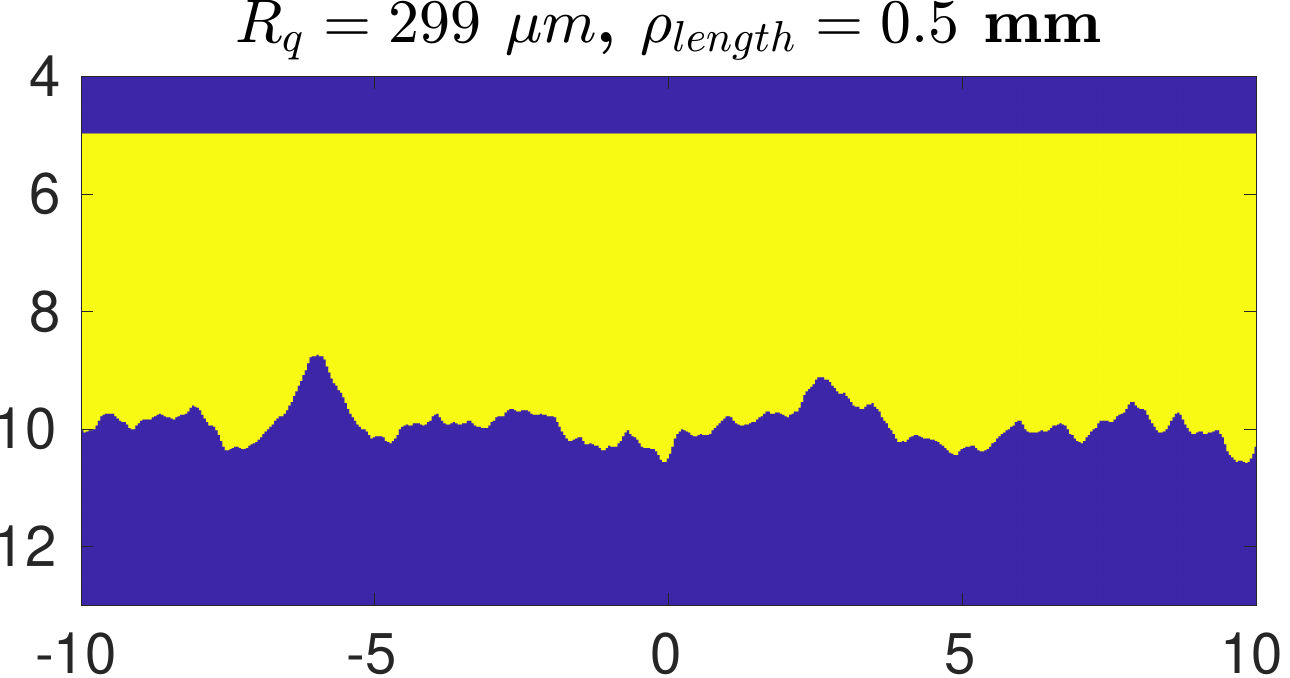}
        \caption{}
    \end{subfigure}
 &\begin{subfigure}{.3\textwidth}
        \includegraphics[width=\linewidth]{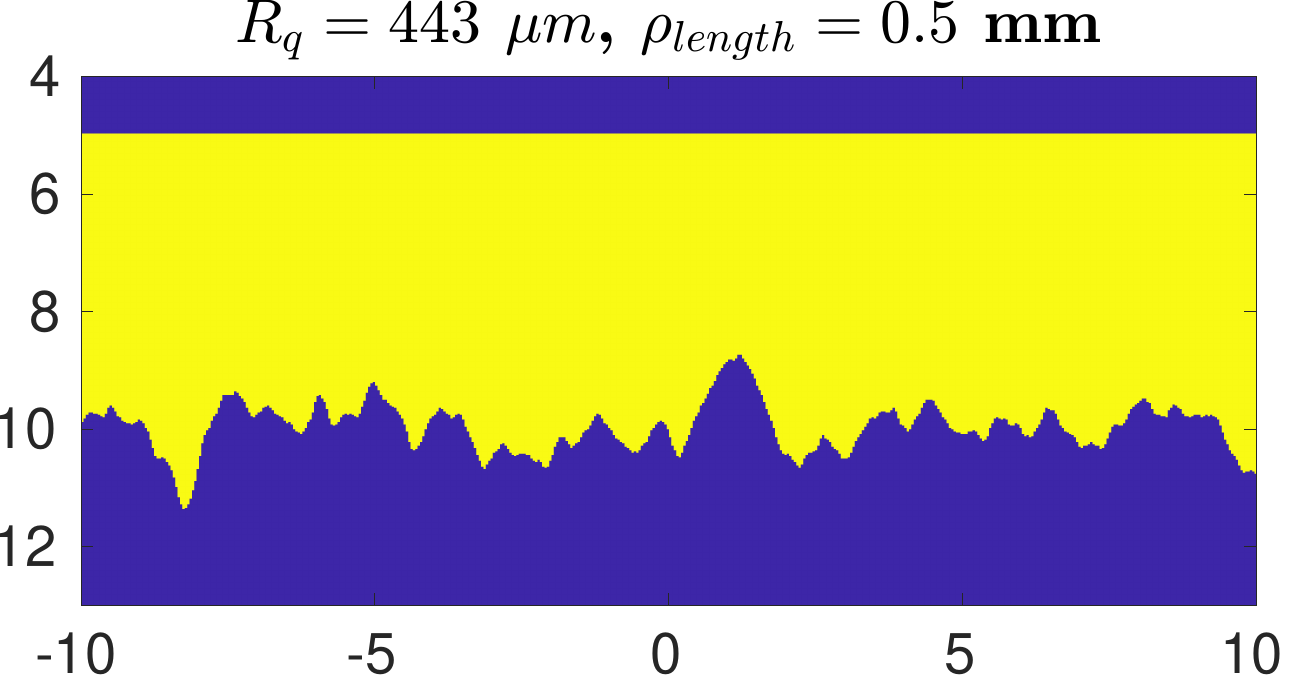}
        \caption{}
    \end{subfigure}
    \\
 &\begin{subfigure}{.3\textwidth}
        \includegraphics[width=\linewidth]{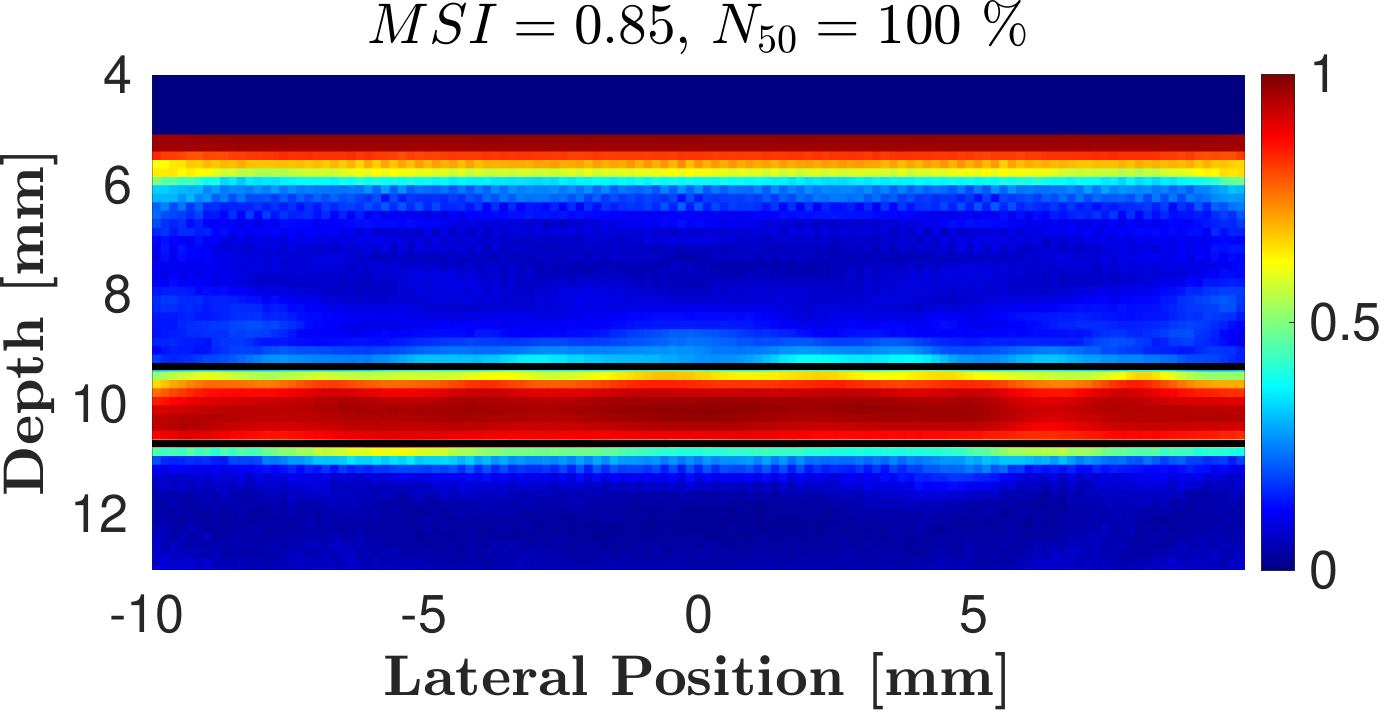}
        \caption{}
    \end{subfigure}
 &\begin{subfigure}{.3\textwidth}
        \includegraphics[width=\linewidth]{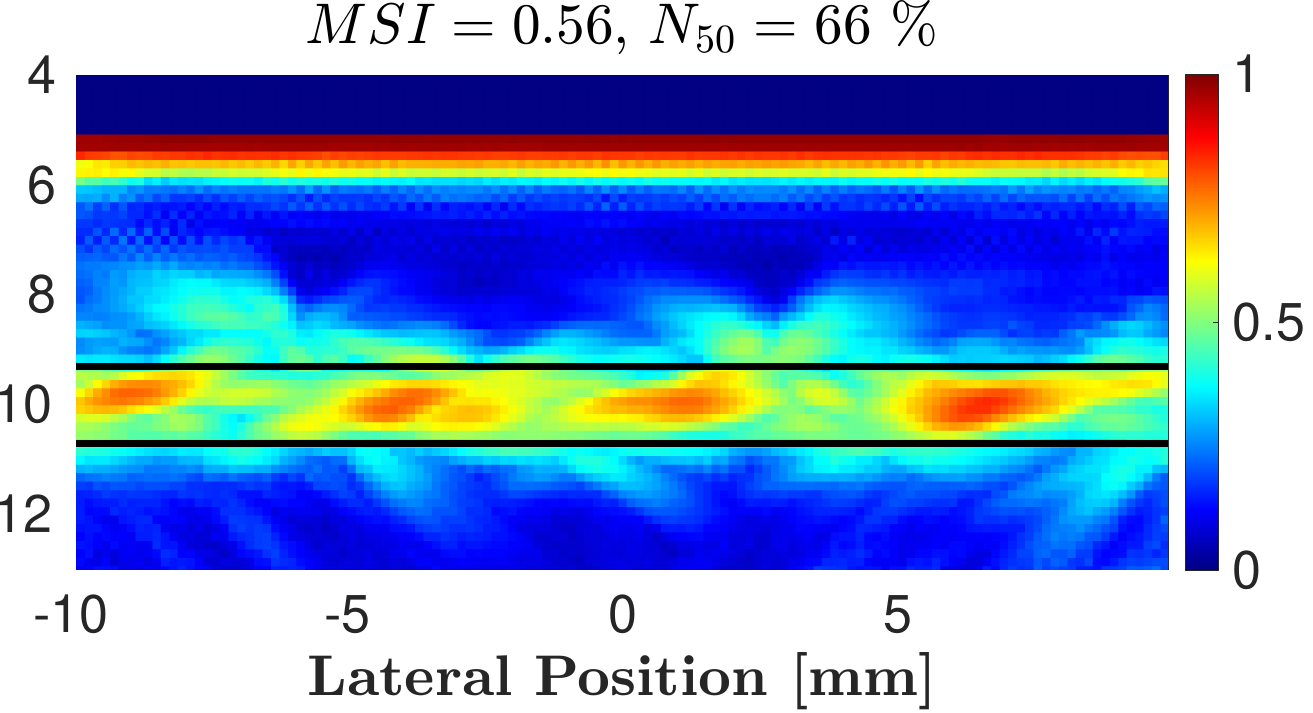}
        \caption{}
    \end{subfigure}
 &\begin{subfigure}{.3\textwidth}
        \includegraphics[width=\linewidth]{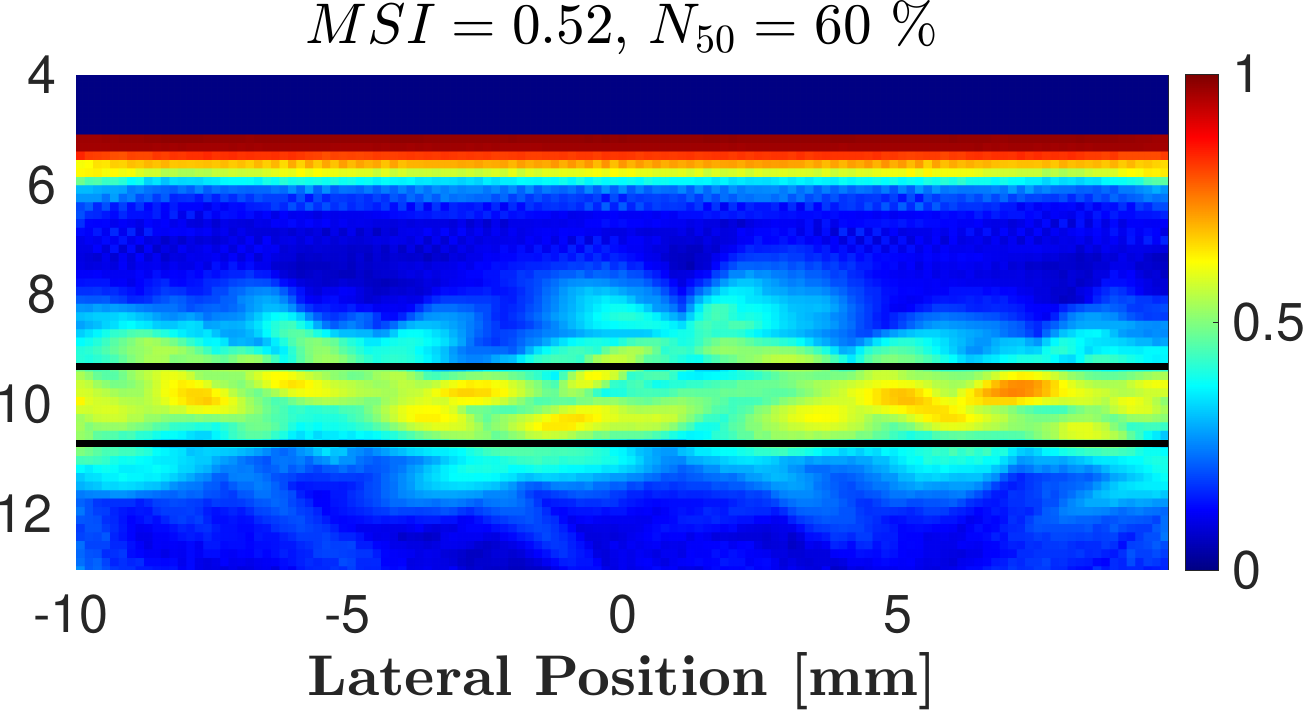}
        \caption{}
    \end{subfigure}
    \\
    \bottomrule
    \end{tabularx}

    \begin{tabularx}{\linewidth}{m{.09\linewidth}|m{.3\linewidth}|m{.3\linewidth}|m{.3\linewidth}}    
    \toprule
    \multirow{2}{*}{$2~$mm}
 &\begin{subfigure}{.3\textwidth}
        \includegraphics[width=\linewidth]{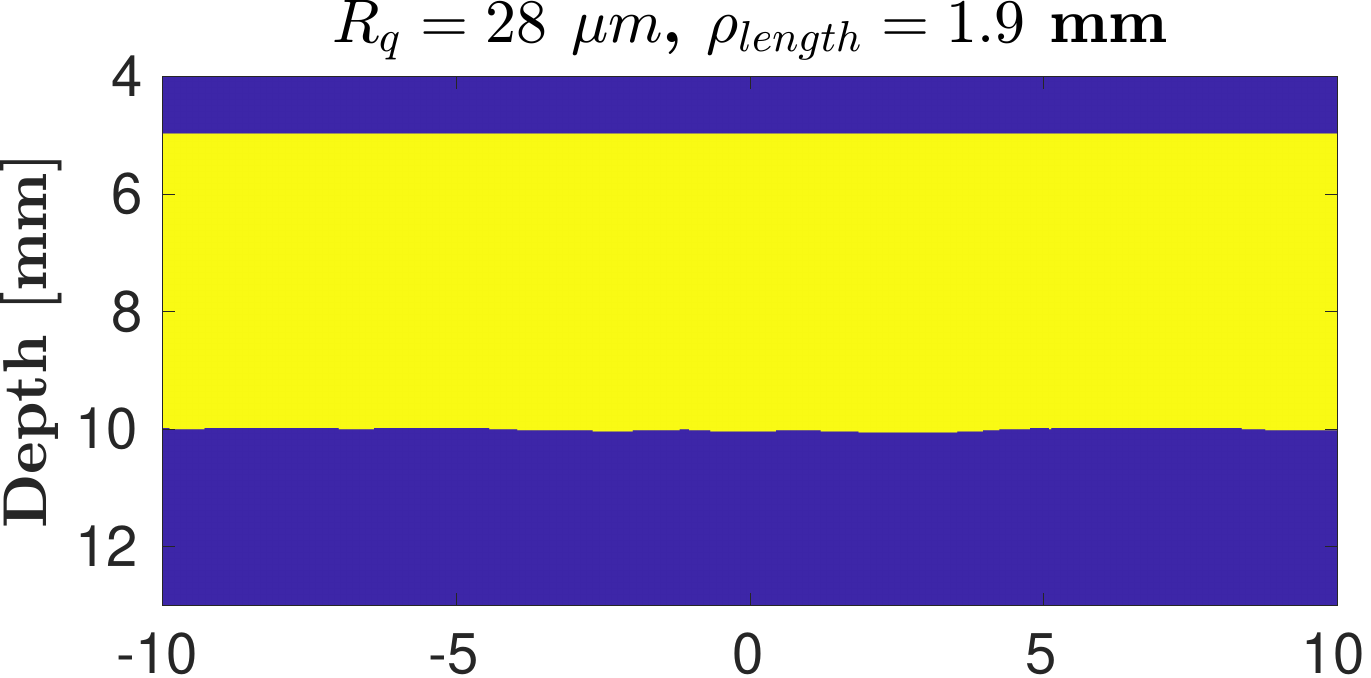}
        \caption{}
    \end{subfigure}
 &\begin{subfigure}{.3\textwidth}
        \includegraphics[width=\linewidth]{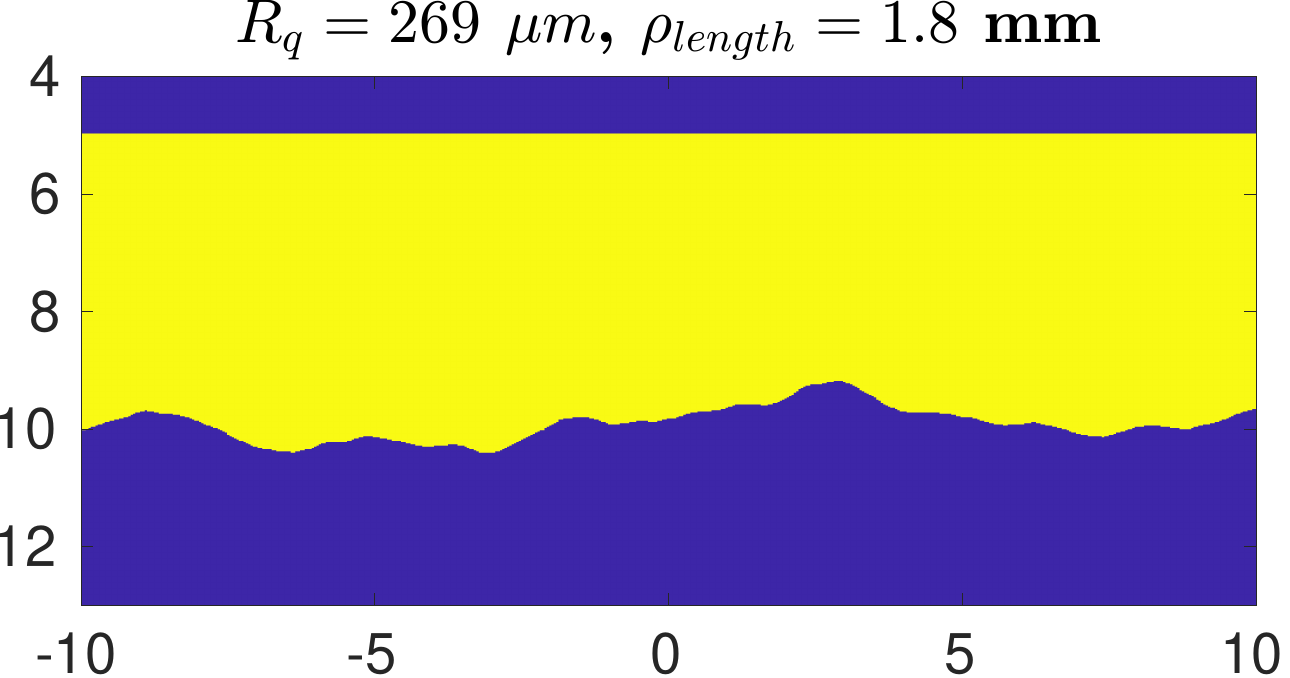}
        \caption{}
    \end{subfigure}
 &\begin{subfigure}{.3\textwidth}
        \includegraphics[width=\linewidth]{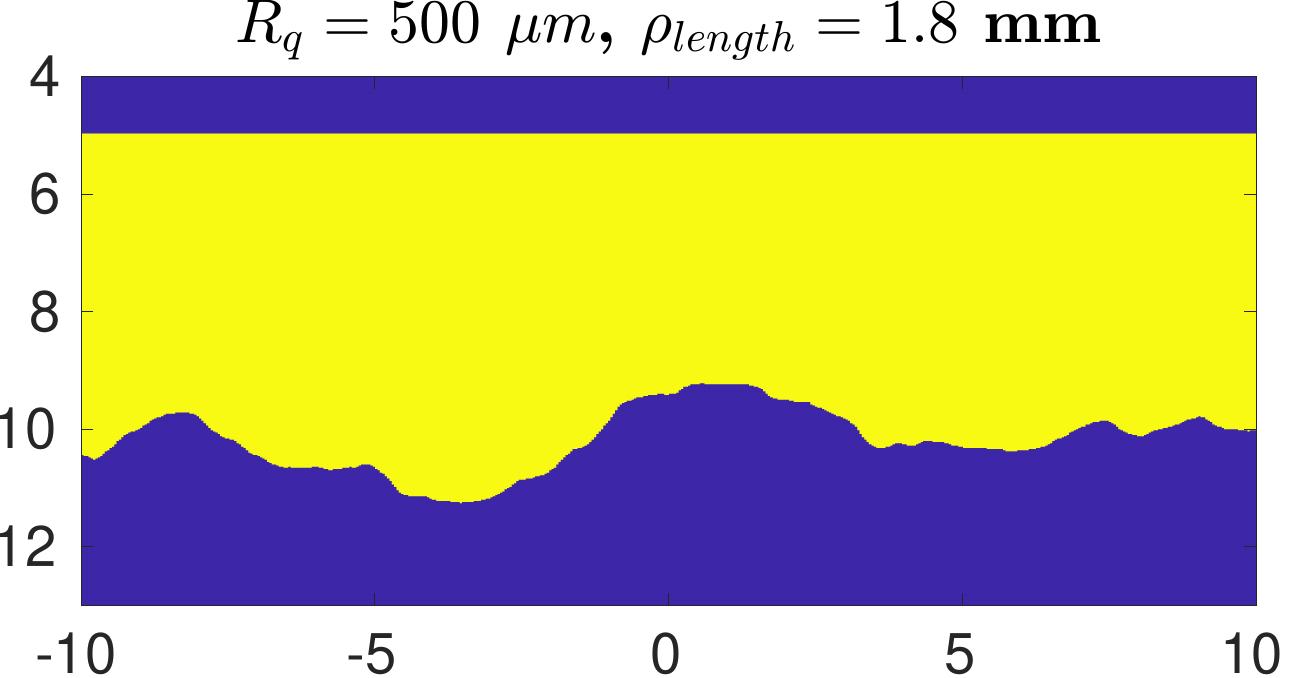}
        \caption{}
    \end{subfigure}
    \\
 &\begin{subfigure}{.3\textwidth}
        \includegraphics[width=\linewidth]{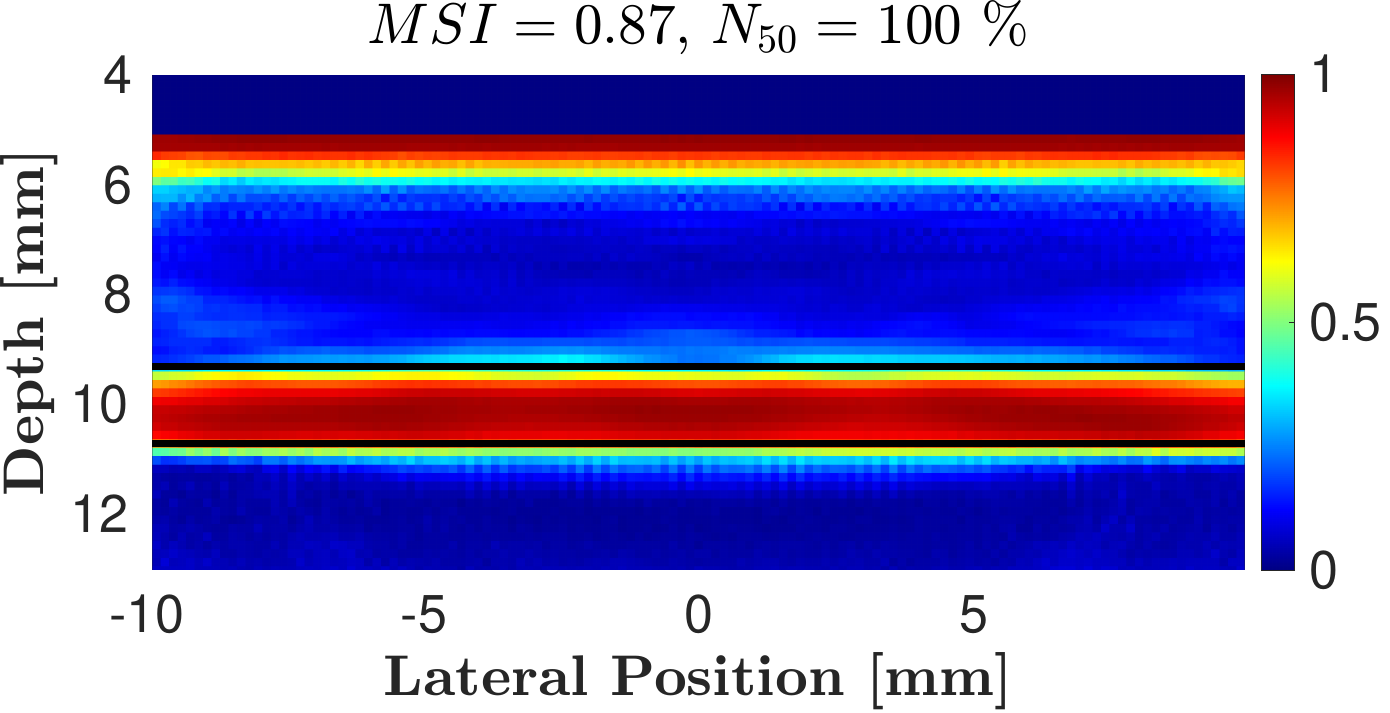}
        \caption{}
    \end{subfigure}
 &\begin{subfigure}{.3\textwidth}
        \includegraphics[width=\linewidth]{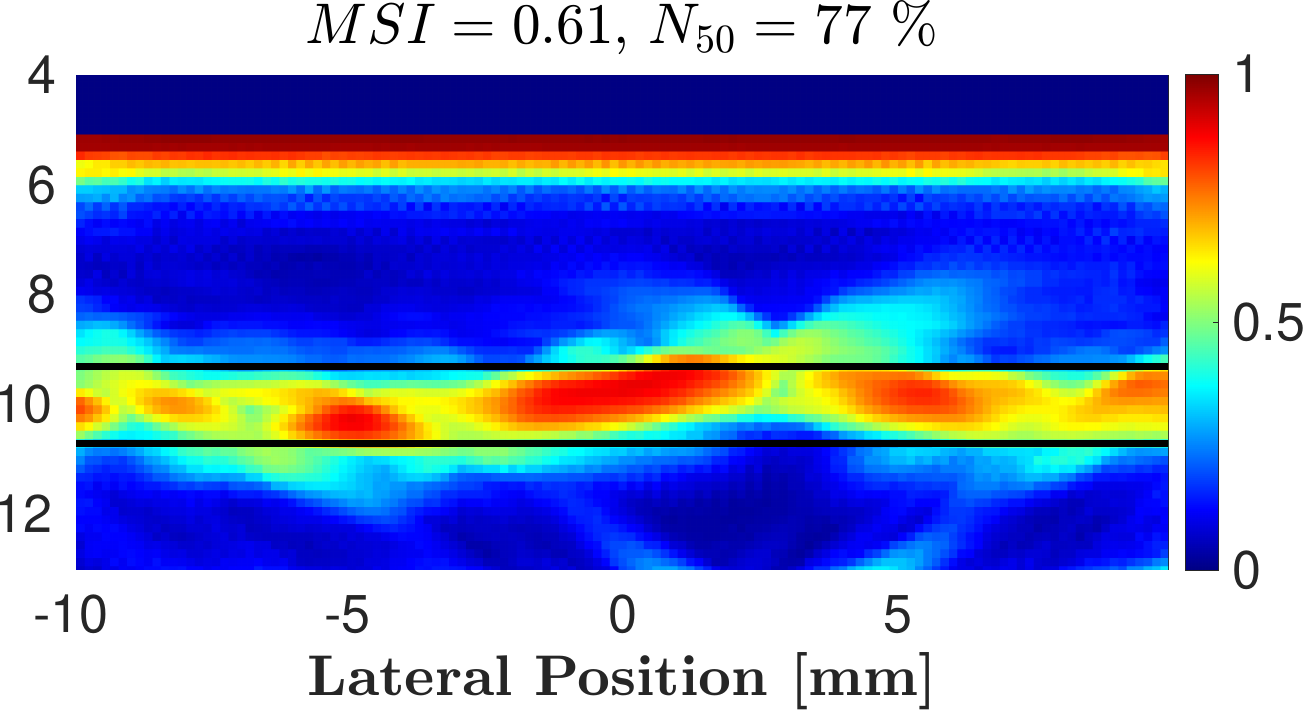}
        \caption{}
    \end{subfigure}
 &\begin{subfigure}{.3\textwidth}
        \includegraphics[width=\linewidth]{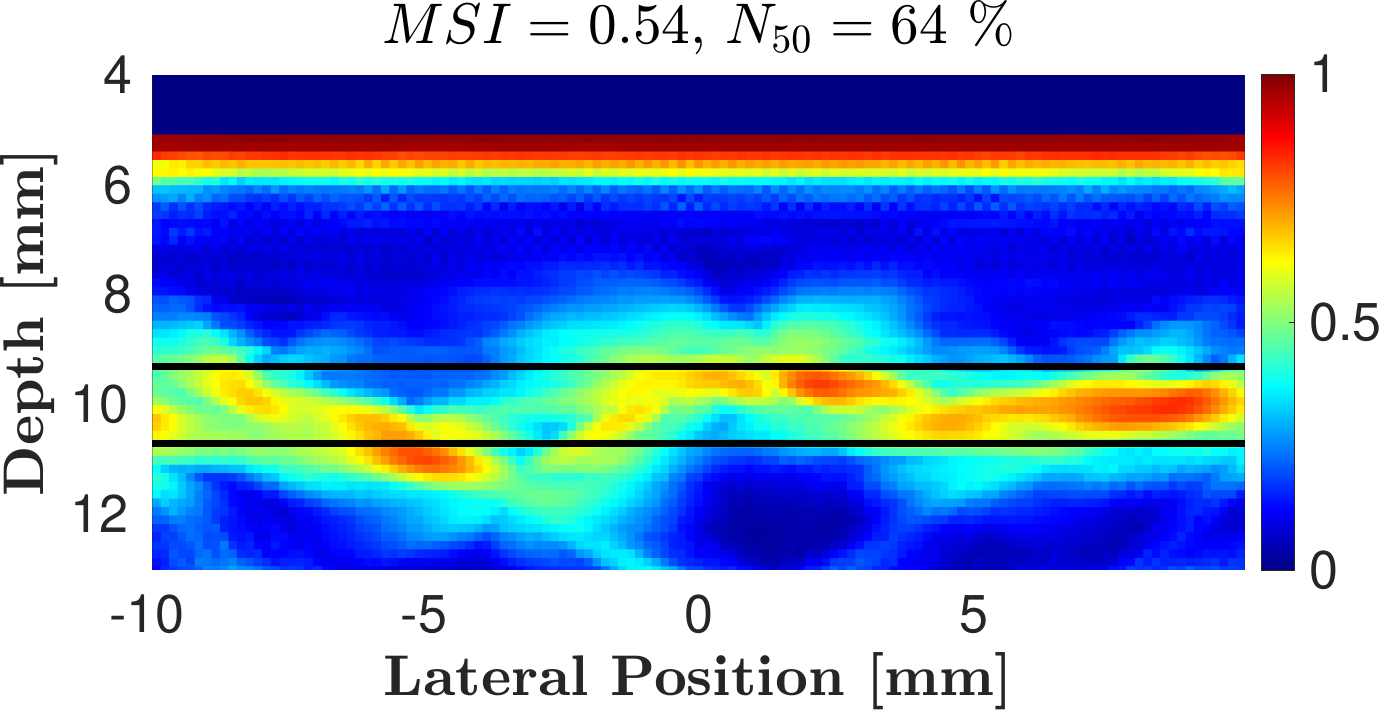}
        \caption{}
    \end{subfigure}     
    \\
    \end{tabularx}

    \caption{Examples of simulated specularity maps for different surface roughness parameters. Panels (a)–(f) correspond to a fixed correlation length $\rho_{\text{length}} = 0.5$~mm with increasing rms height of $Rq = 30~$\textmu m, $300~$\textmu m, and $500~$\textmu m. Panels (g)–(l) show similar rms values for $\rho_{\text{length}} = 2$~mm. The first row in each block (panels (a)–(c) and (g)–(i)) displays the simulated surface configurations, while the second row (panels (d)–(f) and (j)–(l)) shows the corresponding specularity maps.}

    \label{fig:simul_specu_maps}
\end{figure}

Examples of simulated specularity maps for different roughness parameters are shown in Figure~\ref{fig:simul_specu_maps}. As expected, the specularity is close to zero inside bone and reaches its highest values at flat interfaces, including the periosteal surface. The specularity at the endosteal surface is strongly impacted by the properties of the interface and the shape of the interface revealed by the specularity maps matches the gross morphology of the profiles used for the simulations.
For the smallest rms height, $R_q = 30~$\textmu m, the specularity of the endosteal region is high and spatially uniform across the surface. This is quantified by the specularity metrics, with $MSI$ values of 85\% and 87\%, and all pixels in the endosteal region classified as specular ($N_{50} = 100\%$) for both correlation lengths (see panels (d) and (j)).
For $R_q = 300~$\textmu m, the specularity metrics are lower ($MSI$ = 56\% and 61\%; $N_{50} = 66\%$ and $77\%$), and specularity maps become less homogeneous. A similar trend is observed for $R_q = 500$~\textmu m.
Increasing the correlation length from 0.5~mm to 2~mm leads to higher specularity. For instance, more specular pixels are visible in panels (k) and (l) than in panels (e) and (f). The effect is also quantitatively reflected in the specularity metrics. For example, at $R_q = 300~$\textmu m, the $MSI$ and $N_{50}$ values are 56\% and 66\%, respectively, for a correlation length of 0.5~mm, whereas they increase to 61\% and 77\% for a correlation length of 2~mm.

\begin{figure}[ht!]
    \begin{subfigure}{.495\textwidth}
            \includegraphics[width=\linewidth]{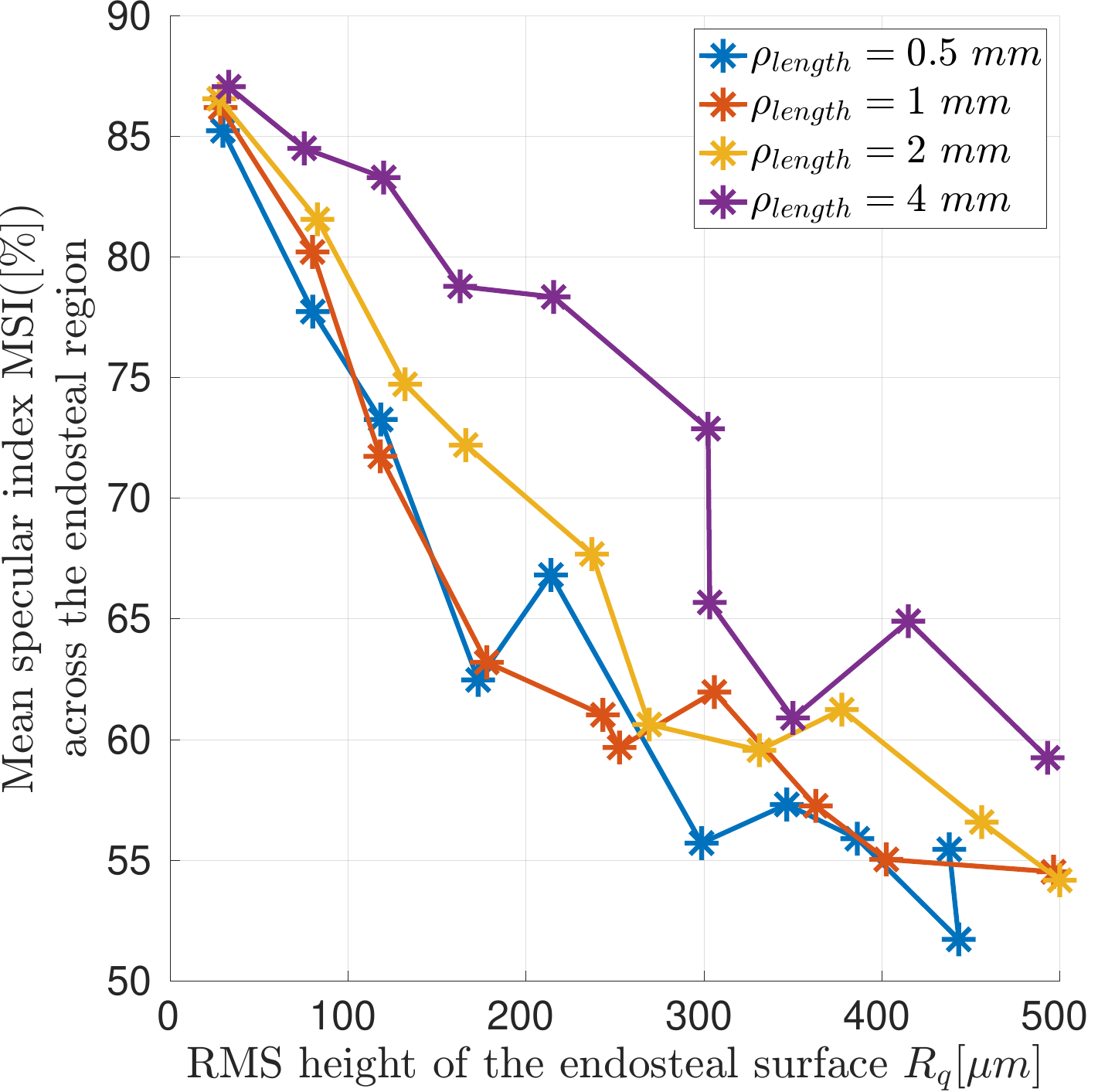}
    \caption{}
    \end{subfigure}
    \begin{subfigure}{.495\textwidth}
            \includegraphics[width=\linewidth]{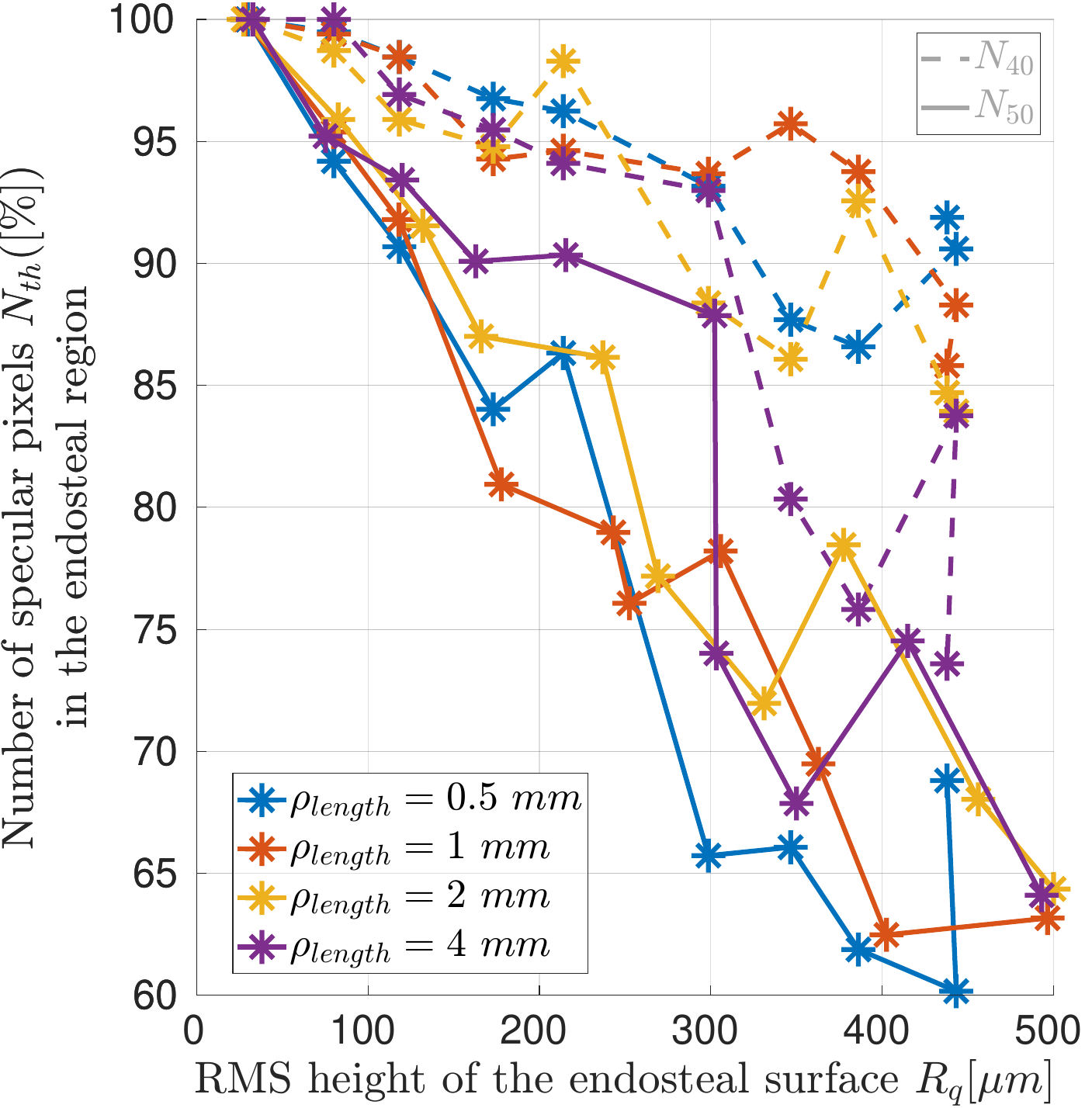}
            \caption{}
    \end{subfigure}    
    \caption{Simulation results showing the evolution of (a) the Mean Specular Intensity ($MSI$) and (b) the number of specular pixels ($N_{50}$ in plain line and $N_{40}$ in dashed line) in the endosteal region as functions of rms height $R_q$ of the endosteal surface, for different correlation lengths $\rho_{\text{length}}$.}
    \label{fig:simul_result_plot_0}
\end{figure}

In figure~\ref{fig:simul_result_plot_0}, the specularity metrics for all simulations are plotted as a function of $R_q$ for different correlation lengths. 
For all tested correlation lengths, there was a marked decrease in both $MSI$ and $N_{50}$ as $R_q$ increased. 
For a given $R_q$, $MSI$ and $N_{50}$ tended to be slightly higher for larger correlation lengths.

Table~\ref{table:TableAdjustedR2Simul} gives the correlations between surface roughness and specularity metrics. Strong negative correlations were observed between $R_q$ and both specularity metrics across all correlation lengths, with correlation coefficients ($\rho$) exceeding -0.9 in every case (p$<$0.0001). 

\begin{table}[ht!]
	\centering
	\begin{tabular}{|c|c|c|c|c|c|c|}
    \hline
 &\multicolumn{3}{c|}{$MSI$ vs $R_q$}&\multicolumn{3}{c|}{$N_{50}$ vs $R_q$}\\
		$\rho_{\text{length}}$(mm) & $\rho$ & Adj-R$^2$ & RMSE \hl{(\%)}& $\rho$ & Adj-R$^2$ & RMSE (\%) \\
\hline
		$0.5$ & -0.94$^{***}$ & 0.87$^{***}$ & 4$^{}$ & -0.95$^{***}$ & 0.90$^{***}$ & 5$^{}$ \\
		$1$ & -0.91$^{***}$ & 0.80$^{***}$ & 5$^{}$ & -0.97$^{**}$ & 0.93$^{**}$ & 3$^{}$ \\
		$2$ & -0.96$^{***}$ & 0.91$^{***}$ & 3$^{}$ & -0.97$^{***}$ & 0.93$^{***}$ & 3$^{}$ \\
		$4$ & -0.96$^{**}$ & 0.91$^{**}$ & 3$^{}$ & -0.93$^{***}$ & 0.84$^{***}$ & 5$^{}$ \\       
\hline
	\end{tabular}

    \caption{Correlations for simulation data. Correlation coefficient ($\rho$), adjusted R-squared (Adj-R$^2$), and root mean square error (RMSE) characterising the relationship between surface roughness ($R_q$) of the endosteal interface and ultrasound specularity metrics ($MSI$ and $N_{50}$), for different correlation lengths ($\rho_{\text{length}}$). Statistical significance: $^*p~<$~0.05; $^{**}p~<$~0.001; $^{***}p~<$~0.0001; n.s.: not significant.}
	
	\label{table:TableAdjustedR2Simul}
\end{table}


\subsection{Ex-vivo results}

\begin{figure}[ht!]
    \begin{subfigure}{.24\linewidth}
        \caption{}
        \includegraphics[width=\linewidth]{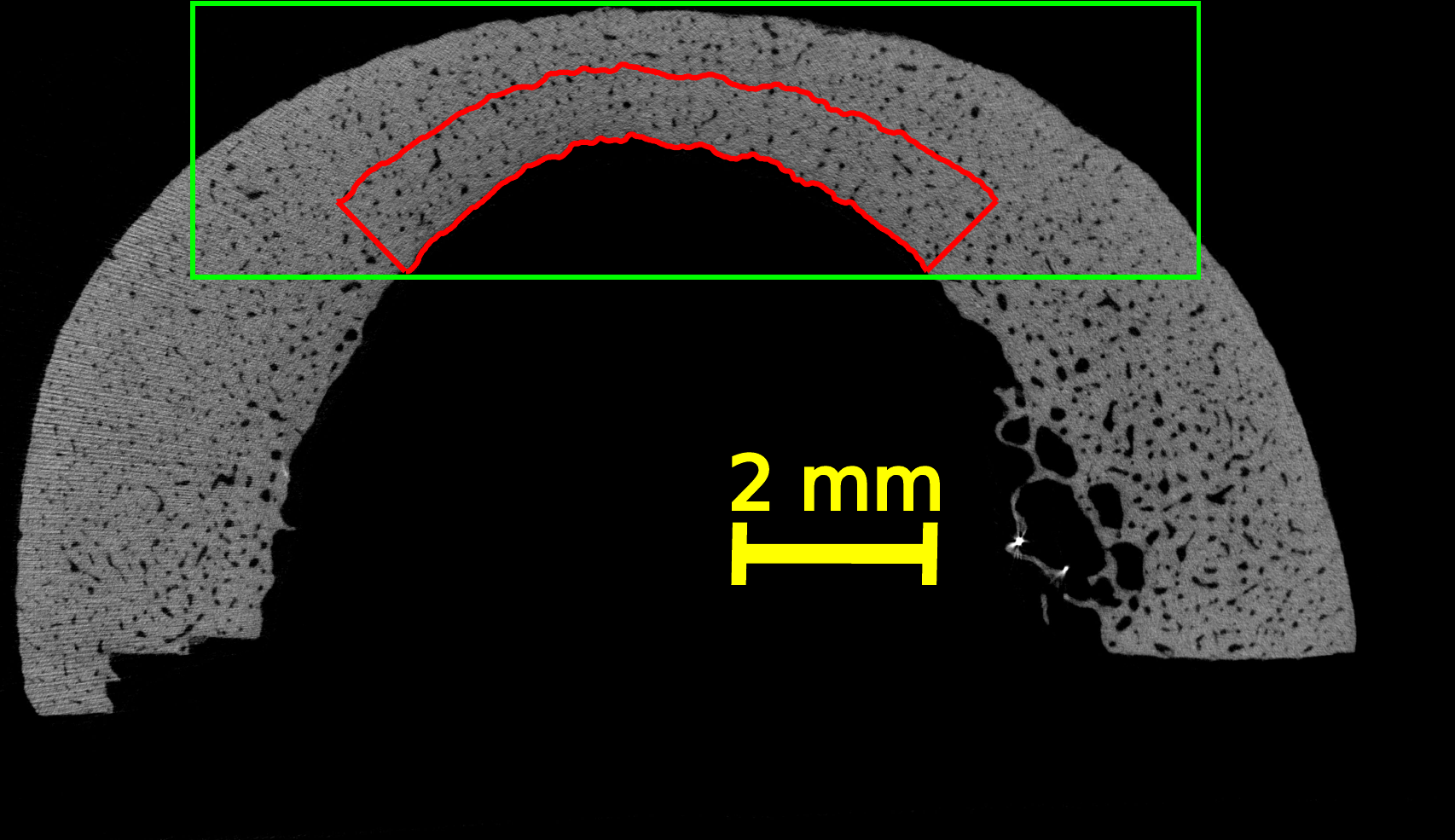}
    \end{subfigure}
    \begin{subfigure}{.24\linewidth}
        \caption{}
        \includegraphics[width=\linewidth]{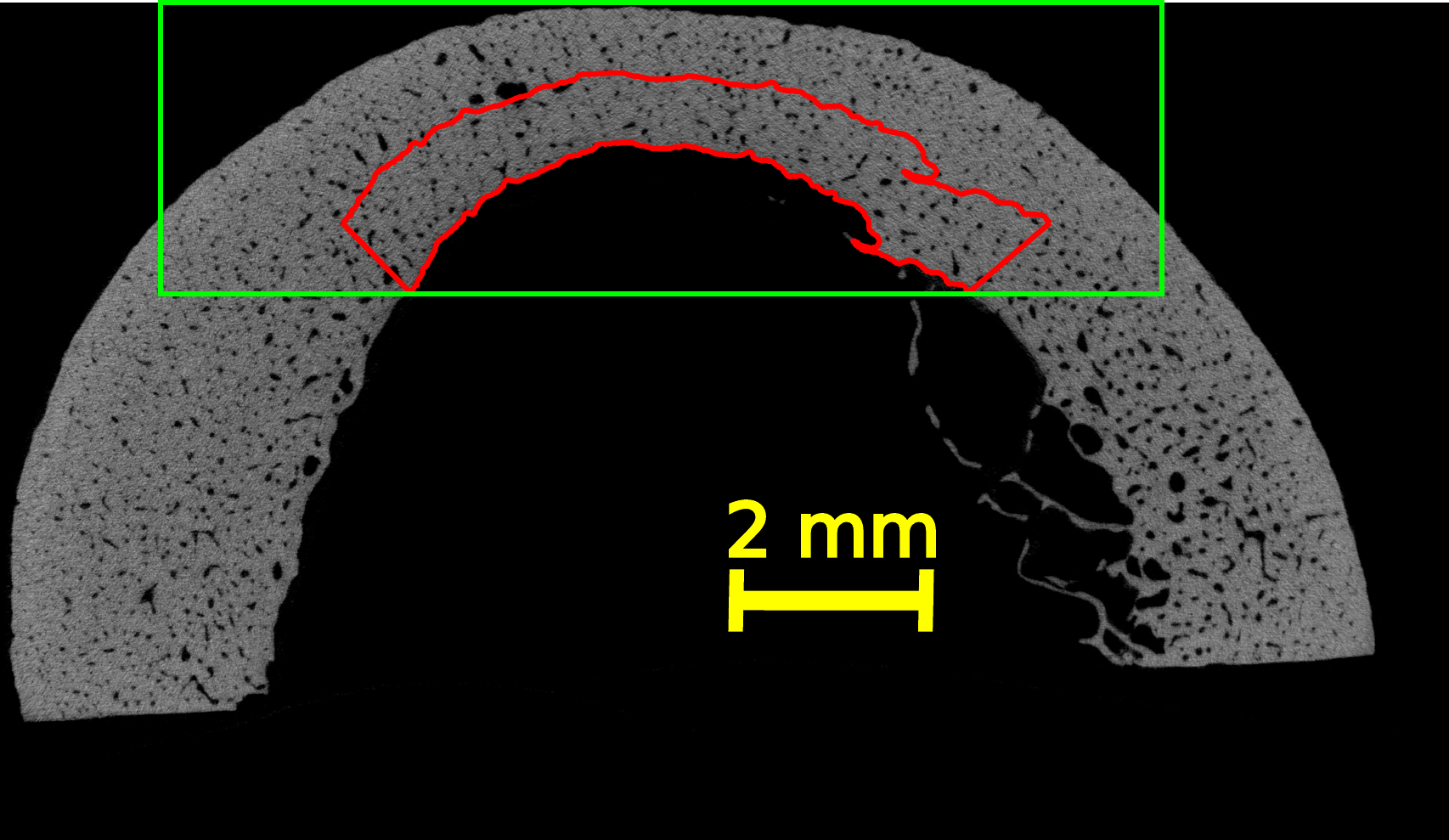}
    \end{subfigure}
    \begin{subfigure}{.24\linewidth}
        \caption{}
        \includegraphics[width=\linewidth]{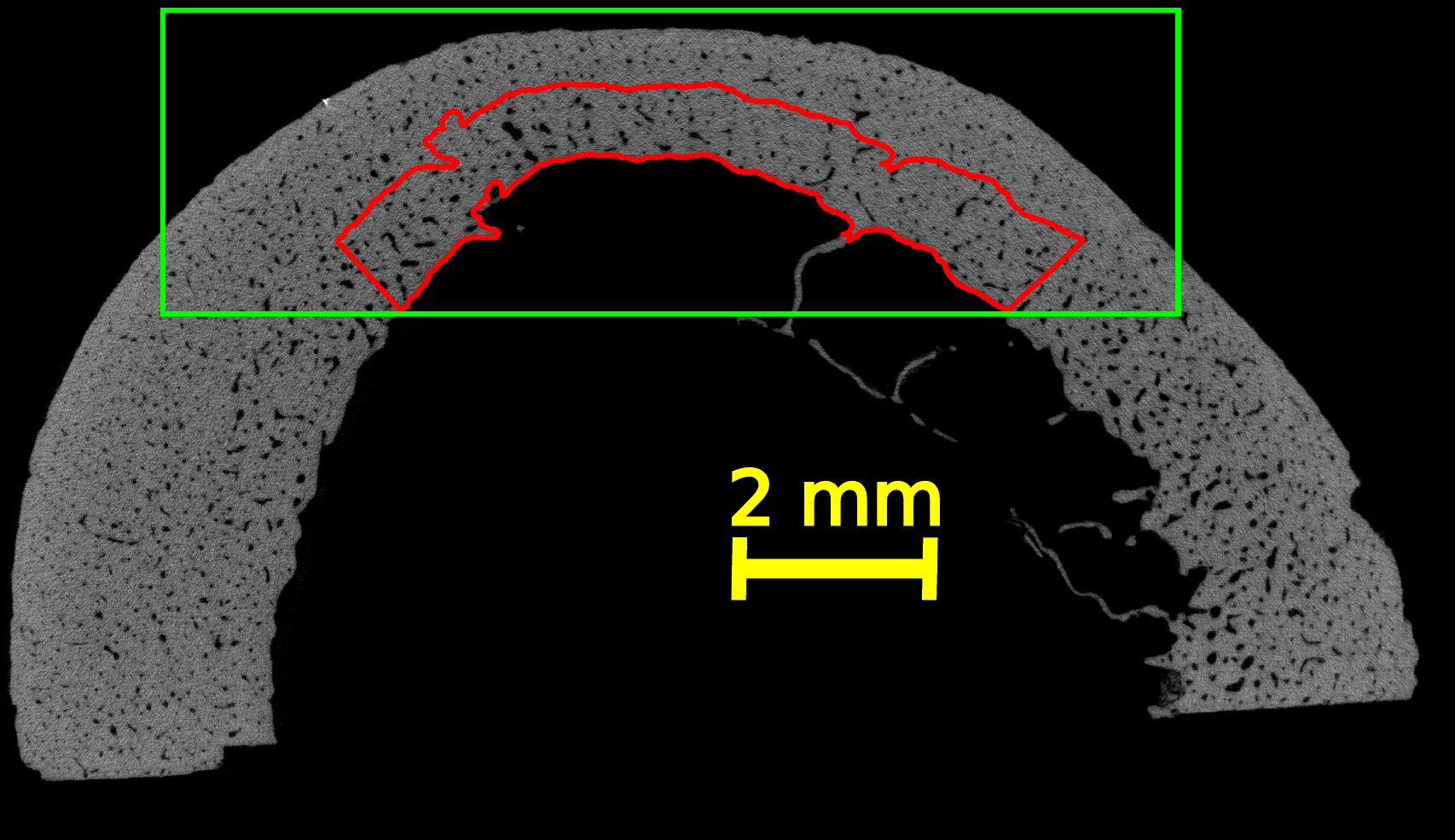}
    \end{subfigure}
    \begin{subfigure}{.24\linewidth}
        \caption{}
        \includegraphics[width=\linewidth]{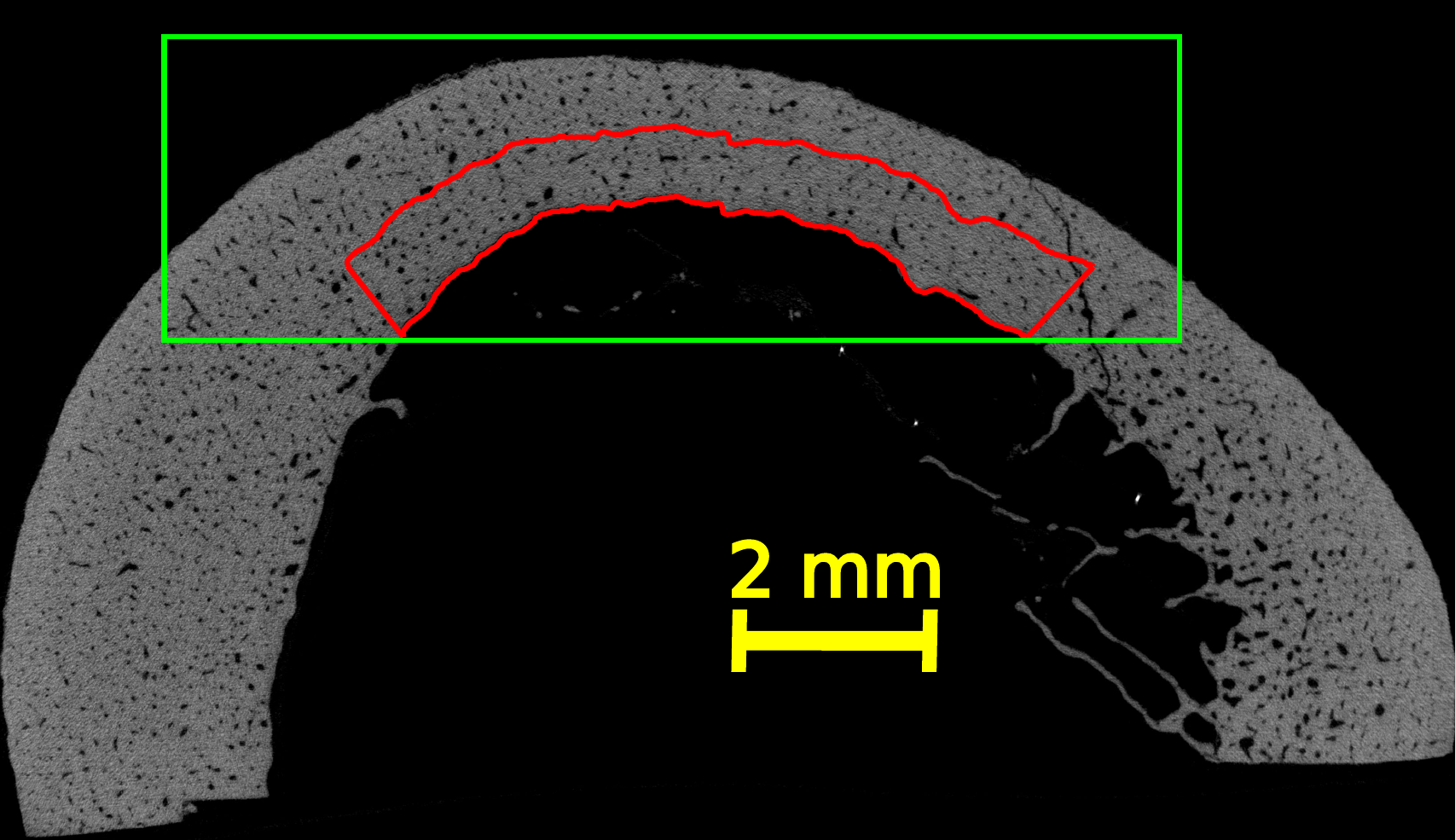}
    \end{subfigure}

    \begin{subfigure}{.23\linewidth}
        \caption{}
        \includegraphics[width=\linewidth]{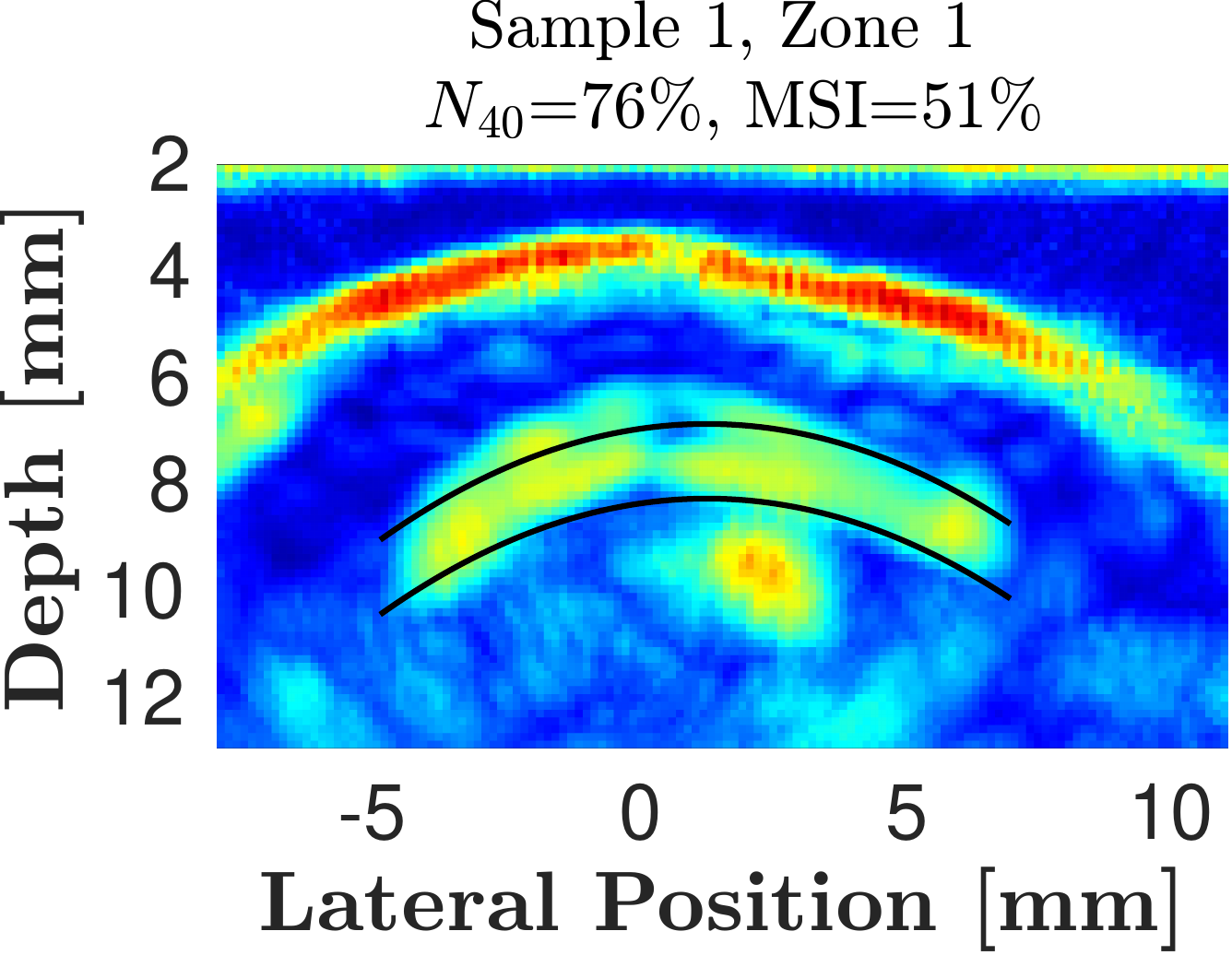}        
    \end{subfigure}
    \begin{subfigure}{.23\linewidth}
        \caption{}
        \includegraphics[width=\linewidth]{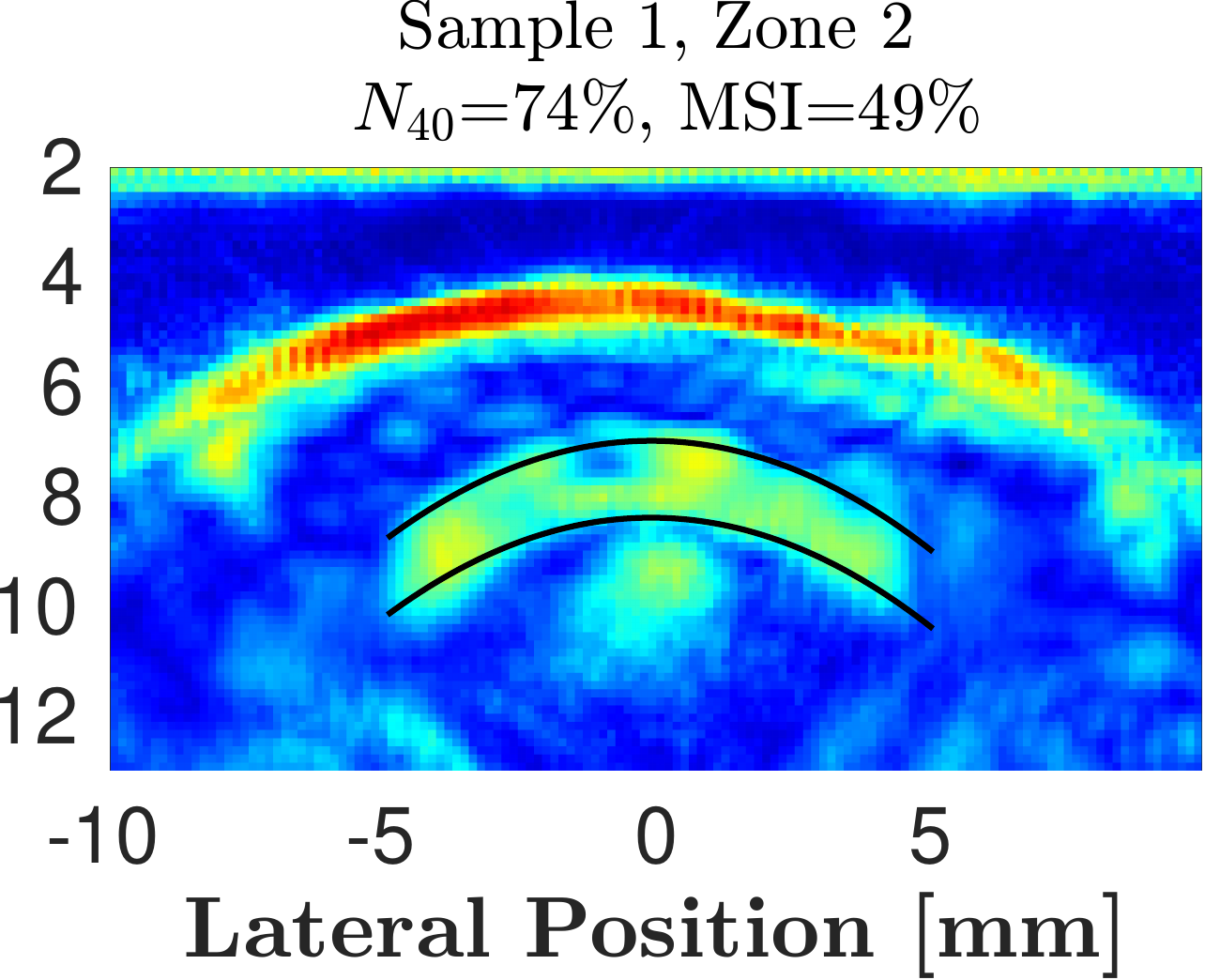}        
    \end{subfigure}
    \begin{subfigure}{.23\linewidth}
        \caption{}
        \includegraphics[width=\linewidth]{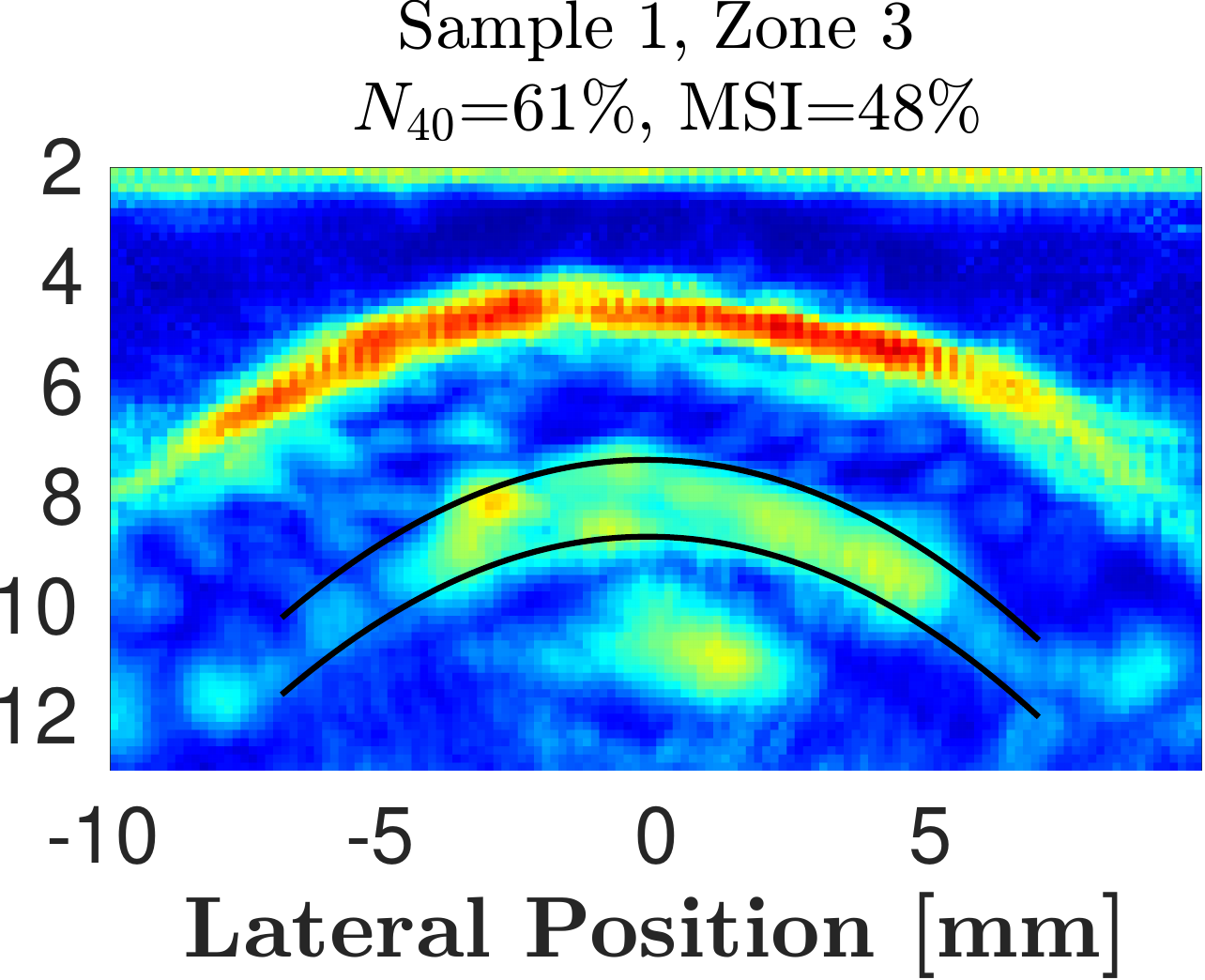}        
    \end{subfigure}
    \begin{subfigure}{.23\linewidth}
        \caption{}
        \includegraphics[width=\linewidth]{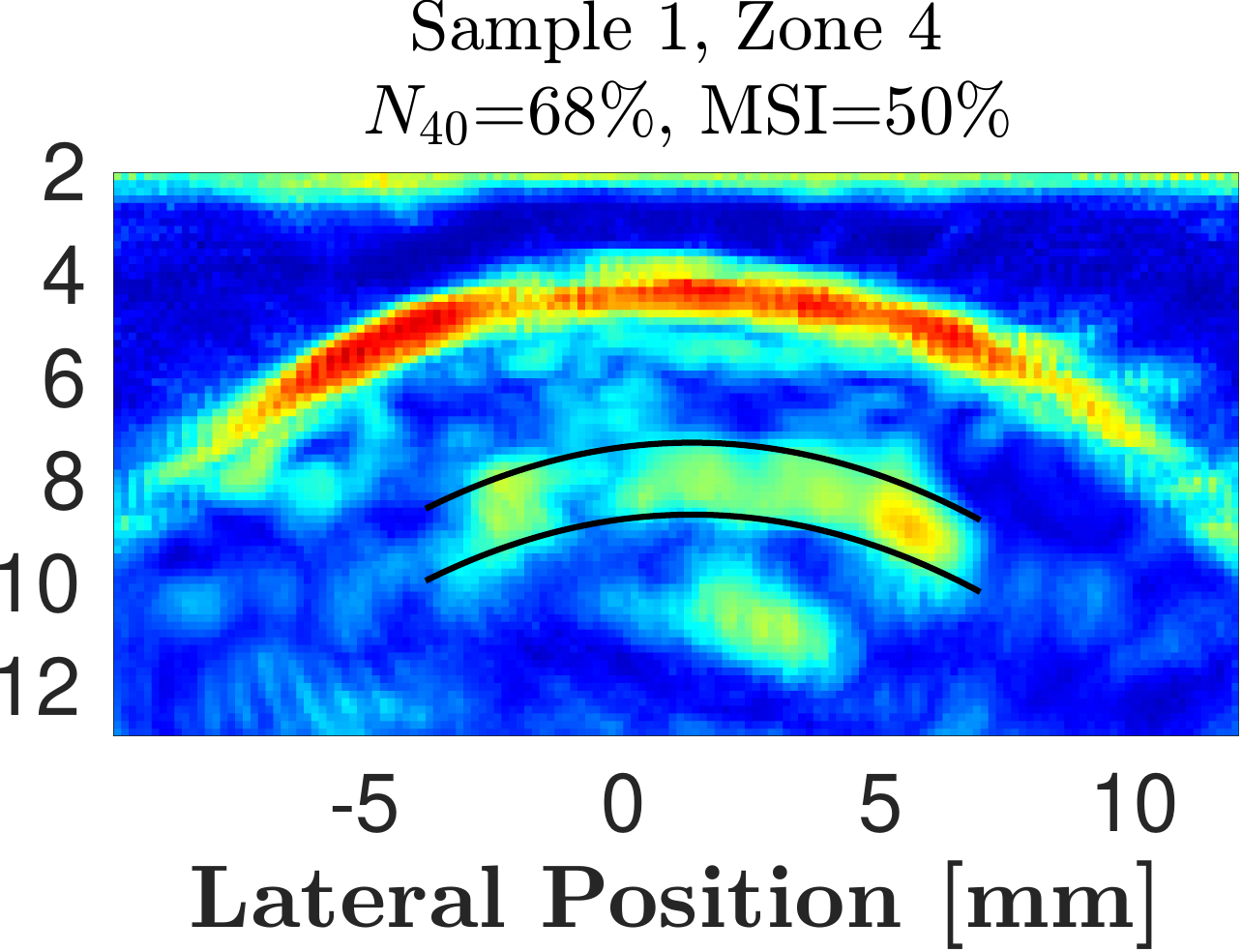}        
    \end{subfigure}
    \begin{subfigure}{.045\linewidth}
        \includegraphics[width=\linewidth]{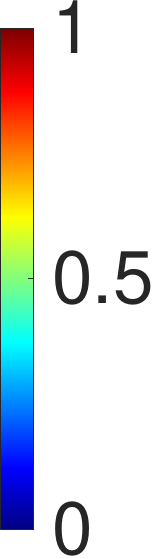}
    \end{subfigure}

\begin{subfigure}{.24\linewidth}
        \caption{}
        \includegraphics[width=\linewidth]{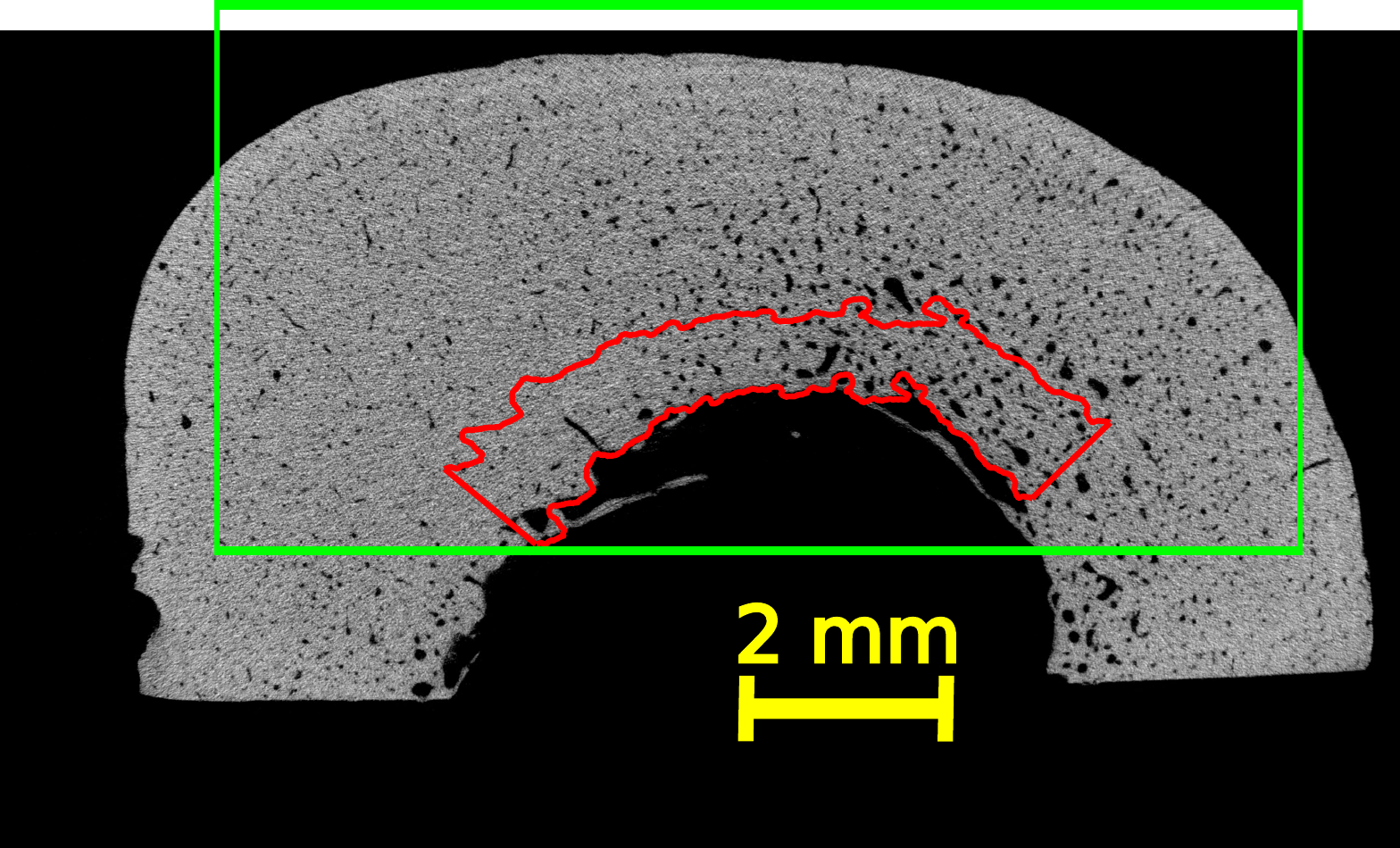}
    \end{subfigure}
    \begin{subfigure}{.24\linewidth}
        \caption{}
        \includegraphics[width=\linewidth]{images/SAMPLE_2_MID_SLICE_Z1_with_eROI_comp.pdf}
    \end{subfigure}
    \begin{subfigure}{.24\linewidth}
        \caption{}
        \includegraphics[width=\linewidth]{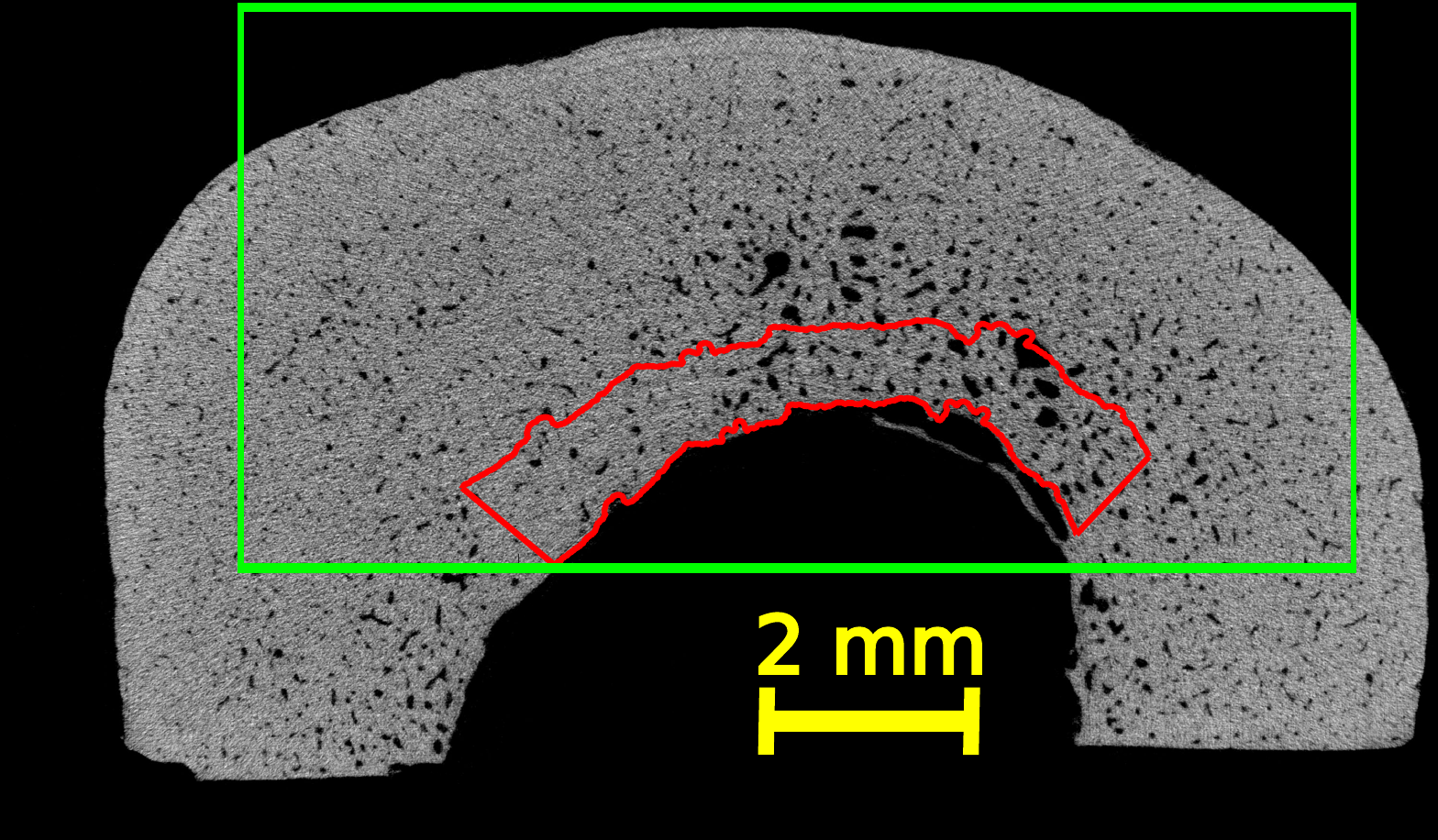}
    \end{subfigure}
    \begin{subfigure}{.24\linewidth}
        \caption{}
        \includegraphics[width=\linewidth]{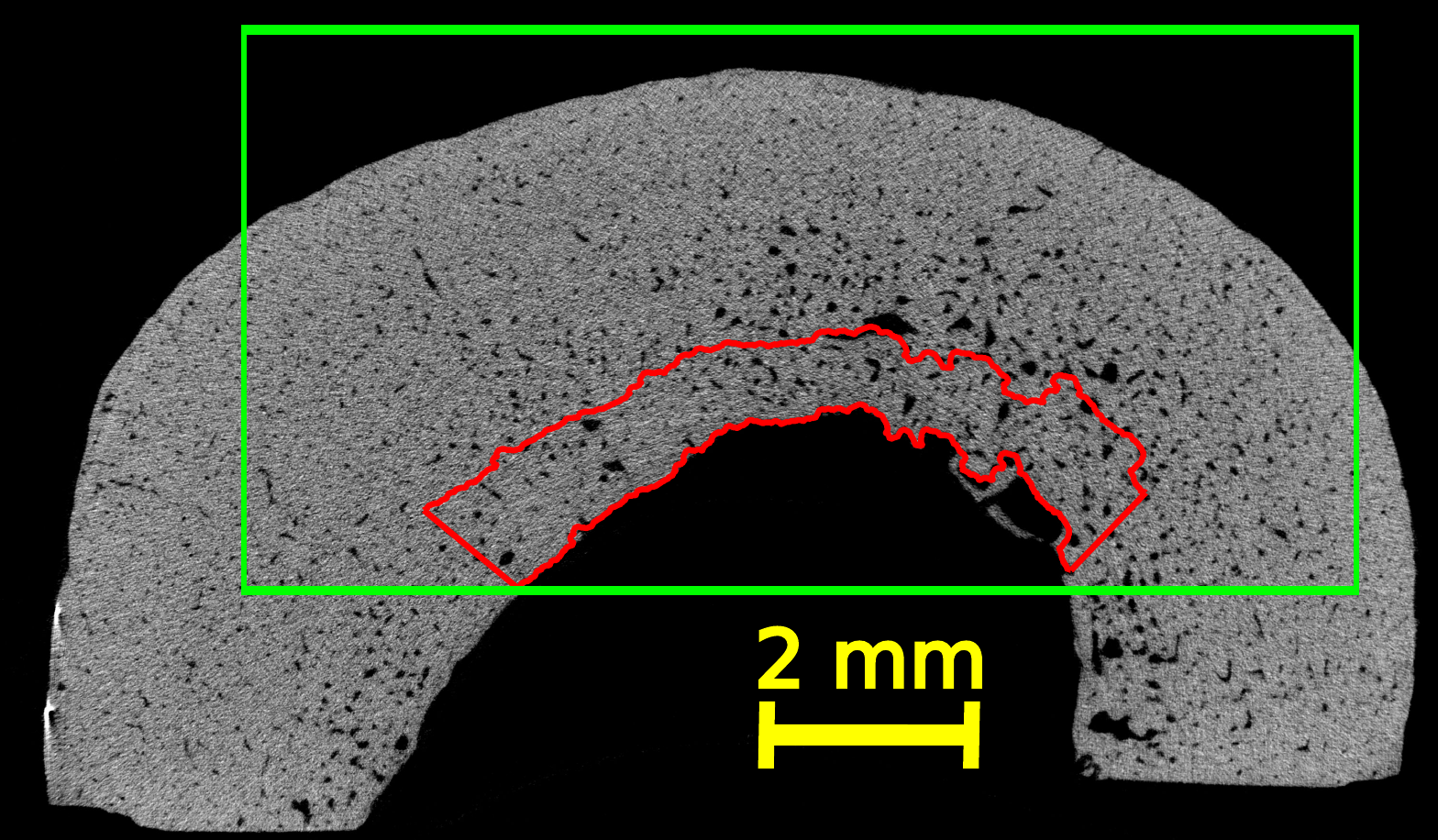}
    \end{subfigure}

    \begin{subfigure}{.23\linewidth}
        \caption{}
        \includegraphics[width=\linewidth]{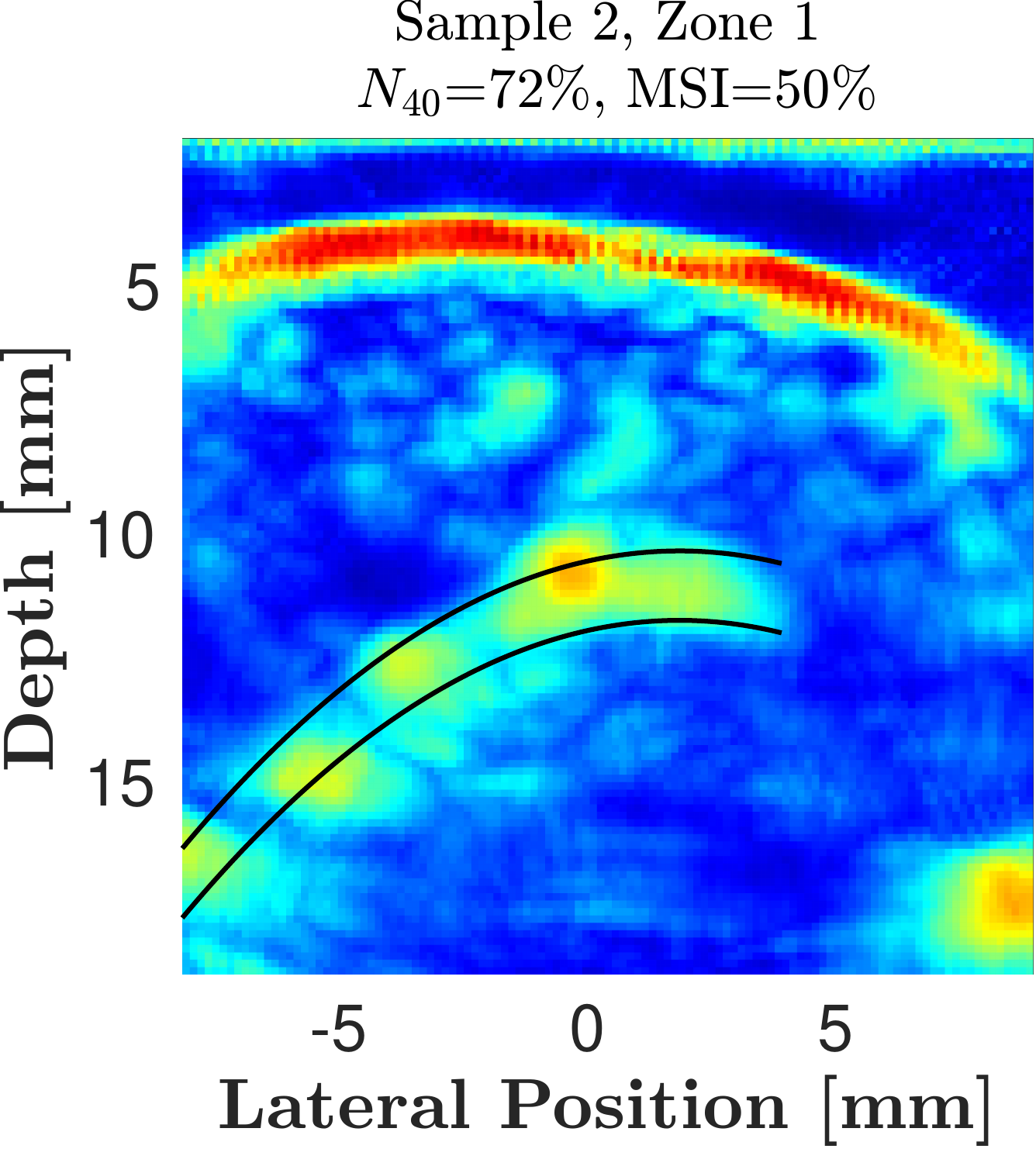}        
    \end{subfigure}
    \begin{subfigure}{.23\linewidth}
        \caption{}
        \includegraphics[width=\linewidth]{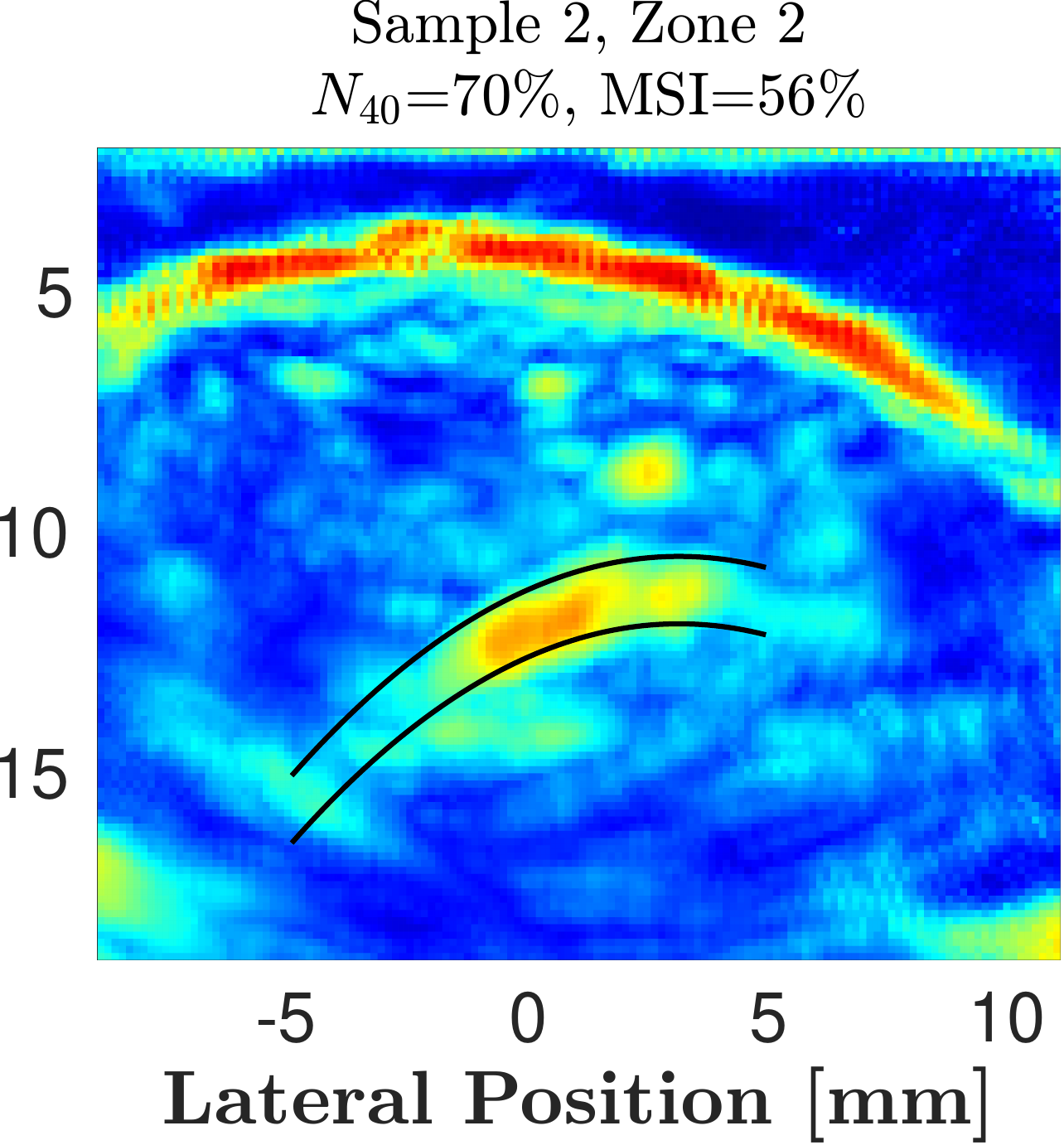}        
    \end{subfigure}
    \begin{subfigure}{.23\linewidth}
        \caption{}
        \includegraphics[width=\linewidth]{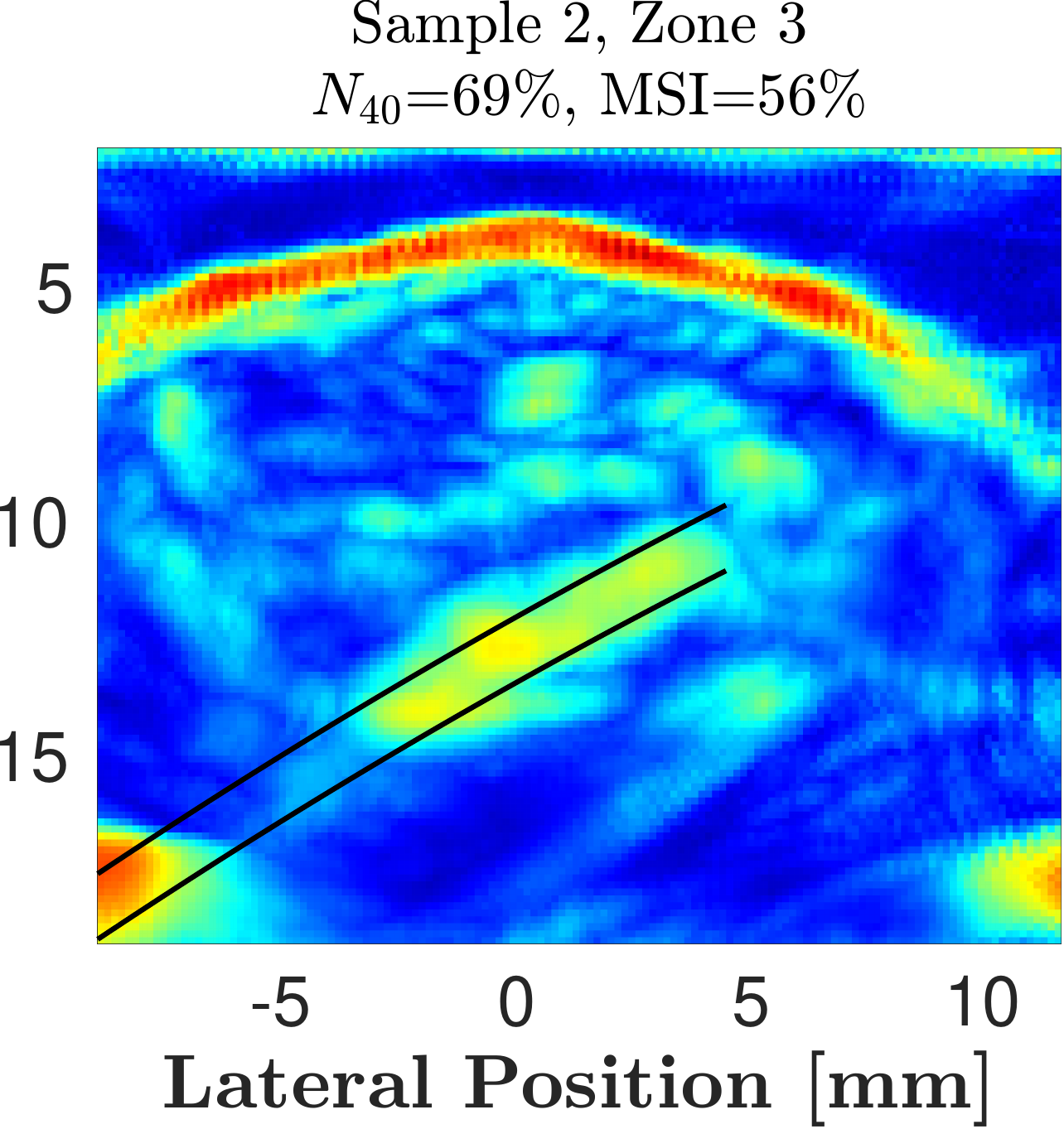}        
    \end{subfigure}
    \begin{subfigure}{.23\linewidth}
        \caption{}
        \includegraphics[width=\linewidth]{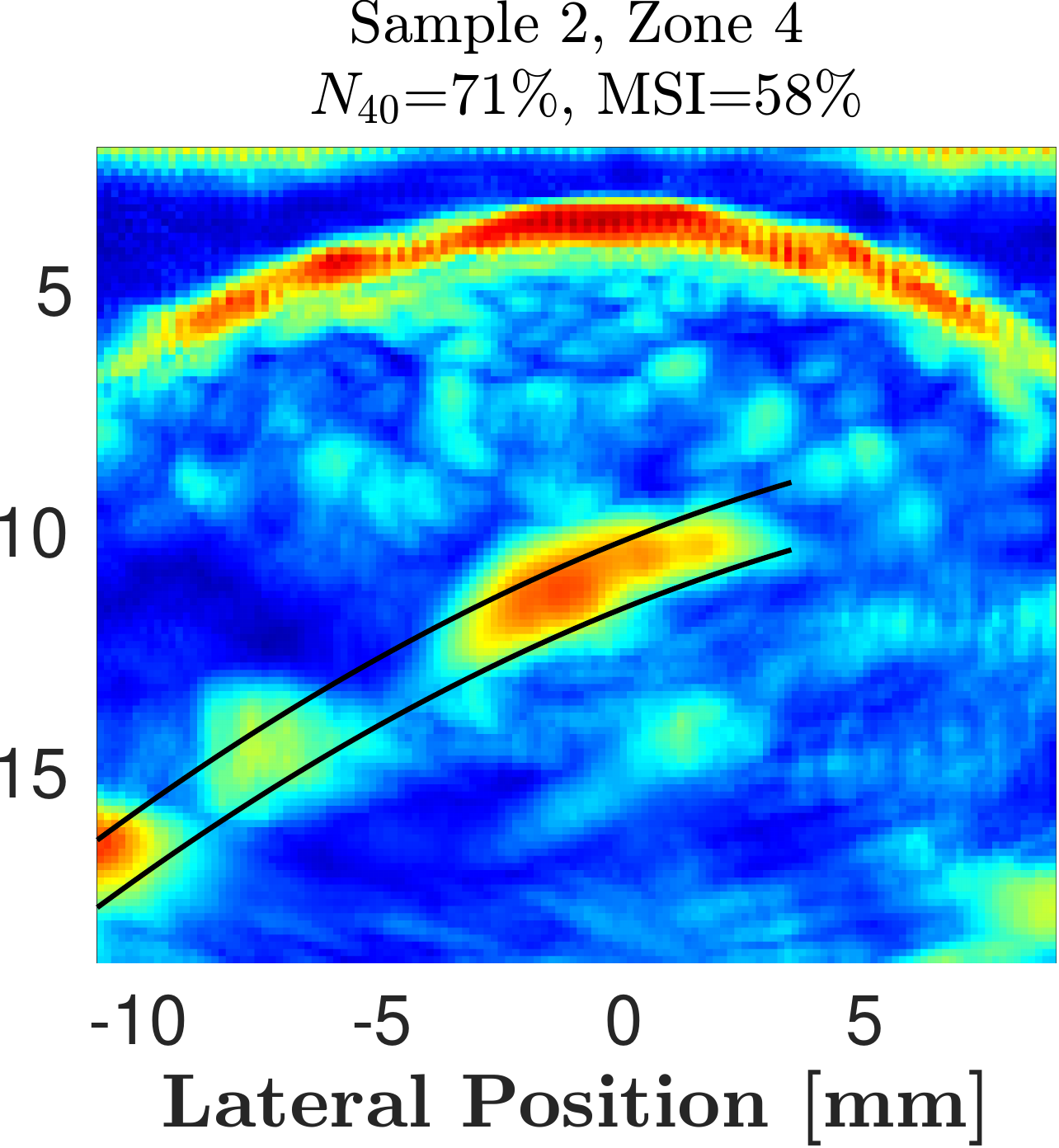}        
    \end{subfigure}
    \begin{subfigure}{.045\linewidth}
        \includegraphics[width=\linewidth]{images/colorbar_axis.pdf}
    \end{subfigure}

\begin{subfigure}{.24\linewidth}
        \caption{}
        \includegraphics[width=\linewidth]{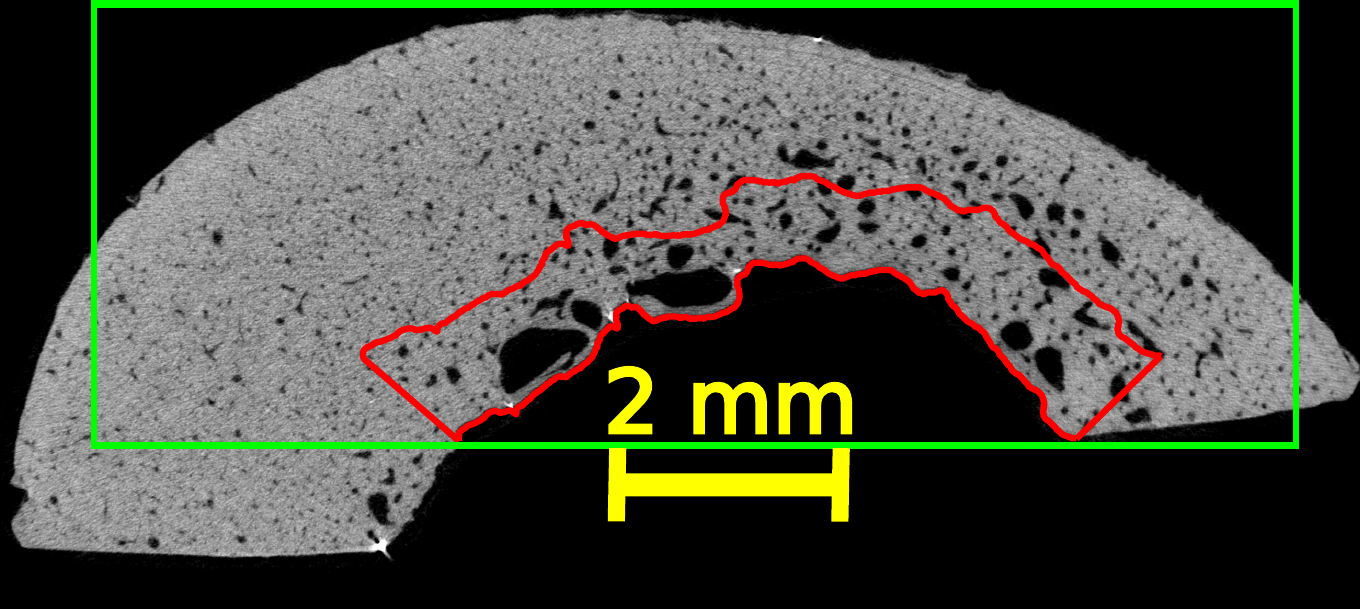}
    \end{subfigure}
    \begin{subfigure}{.24\linewidth}
        \caption{}
        \includegraphics[width=\linewidth]{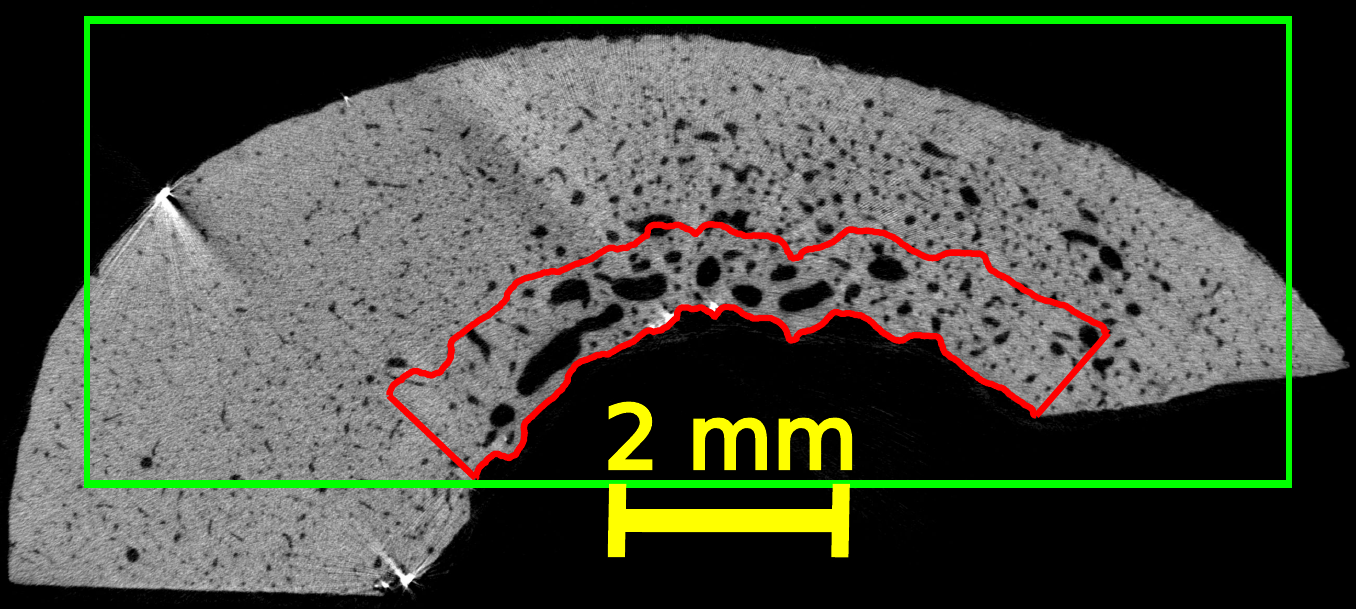}
    \end{subfigure}
    \begin{subfigure}{.24\linewidth}
        \caption{}
        \includegraphics[width=\linewidth]{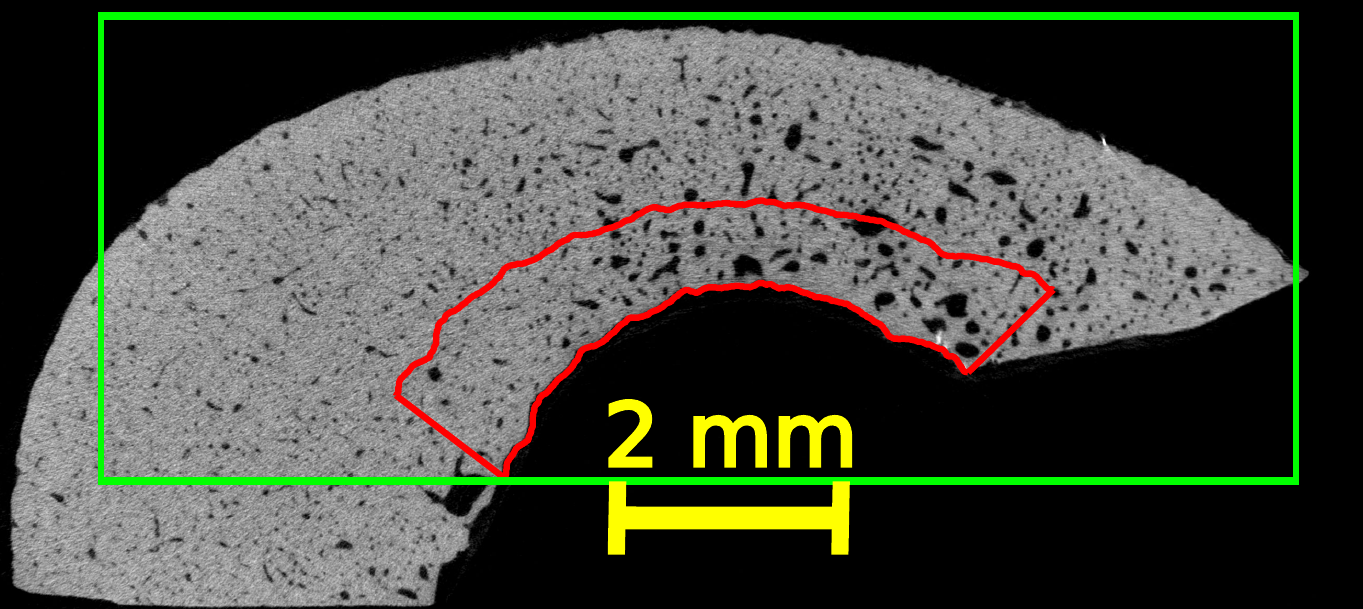}
    \end{subfigure}
    \begin{subfigure}{.24\linewidth}
        \caption{}
        \includegraphics[width=\linewidth]{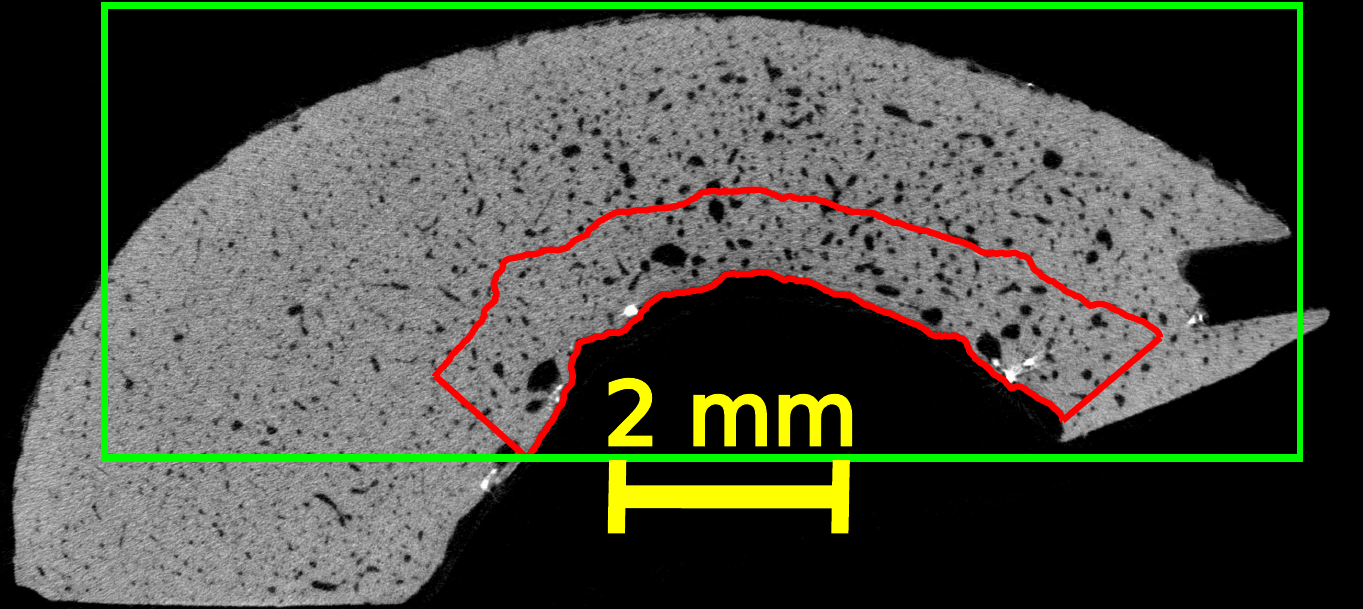}
    \end{subfigure}

    \begin{subfigure}{.23\linewidth}
        \caption{}
        \includegraphics[width=\linewidth]{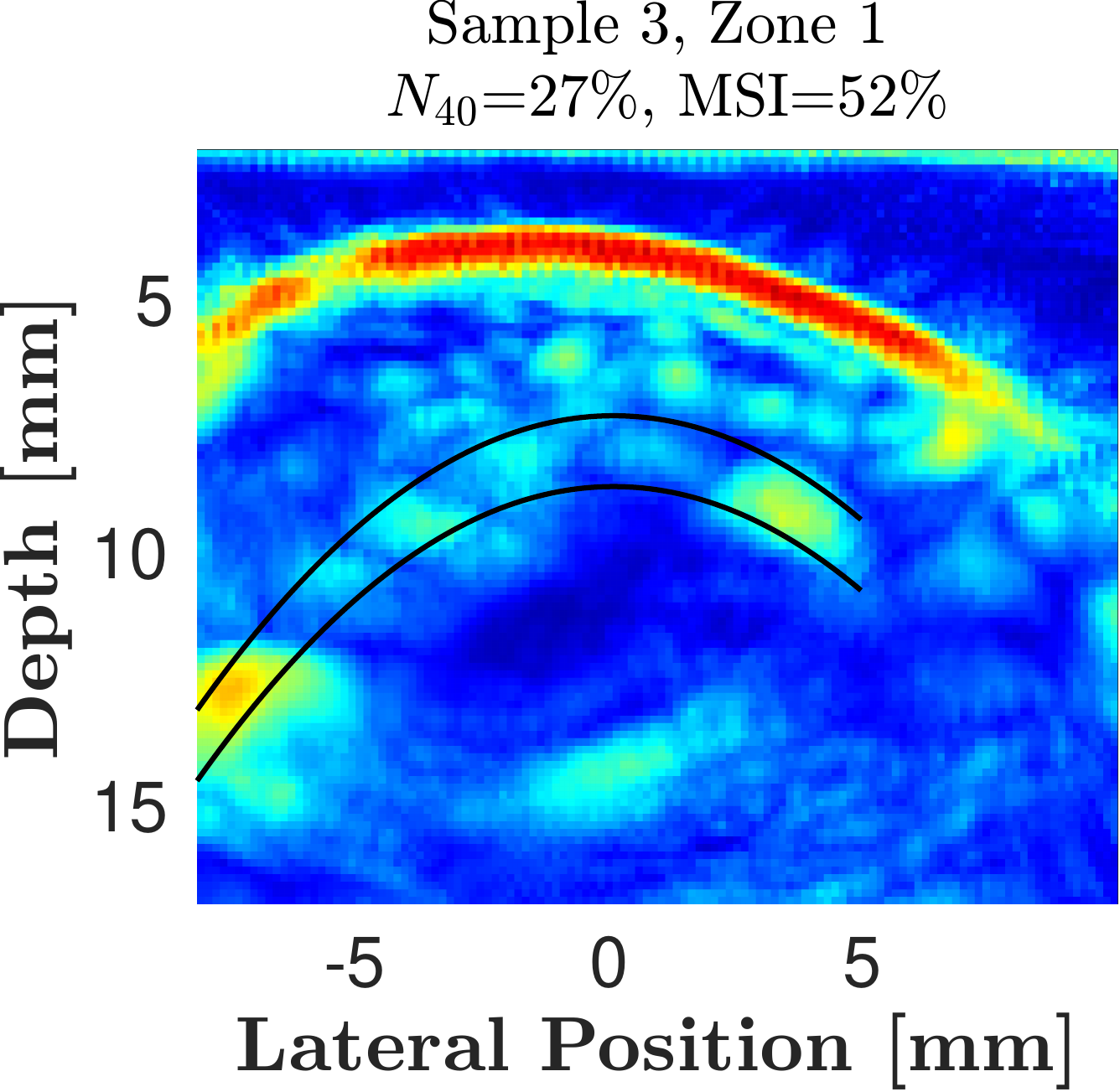}        
    \end{subfigure}
    \begin{subfigure}{.23\linewidth}
        \caption{}
        \includegraphics[width=\linewidth]{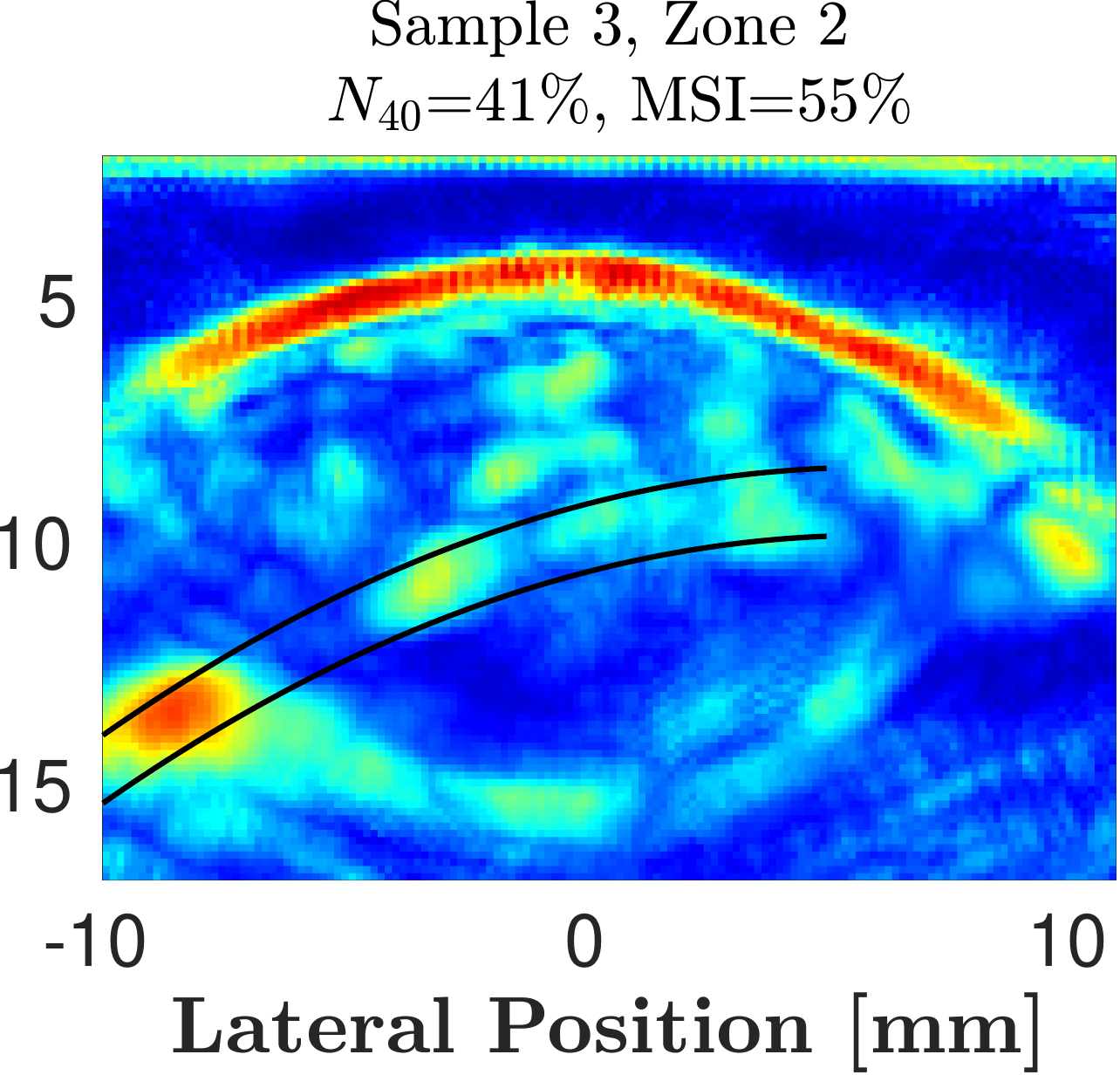}        
    \end{subfigure}
    \begin{subfigure}{.23\linewidth}
        \caption{}
        \includegraphics[width=\linewidth]{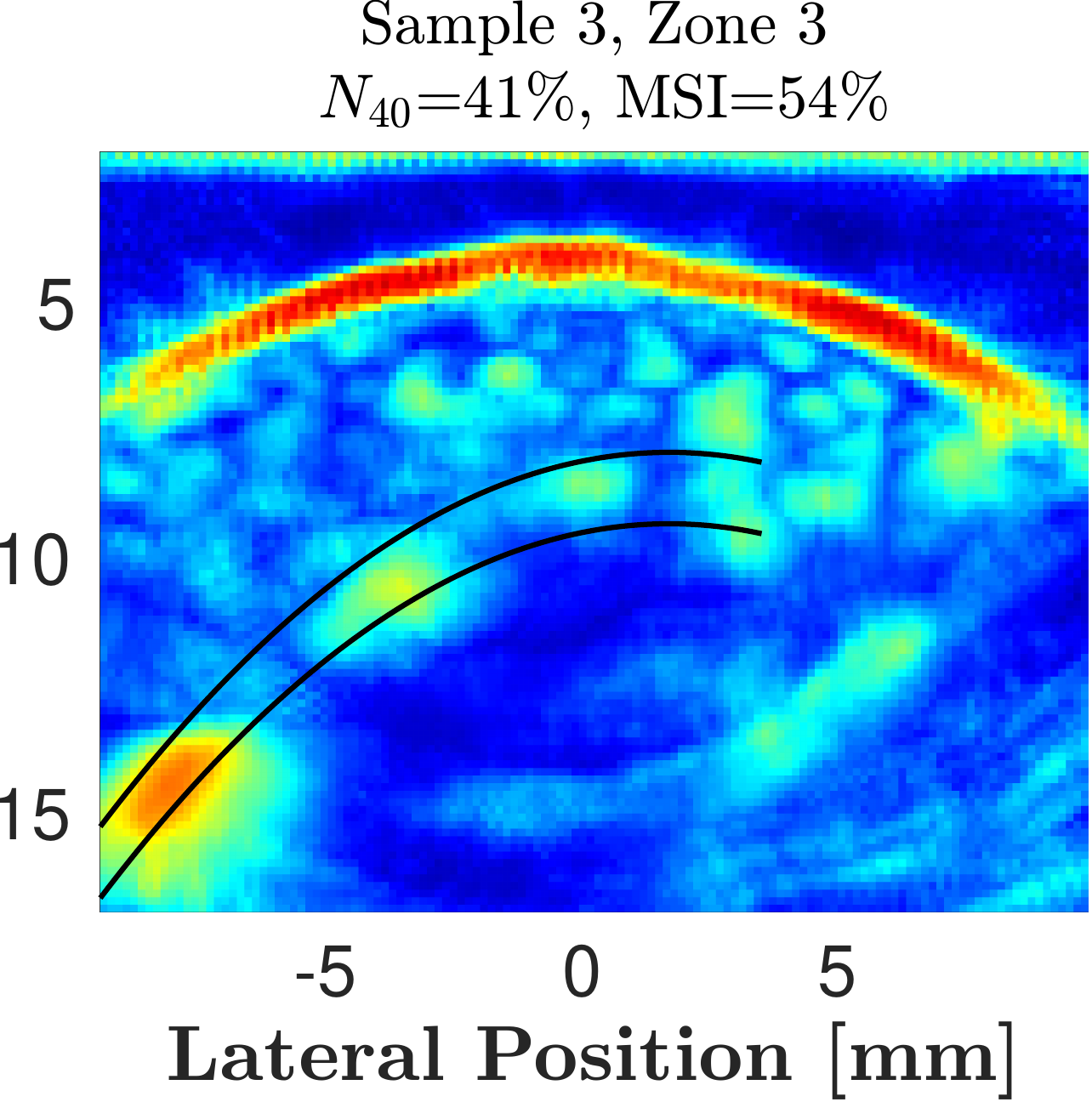}        
    \end{subfigure}
    \begin{subfigure}{.23\linewidth}
        \caption{}
        \includegraphics[width=\linewidth]{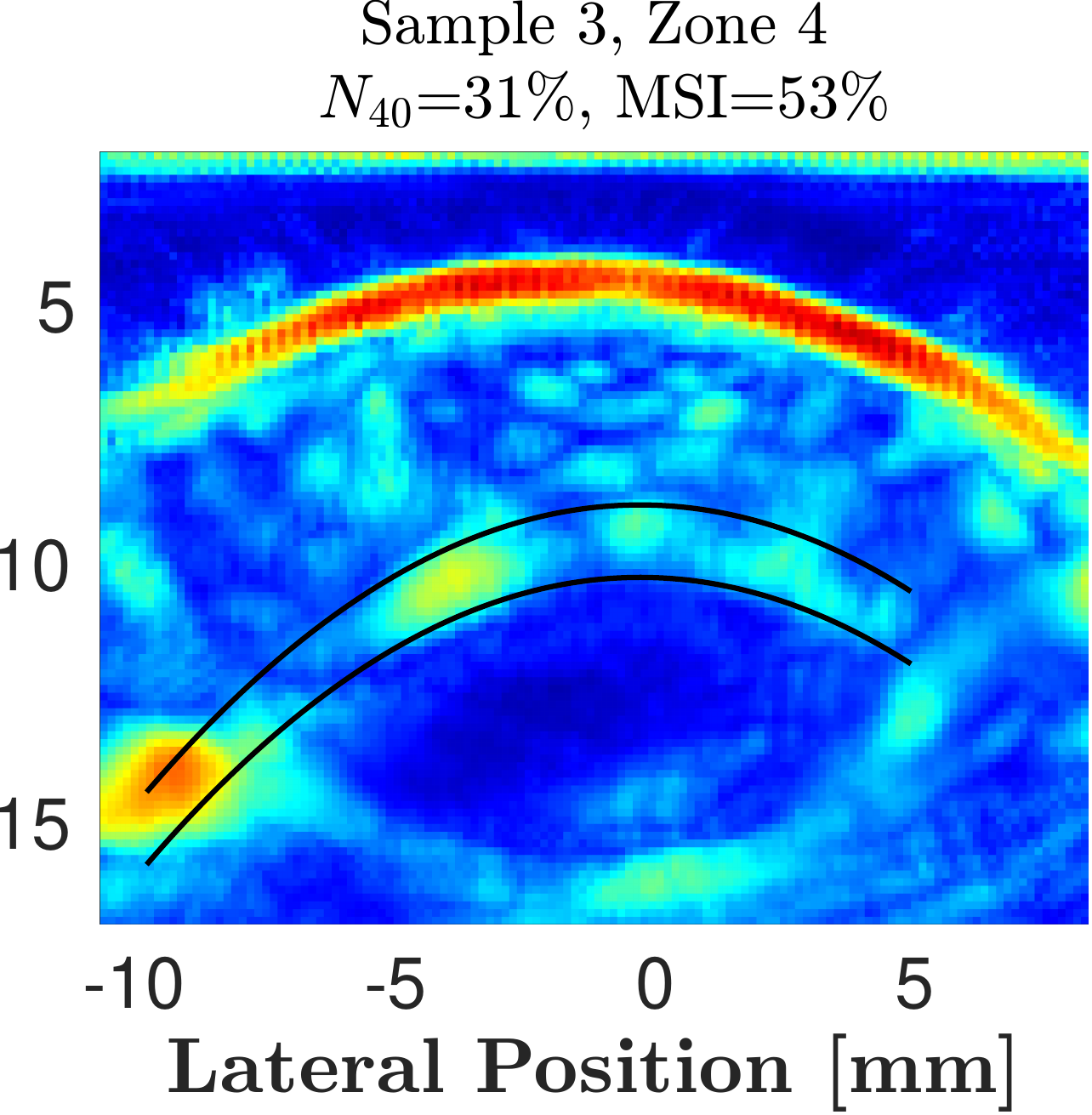}        
    \end{subfigure}
    \begin{minipage}[t]{.045\linewidth}
        \includegraphics[width=\linewidth]{images/colorbar_axis.pdf}
    \end{minipage}
\caption{HR-\textmu CT and ultrasound images for all VOIs across the three samples. 
The first row (panels (a)–(d)) shows HR-\textmu CT images of the four VOIs from sample 1. The green rectangle indicates the approximate region probed by ultrasound, while the red box delineates the endosteal volume of interest (EVOI) used for parameter computation. The second row (panels (e)–(h)) presents the corresponding ultrasound specularity maps, with the detected endosteal surface highlighted (using Dijkstra’s algorithm).
This layout is repeated for samples 2 and 3 in the subsequent rows.
Specularity metrics $N_{40}$ and MSI are given for each measurement zone.
}

\label{fig:bigfig_specularity}

\end{figure}

The values of endosteal properties and specularity metrics are summarized in Table ~\ref{table:descriptive_stat}.
Sample 1 had the lowest endosteal porosity, E.Por ranging from 6.2\% to 8.6\%, with minimal variation across the slices (IQR.E.Por = 0.9–1.7\%). Sample 2 showed a moderately higher endosteal porosity (11.6–13.7\%), with slight variation across slices for three of the VOIs (IQR.E.Por = 1.1–1.6\%), and a higher variation in one VOI (IQR.E.Por = 5.4\%).
Samples 1 and 2 had relatively small endosteal pore diameters, ranging from 14–17~\textmu m, with very low intra-VOI variability (IQR.E.Po.Dm = 0–3~\textmu m). Sample 3, on the other hand, had relatively high endosteal porosity (20.2–27.1\%), with a more pronounced variation across the slices (IQR.E.Por = 1.4–8.5\%). It also exhibited the largest endosteal pore diameters (29–45~\textmu m), which were clearly visible in the corresponding X-ray images in Figure~\ref{fig:bigfig_specularity}.

Rms heights ranged from $R_q=124$ to $R_q=185$~\textmu m. Sample 1 showed a relatively longer correlation length ($\rho_{\text{length}}=0.8-1.4$~mm) compared to samples 2 and 3  ($\rho_{\text{length}}=0.6-0.9$~mm).

\paragraph{Maps of specularity}
Figure~\ref{fig:bigfig_specularity} shows the specularity maps for one representative measurement (measurements were repeated 10 times) for each measurement zone. One 2D slice of each corresponding VOIs is also shown. 
For sample 1 (panels (e)–(h)), the specularity is high at both the periosteal and endosteal interfaces, and low within the cortical region. For this sample, $N_{40}$ was larger than 60\% for all measurement zones. These observations are consistent with the relatively homogeneous endosteal microstructure observed in this sample (panels (a)–(d)), characterized by low porosity and uniform pore distribution.

The trend is similar for sample 2 (panels (m)–(p)), despite a more complex bone geometry (panels (i)–(l)). Specular reflections at the periosteal and endosteal surfaces remain strong, with $N_{40}$ slightly higher than in sample 1, ranging from 69\% to 72\%. 

In contrast, sample 3 (panels (u)–(x)) exhibits markedly reduced specularity at the endosteal surface, with $N_{40}$ values not exceeding 41\%. This decrease in specularity is consistent with the higher endosteal porosity and larger pore diameters observed in panels (q)–(t). 


\begin{table}[ht!]
	\centering
	\begin{tabular}{|l|cccc|cccc|cccc|}
\hline
 & \multicolumn{4}{c|}{Sample 1} & \multicolumn{4}{c|}{Sample 2} & \multicolumn{4}{c|}{Sample 3} \\
    VOI / Measurement zone &  1 &  2 &  3 &  4 &  1 &  2 &  3 &  4 &  1 &  2 &  3 &  4 \\
    \hline
		$R_q$ [$\upmu$m] & 149 & 166 & 176 & 170 & 156 & 163 & 139 & 168 & 185 & 184 & 124 & 162 \\
		$\rho_{\text{length}}$ [mm] & 1.4 & 1.3 & 1 & 0.8 & 0.6 & 0.6 & 0.6 & 0.6 & 0.6 & 0.8 & 0.8 & 0.9 \\
		E.Por [\%] & 6.2 & 8.6 & 8.6 & 7.8 & 11.6 & 12.1 & 11.6 & 13.7 & 27.1 & 28.4 & 20.2 & 23.5 \\
		IQR.E.Por [\%] & 1.6 & 1.7 & 1.1 & 0.9 & 1.6 & 1.1 & 1.5 & 5.4 & 8.5 & 5.1 & 1.4 & 3.9 \\
		E.Po.Dm [$\upmu$m] & 17 & 14 & 14 & 14 & 17 & 17 & 14 & 14 & 45 & 34 & 37 & 29 \\
		IQR.E.Po.Dm [$\upmu$m] & 0 & 3 & 0 & 0 & 3 & 3 & 3 & 0 & 8 & 6 & 6 & 3 \\
		$N_{40}$ [\%] & 65 & 63 & 56 & 62 & 63 & 61 & 62 & 65 & 31 & 37 & 48 & 44 \\
		$MSI$ [\%] & 44 & 42 & 41 & 43 & 44 & 45 & 44 & 47 & 36 & 39 & 41 & 40 \\
        \bottomrule
    
	\end{tabular}
    \caption{Summary of endosteal microstructural variables and specularity metrics for all VOIs of all samples.}
    \label{table:descriptive_stat}
\end{table}

\paragraph{Specularity reflects microstructural properties}
Table~\ref{table:corr_coef} presents the Spearman correlation coefficients between all variables. 
Most of the microstructural properties are correlated, with the 
strongest correlations obtained between endosteal porosity ($E.Por$) and endosteal pore diameter ($E.Po.Dm$) ($\rho = 0.89, p < 0.0001$), indicating that greater endosteal porosity tends to be associated with larger endosteal pore diameters.

Both specularity metrics were negatively correlated to most microstructural variables with $N_{40}$ exhibiting stronger correlations. This can be attributed to the fact that the range of variation in $MSI$ (Table~\ref{table:descriptive_stat}) is relatively narrower (from 36.1 to 46.8) compared to $N_{40}$ (ranging from 31.1 to 65.3), limiting its sensitivity to microstructural differences.
Higher porosity and larger pores were strongly associated with lower specularity ($N_{40}$ and $MSI$). 
Specularity metrics were less correlated with variables related to sample heterogeneity in the probe elevation direction (across VOI slices) with $\rho$=-0.7 and $\rho$=-0.8 between $N_{40}$ and IQR.E.Por or IQR.E.Po.Dm, respectively.
Contrary to simulation results, none of the specularity metrics showed a significant correlation with the surface roughness parameters ($R_q$, $\rho_{\text{length}}$). This suggests that in the presence of porosity, surface roughness does not significantly impact specularity.

\begin{figure}[htb!]
\begin{subfigure}{.49\linewidth}
    \includegraphics[width=\linewidth]{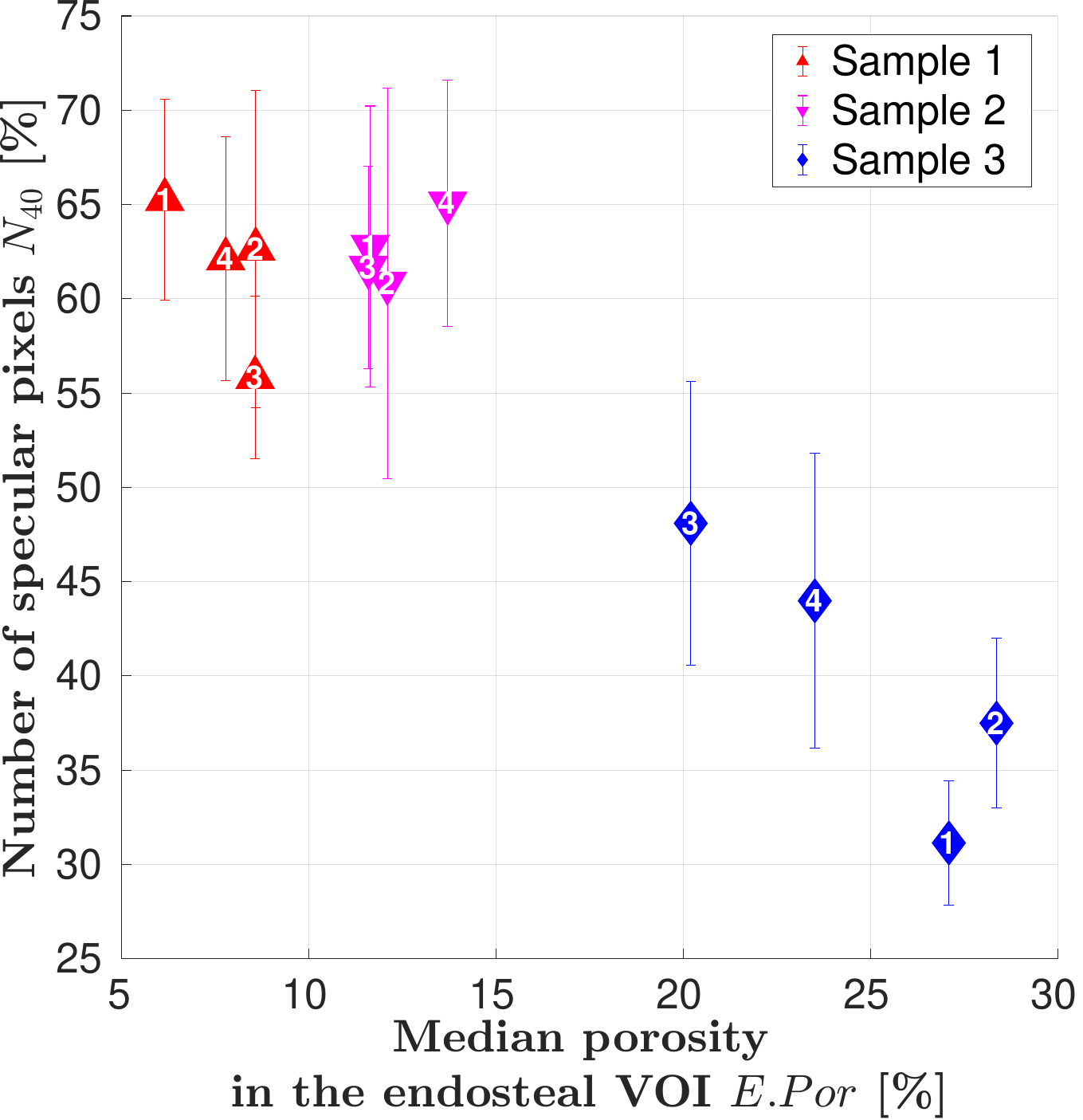}  
    \caption{}
\end{subfigure}
\begin{subfigure}{.49\linewidth}
    \includegraphics[width=\linewidth]{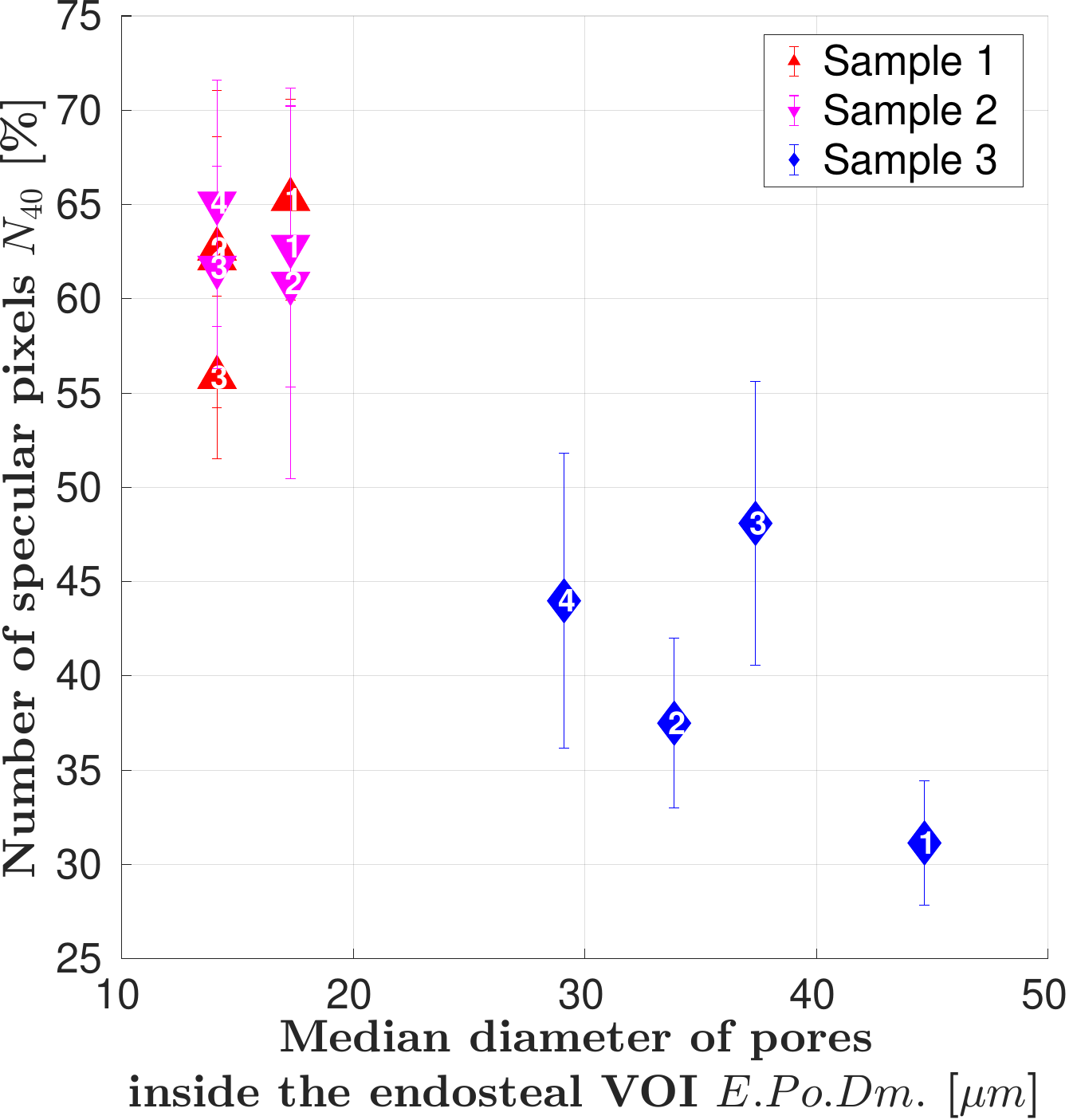}  
    \caption{}
\end{subfigure}
    \caption{Number of specular pixels $N_{40}$ for all measurement zones as a function of endosteal microstructural parameters. (a) $N_{40}$ as a function of endosteal porosity E.Por. (b) $N_{40}$ as a function of endosteal pore diameter E.Po.Dm. The error bars correspond to the standard deviations of $N_{40}$ across the 10 measurement repetitions. Labels 1 to 4 inside the triangle and diamond symbols correspond to measurement zones.}
    \label{fig:N40vsMED}
\end{figure}

Figure~\ref{fig:N40vsMED} shows $N_{40}$ as a function of E.Por and E.Po.Dm. 
In both plots, the data from Sample 3 are clearly separated from those of Samples 1 and 2, consistent with the microstructural differences observed in Figure~\ref{fig:bigfig_specularity}. Specifically, the higher porosity and larger pore diameters in Sample 3 are associated with markedly lower $N_{40}$ values essentially below 50\%. In contrast, VOIs from Samples 1 and 2—characterized by lower porosity and smaller pore diameters—consistently exhibit $N_{40}$ values above 50\%. VOIs from Sample 1 have porosity values below 10\%, while the porosity of VOIs from Sample 2 range between 10\% and 15\%. Despite this difference in porosity, Samples 1 and 2 are not distinguished by their $N_{40}$ values, likely due to their similar pore diameter distributions.

Interestingly, Samples 1 and 2 have similar microstructure but despite the greater thickness and more pronounced curvature of the endosteal surface of Sample 2, both samples exhibit similar specularity. This suggests that the calculation of specularity is not biased by sample thickness or gross bone shape.

\begin{table}[ht!]
	\centering
	\begin{tabular}{lllllllll}
\hline
		 & $R_q$ & $\rho_{\text{length}}$ & E.Por & IQR.E.Por & E.Po.Dm & IQR.E.Po.Dm & $N_{40}$ & $MSI$ \\
\hline
		$R_q$ & - & - & - & - & - & - & - & - \\
		$\rho_{\text{length}}$ & n.s. & - & - & - & - & - & - & - \\
		E.Por & n.s. & n.s. & - & - & - & - & - & - \\
		IQR.E.Por & n.s. & n.s. & 0.75$^{*}$ & - & - & - & - & - \\
		E.Po.Dm & n.s. & n.s. & 0.89$^{***}$ & 0.66$^{*}$ & - & - & - & - \\
		IQR.E.Po.Dm & n.s. & n.s. & 0.81$^{*}$ & n.s. & 0.86$^{**}$ & - & - & - \\
		$N_{40}$ & n.s. & n.s. & -0.93$^{***}$ & -0.70$^{*}$ & -0.92$^{***}$ & -0.80$^{*}$ & - & - \\
		$MSI$ & n.s. & n.s. & -0.69$^{*}$ & n.s. & -0.79$^{*}$ & -0.67$^{*}$ & 0.90$^{***}$ & - \\
	\end{tabular}
        \caption{Correlation coefficients ($\rho$) between variables. Significance levels: $^*p~<$ 0.05; $^{}p~<$ 0.001; $^{***}p~<$ 0.0001; n.s.: not significant.}	
        \label{table:corr_coef}
\end{table}

\paragraph{Linear regression analysis}
In the univariate analysis, pore diameter and porosity emerged as the microstructural variables showing the strongest individual associations with specularity. To further quantify their relationship with specularity, simple linear regression models were fitted separately between each microstructural parameter and specularity (Table~\ref{table:TableAdjustedR2Median}).  Given its stronger correlation with microstructural properties, $N_{40}$ was selected as the specularity metric for regression analysis.

Among porosity-related variables, E.Por showed the highest predictive performance, with an adjusted $R^2$ of 84\% and a RMSE of 5\%, whereas IQR.E.Por demonstrated a weaker explanatory power. Similarly, for pore diameter-related variables, E.Po.Dm was a stronger predictor of $N_{40}$ than IQR.E.Po.Dm, yielding a higher adjusted $R^2$ and lower RMSE.  

Overall, mean pore diameter (E.Po.Dm) and mean porosity (E.Por) were the individual microstructural parameters most strongly associated with variations in $N_{40}$.
\begin{table}[ht!]
	\centering
	\begin{tabular}{c|l|l|c}
\hline
		&Quantity & Adj-R$^2$ & RMSE (\%) \\
\hline
\multirow{3}{*}{Porosity model}&
		E.Por & 84$^{***}$ & 5$^{}$ \\
		&IQR.E.Por & 44$^{*}$ & 9$^{}$ \\
\hline
		\multirow{3}{*}{Pore diameter model}&E.Po.Dm & 84$^{***}$ & 5$^{}$ \\
		&IQR.E.Po.Dm & 60$^{*}$ & 7$^{}$ \\
\hline
	\end{tabular}
	\caption{Adjusted R$^2$ and RMSE of linear regression models predicting 
$N_{40}$ from endosteal microstructural variables. Statistical significance: $^*p~<$ 0.05; $^{**}p~<$ 0.001; $^{***}p~<$ 0.0001; n.s.: not significant.}
	\label{table:TableAdjustedR2Median}
\end{table}

\clearpage


\section{Discussion}
In this work we combined a signal processing technique originally developed to enhance specular reflectors in ultrasound images of soft tissues \cite{rodriguez-molares_specular_2017} with a bone imaging method which extends B-mode imaging into bone by applying refraction correction at the bone–soft tissue interfaces \cite{renaud_vivo_2018, renaud_measuring_2020}. 
Our approach produces specularity maps that are anatomically aligned with the cortical bone structure, with both periosteal and endosteal surfaces clearly delineated.

A reduced specularity at bone interfaces can, in principle, result from several factors: 1)~Specularly reflected rays may fail to reach the receiving surface of the probe. This can occur if the inclination of the bone surface relative to the probe surface is too large, either within the image plane or along the elevation direction of the probe; 2)~The interface may exhibit irregularities or roughness with a characteristic size smaller than the wavelength, leading to diffuse scattering; 3)~Scattering sources such as pores may be present near the interface, at a distance less than approximately one wavelength.

The aim of this study was to evaluate whether specularity maps can serve as a quantitative tool for assessing microstructural degradation associated with osteoporosis--specifically, more numerous and enlarged pores adjacent to the endosteal surface, and pores merged with this surface contributing to interface roughness. 
Specularity maps from \exvivo~measurements were analyzed in the light of HR-\textmu CT images which provided ground truth for bone microstructure, offering quantitative measures of endosteal surface roughness, porosity (E.Por), and pore size distribution (notably median pore diameter, E.Po.Dm) in the endosteal region.

The implementation of the calculation of specularity in a two-layer medium was validated using numerical simulations. With the simulations, we could verify that the specularity is a quantity sensitive to the microstructure patterns (roughness) at the  endosteal interface.
In the simulations, planar interfaces were considered. As shown in Figure~\ref{suppfig:specularity_vs_curvature} in the supplementary material, specularity maps were very similar for planar and curved interfaces when the microstructure was identical. Due to the finite aperture of the probe, regions of the endosteal interface that are locally more curved contribute less to the received specular signal. This effect mainly influences the overall signal amplitude rather than the relationship between specularity metrics and microstructural properties. Therefore, the use of a planar interface in the simulations is justified as a simplified configuration that isolates the effect of microstructure.

In these simulations, specularity was equal to one for perfectly flat interfaces, while specularity was highly reduced at rough interfaces. To quantify the specularity in the region of the interface, we introduced two specularity metrics (MSI and $N_{th}$) which were found to be highly sensitive to changes in the roughness.
This sensitivity to surface roughness was not observed in the \exvivo~experimental data. One possible explanation is the relatively small variation in endosteal roughness values ($R_q$) observed ex vivo, compared to the wider range explored in simulations. Another contributing factor was the presence of pores near the endosteal surface. Pores are known to contribute to additional scattering \cite{dia_influence_2023,iori_estimation_2020}, compounding that caused by surface roughness. This disrupts coherent reflection at the interface and reduces specularity in the endosteal region. 
This phenomenon was also observed in preliminary simulations performed with a single pore size of 50 and 100~\textmu m, while varying the endosteal interface profile and a porosity of 7 and 14\% in the region adjacent to the interface. When porosity was included in the simulations, we observed a marked reduction in specularity and a loss of the strong correlation between specularity and surface roughness \hl{as illustrated in Figure~\ref{suppfig:specularity_vs_rugo_por} of supplementary materials}. These simulation findings indicate that, at frequencies of the order of 2.5~MHz used for bone imaging, the acoustic signature of the endosteal interface is closely related to the adjacent porous microstructure.

Correlation analyses of \exvivo~experimental data revealed that both E.Por and E.Po.Dm were strongly associated with specularity. 
In simple linear regression analyses, each of these variables individually explained a large proportion of the variability in $N_{40}$. 
We also tested the hypothesis that heterogeneity in bone microarchitecture along the elevation direction of the probe (i.e., along the bone axis) could influence observed specularity. This effect was found to be minor. 
As noted above, E.Por and E.Po.Dm alone accounted for the observed variability in $N_{40}$. This observation is consistent with previous findings, as both porosity and pore size have been identified as significant determinants of scattering strength in bone~\cite{dia_influence_2023, iori_estimation_2020}. 
Note that E.Por and E.Po.Dm exhibit a strong correlation, which prevents disentangling their respective independent contributions to the variability of the specularity metrics given the current limited dataset.
\hl{It should also be noted that the reported $R^2$ may be inflated due to between-donor clustering; indeed the 12 ROIs are nested within only 3 donors and the calculated $R^2$ may be largely driven by a two-group separation (Samples 1 and 2 vs. Sample 3).}

The threshold used in the \textit{ex vivo} experiments differed from the 50\% threshold used in simulations. To assess the sensitivity of the specularity metrics to this choice, additional analyses were performed using thresholds of 30\% and 50\%, alongside the 40\% threshold used in the main analysis. The results are reported in Tables~\ref{supptable:spearman-rho-onewl} and~\ref{supptable:AdjustedR2-onewl} in the supplementary material. 
While all three thresholds yielded significant correlations with endosteal porosity and pore diameter, the 40\% threshold consistently provided the strongest associations and the highest explained variance in regression models. Lowering the threshold (30\%) slightly reduced specificity by including weaker specular contributions, whereas increasing it (50\%) likely excluded relevant specular regions, leading to a decrease in correlation strength. This indicates that the 40\% threshold provides an effective balance between robustness to noise and sensitivity to meaningful microstructural variations.  However this analysis is of limited outreach because it is limited to our dataset. Obviously such analysis should be conducted in vivo to determine the optimal specularity threshold in patients.

We hypothesized that specularity metrics, computed within a ROI surrounding the interface in specularity maps, are sensitive to the bone properties in the immediate vicinity of that interface. To investigate this, we correlated the specularity metrics with bone microstructural parameters measured within a ROI (in the HR-\textmu CT images) of 1.4~mm thickness (corresponding to one wavelength in bone) located above the endosteal interface (Figure~\ref{meth:e_microstruct}). 
Additionally, we evaluated these correlation using a smaller ROI (in the HR-\textmu CT images) of 0.7~mm thickness(corresponding to approximately half the ultrasonic wavelength in cortical bone at the operating frequency). Spearman rank correlation coefficients for this ROI are reported as supplementary material in Table~\ref{supptable:spearman-rho-halfwl}. Correlations obtained with the smaller ROI were slightly stronger. While the 0.7~mm ROI provides a more localized descriptor of microstructure, it is smaller than the axial resolution of the system. Therefore, we primarily reported the results for the 1.4~mm ROI because it can be interpreted as the bone region contributing to the measured backscattered signal. 
A more extensive study, including a larger and morphologically more diverse sample set, would be required to more precisely delineate the region of bone contributing to the observed specularity at the interface. Nonetheless, our findings support the conclusion that the specularity metrics are sensitive to bone properties located within a sub-wavelength distance from the interface.

Overall, our results suggest that a threshold could be established to differentiate bone with non-pathological versus highly remodeled microstructure near the endosteal interface. Specifically, in our dataset, $N_{40}$ was larger than 50\% for Samples 1 and 2, which had a homogeneous microstructure with no large pores, while $N_{40}$ was below 50\% for Sample 3, which was highly remodeled and exhibited large pores in the endosteal region (see Figure~\ref{fig:N40vsMED}).

The sensitivity of specularity metrics to changes in bone microstructure can be compared to the sensitivity of the speed of sound, which has previously been proposed in the literature as an indicator of bone quality. As shown in Figure~\ref{fig:N40vsMED}(a), an increase in endosteal porosity of approximately 10\%—typically associated with a 20\% reduction in bone strength, according to Peralta et al. \cite{peralta_bulk_2021}—results in a decrease of roughly 20\% in the number of specular pixels ($N_{40}$). In comparison, prior work on the same bone samples \cite{dia_ultrasound_2026} found that a comparable porosity increase across the full cortex leads to a reduction in radial speed of sound by approximately 200 m/s, or about 6\% from a baseline cortical value of 3.500 m/s. Furthermore, since speed of sound, as measured by existing approaches, represents an average measurement across the entire cortex, it may be blind to localized degradation, particularly near the endosteal surface.
We therefore hypothesize that the specularity metrics may be more sensitive than speed of sound to bone microstructure deterioration, especially in the early stages of osteoporosis. This remains to be confirmed in a larger and more diverse dataset.


This study has some limitations. 
Our simulation's findings are only valid in the absence of porosity seen in real tissue. 
With the simulations, we did not try to model a realistic porosity and as a consequence we could not reproduce the experimental trend indicating that the pores near the endosteal interface have a dominant effect on the specularity at the expense of interface roughness. 
Simulating a realistic porosity (i.e. yielding a realistic scattering) is challenging. To obtain statistically reliable results from simulations would require to dispose of a large set of (HR-\textmu CT) images or generate synthetic microstructures from a postulated distribution of pore location and sizes.  Although the vast majority of simulations of ultrasound propagation in cortical bone has been conducted in 2D configurations in the plane transverse to osteons \cite{nguyen_minh_estimation_2020, mohanty_artificial_2019}, the validity of this configuration has not been \hl{established}. Haversian canals are not infinite cylinders  aligned along well-defined direction Volkmann canals, which run nearly perpendicular to Haversian canals, are not modeled in 2D configurations. More realistic simulation should be conducted in a 3D framework but the numerical cost may be prohibitive for the relatively large bone volume probed in our experimental configuration.  

%
The present study was not designed to determine whether porosity or surface roughness has the predominant impact on specularity. Based on our data, we observe that both parameters are associated with specularity metrics. A more rigorous assessment of their respective effects should be addressed in future work, for instance through experiments using phantoms with independently controlled surface roughness and porosity, or by reproducing the study in a substantially larger cohort to ensure sufficient statistical power.
%
Only three bones from three human donors were used, and thus our data do not capture the full variability of human bone across different stages of osteoporosis.  Nevertheless, our conclusions are supported by a diverse dataset comprising 12 independently measured and processed bone regions, obtained by defining four non-overlapping measurement zones in each bone. Our results will need to be confirmed by a study involving a larger number of samples.
Another limitation is that we have used a phased-array probe with an elevation aperture of 14~mm to perform two-dimensional imaging. This approach may miss reflections occurring outside the imaging plane and could potentially lead to an underestimation of both diffuse scattering and specular contributions. 
It should be noted that a central frequency of 2.5 MHz was used. This is important because the scattering strength of pores depends on the ratio between the ultrasound wavelength and the characteristic size of the microstructure. Further studies are required to evaluate the influence of frequency on the specularity metrics.
Extending the methodology to three-dimensional imaging would allow for a more comprehensive assessment of ultrasound interactions with the interface morphology. Nevertheless, our results suggest that two-dimensional imaging can provide aspecularity index effectively related to bone microstructure. 

\subsection{Conclusion}
The results of this study suggest that focal degradation of bone microstructure near the endosteal surface can be assessed by analyzing the balance between specular and diffuse ultrasound reflection in this region. 
These results will need to be confirmed in a larger and more diverse dataset, including males and females and individuals across a wider age range, and with varying degrees of osteoporosis. 
Implementing the approach presented in this study in patients should be straightforward, as the approach only involves an additional signal processing step to the bone imaging method proposed by Renaud et al., which is already operational in an in vivo setting \cite{renaud_vivo_2018}.
\hl{Nevertheless, appropriate in vivo specularity thresholds remain to be determined.}
Preliminary in vivo specularity maps have been obtained by \cite{dia_quantitative_2024}. The next step to validate the technique in vivo would be to compare in vivo specularity indices with microstructural features extracted from HR-pQCT.
The current implementation of cortical bone imaging is already real-time \cite{renaud_measuring_2020}. Computing the specularity map involves independent processing of each received RF signal at each image pixel, which is highly parallelizable. Therefore, real-time visualization (a few images per second) of the specularity map during patient scanning could likely be achieved using a GPU-based implementation.

The acquisition protocol could be further improved by increasing the number of measurements averaged to calculate the specular index for a given bone region, for example, by using a probe with a mechanical sweep of the transducer array perpendicular to the bone axis~\cite{armbrecht_pore-size_2021} (in this study, 10 measurements were used with manual positioning).
This approach could potentially provide a novel biomarker of unbalanced cortical bone remodeling more sensitive than anatomical measures (i.e. cortical thickness measurement) or material properties such as speed of sound averaged over a large bone volume, and complement the current set of tools available for osteoporosis diagnosis.

\clearpage
\bibliographystyle{unsrt}
\bibliography{biblio_exvivo}
\end{document}


\journal{Ultrasonics}
\maketitle
\section*{Supplementary material}

This supplementary material provides additional simulation results and statistical analyses supporting the findings reported in the main manuscript.

\begin{figure}[ht!]
    \centering
\begin{tabular}{c|c|c}
    \multirow{2}{*}{Cortical porosity} &  Planar interfaces & Curved interfaces\\
    &\includegraphics[width=.4\linewidth]{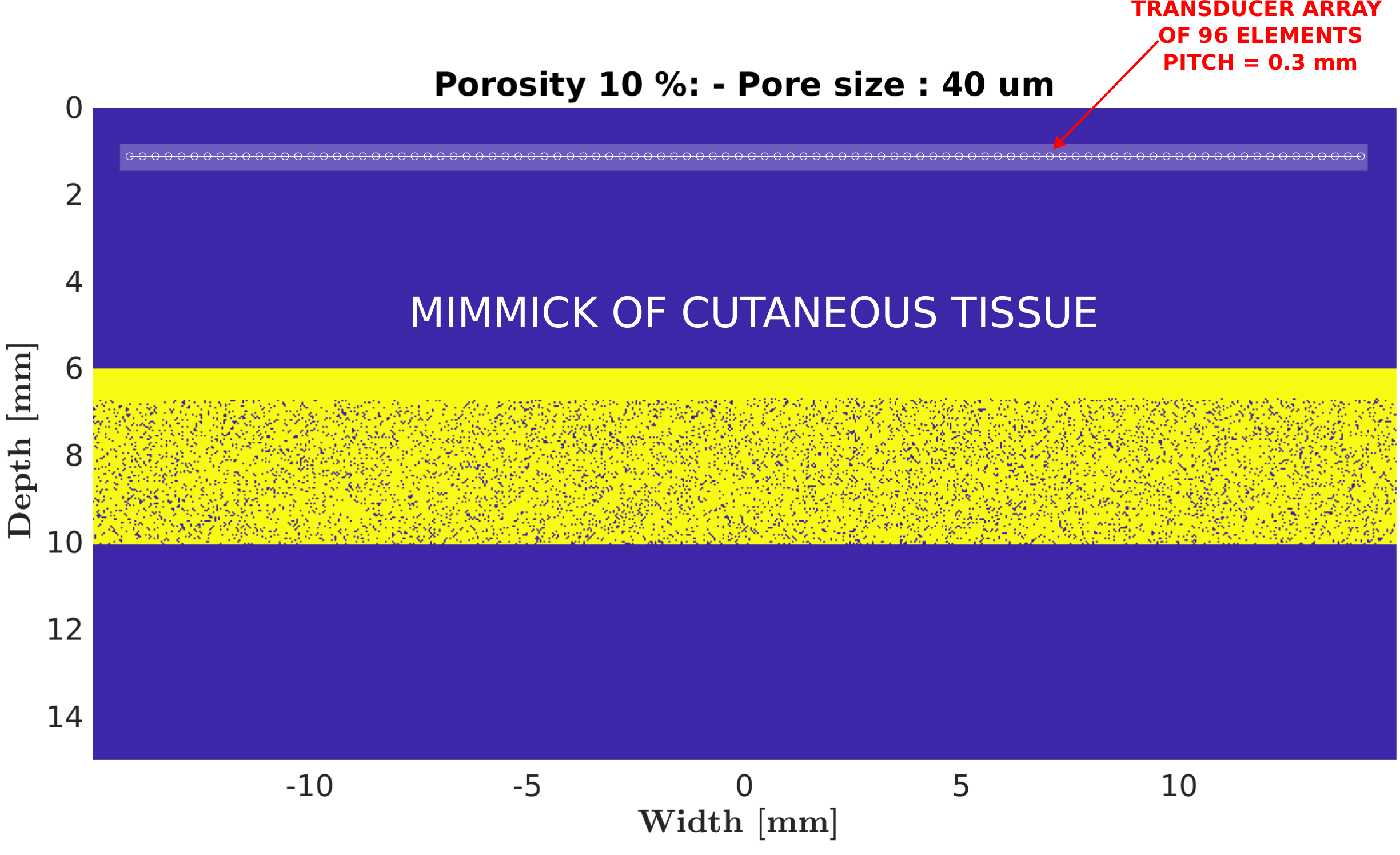}
    &\includegraphics[width=.4\linewidth]{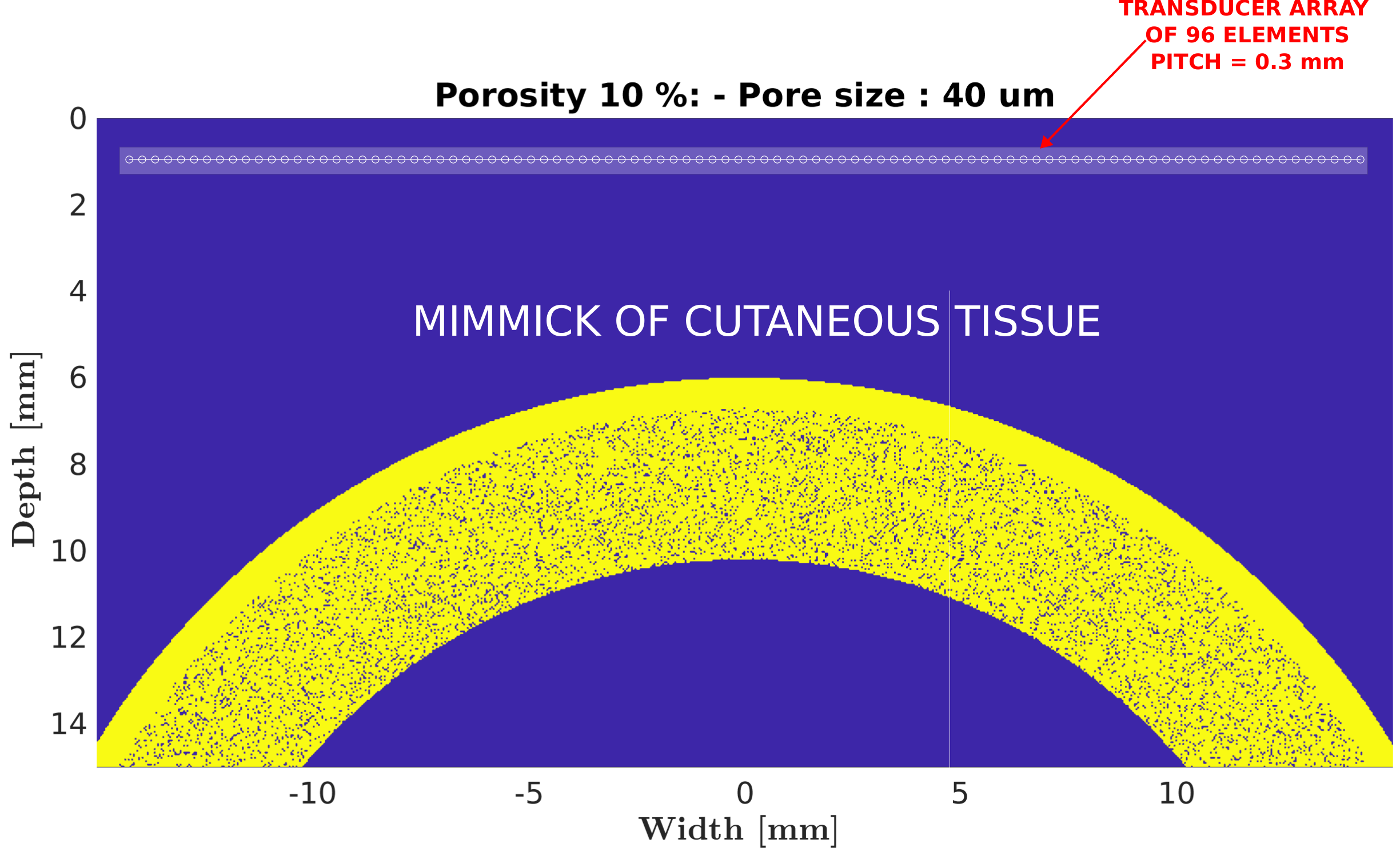}\\\hline
    \multicolumn{1}{m{2cm}|}{6\%} &\includegraphics[trim={0 90 0 90}, clip,width=.4\linewidth]{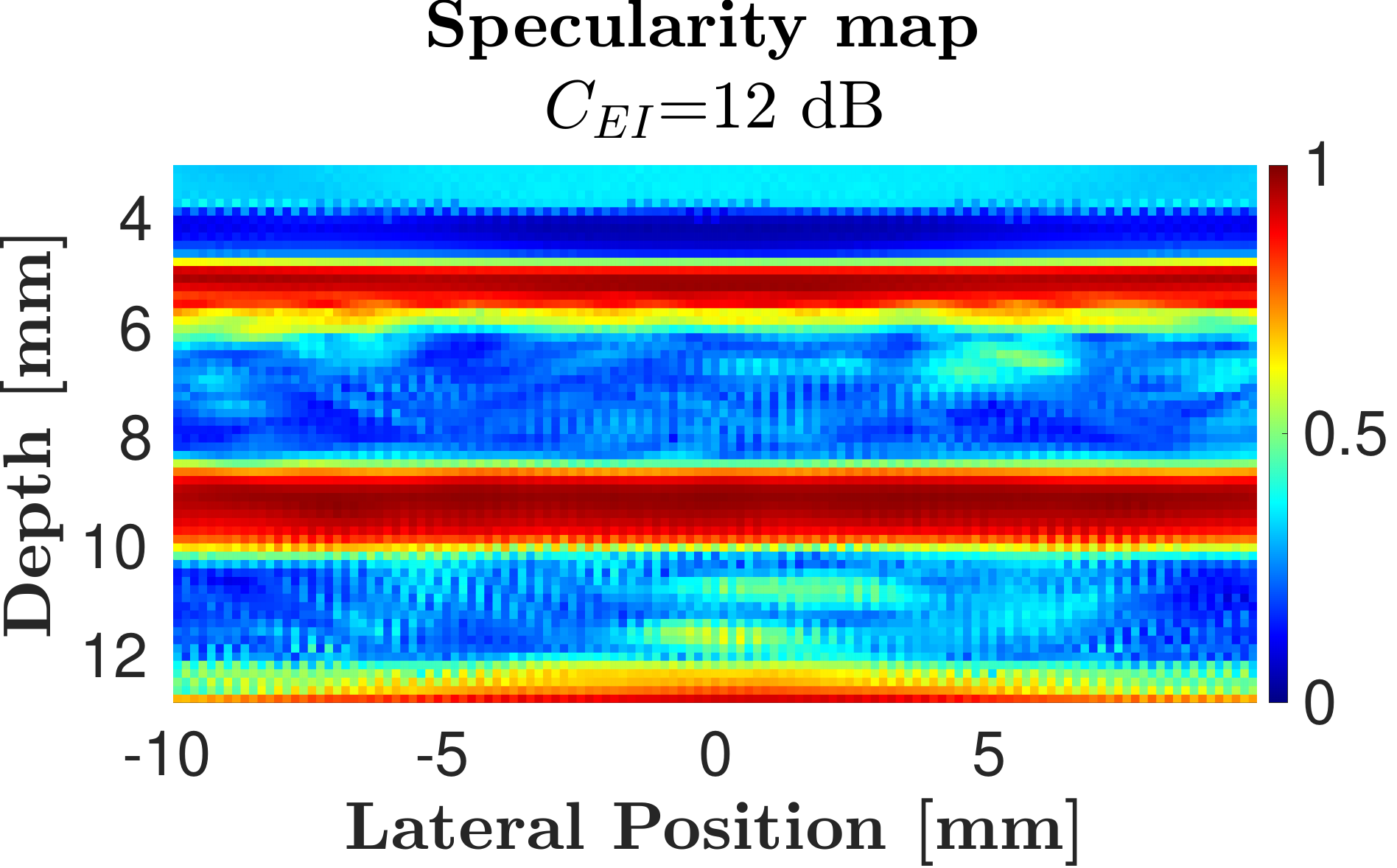}
    &\includegraphics[trim={0 90 0 90}, clip,width=.4\linewidth]{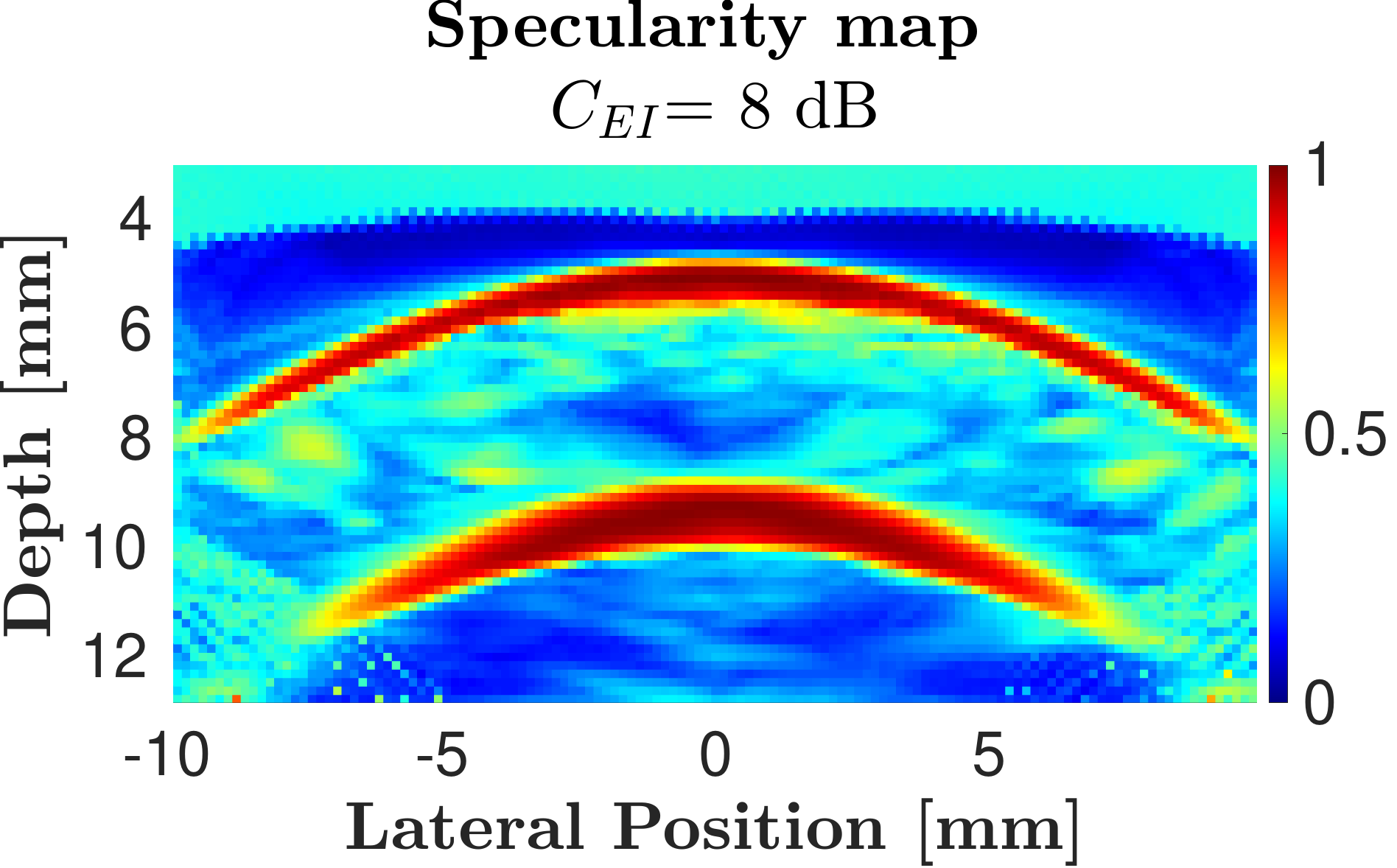}\\\hline
    \multicolumn{1}{m{2cm}|}{10\%} &\includegraphics[trim={0 90 0 90}, clip, width=.4\linewidth]{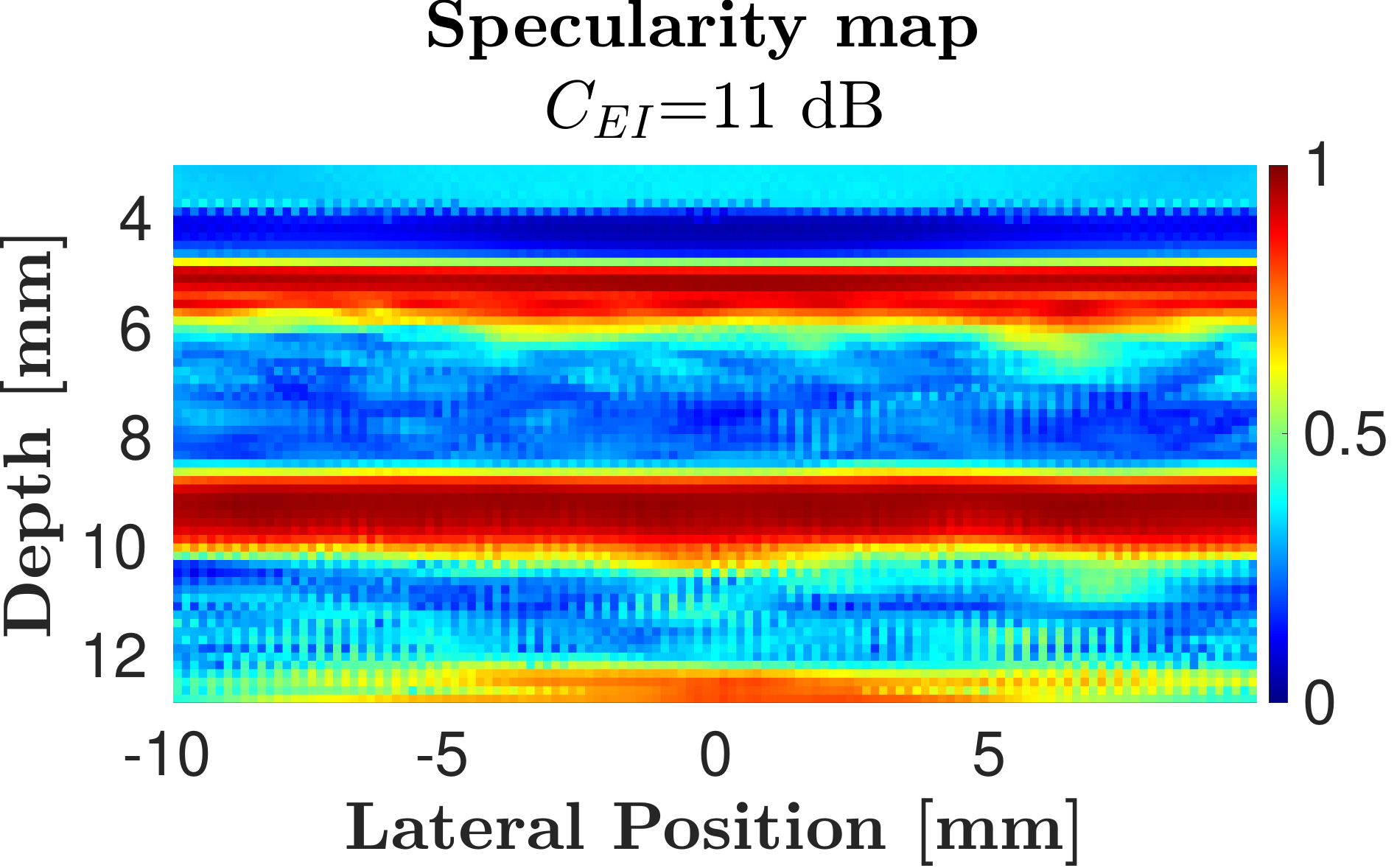}
    &\includegraphics[trim={0 90 0 90}, clip,width=.4\linewidth]{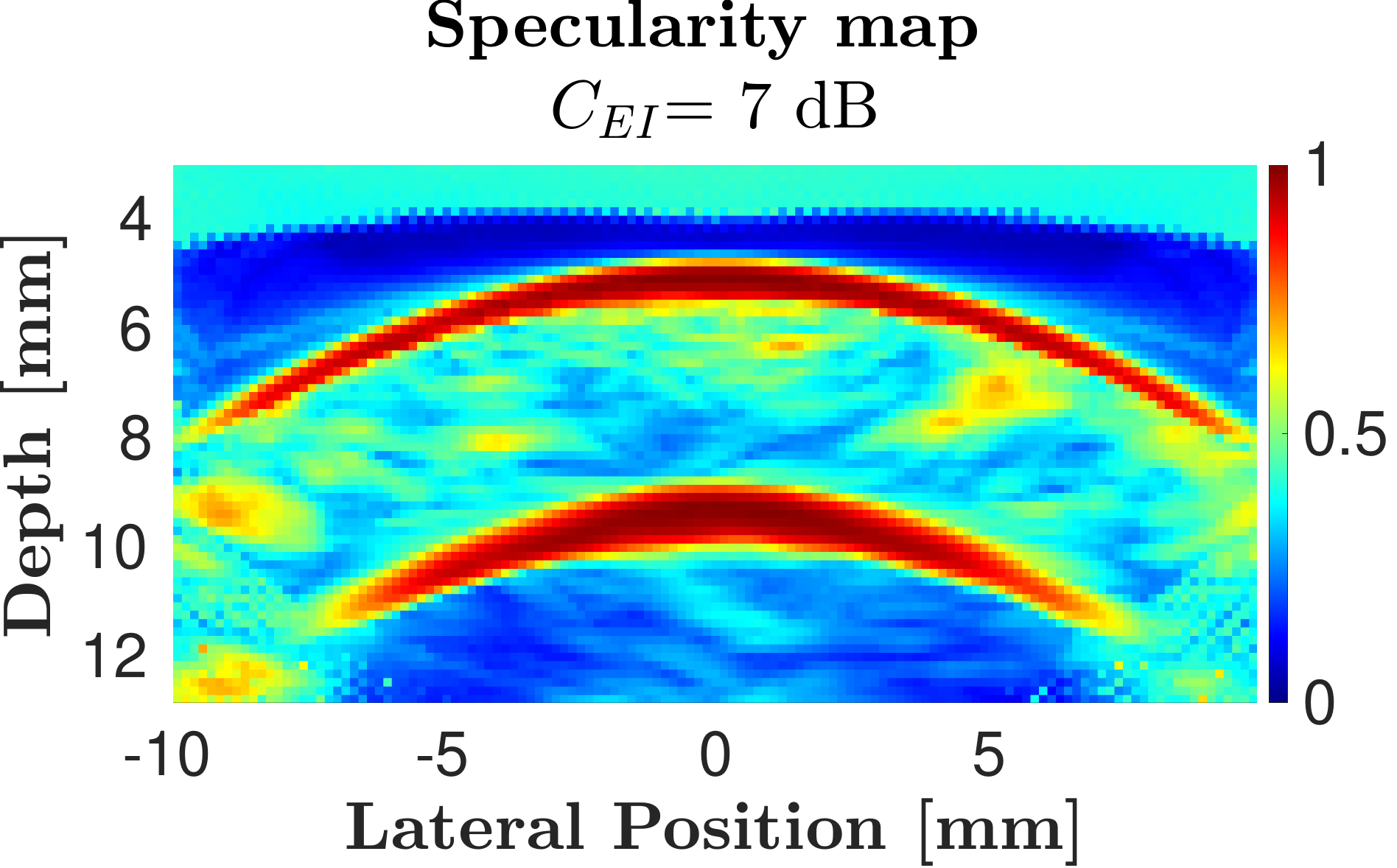}\\\hline
    \multicolumn{1}{m{2cm}|}{14\%} &\includegraphics[trim={0 0 0 90}, clip,width=.4\linewidth]{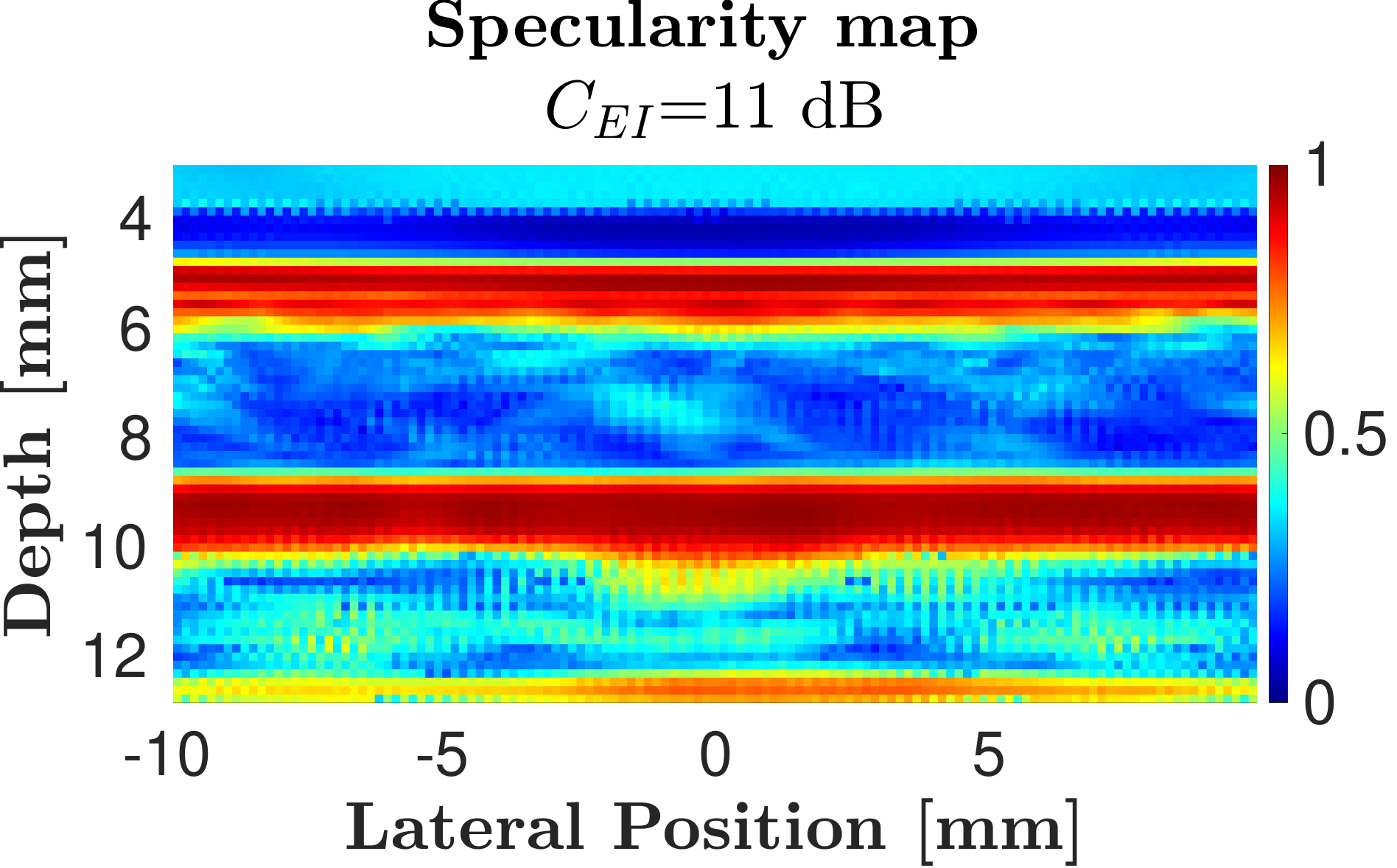}
    &\includegraphics[trim={0 0 0 90}, clip,width=.4\linewidth]{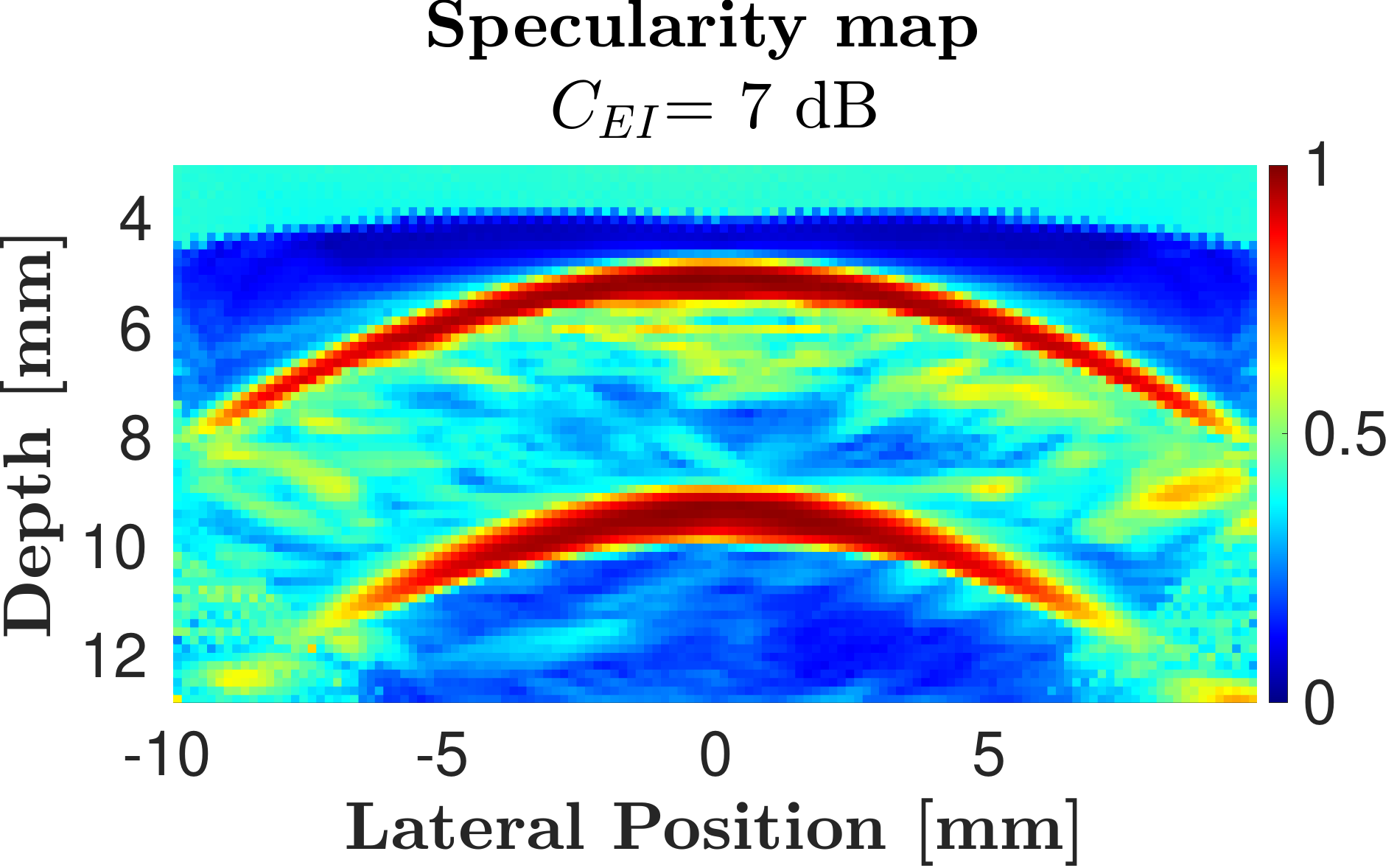}  \\\hline
\end{tabular}
\caption{Comparison of simulated configurations with planar and curved bone interfaces for identical microstructural properties (each row corresponds to porosity values of 6\%, 10\%, and 14\%, respectively, with a mean pore diameter of 10 \textmu m). Top: simulation geometries. Bottom: corresponding specularity maps. Similar spatial distributions and specularity magnitudes are observed for both configurations, indicating that moderate interface curvature has a limited influence on specularity compared with local microstructural features. \hl{The simulations were performed for a plate thickness of 4 mm and a fixed curvature.} }
\label{suppfig:specularity_vs_curvature}
\end{figure}

Figure~\ref{suppfig:specularity_vs_curvature} compares planar and curved endosteal interface configurations for identical microstructural properties. The similarity of the specularity maps in both cases indicates that moderate interface curvature has a limited effect on the specular response compared with the influence of the adjacent porous microstructure.

\begin{figure}
    \centering
    \includegraphics[width=0.75\linewidth]{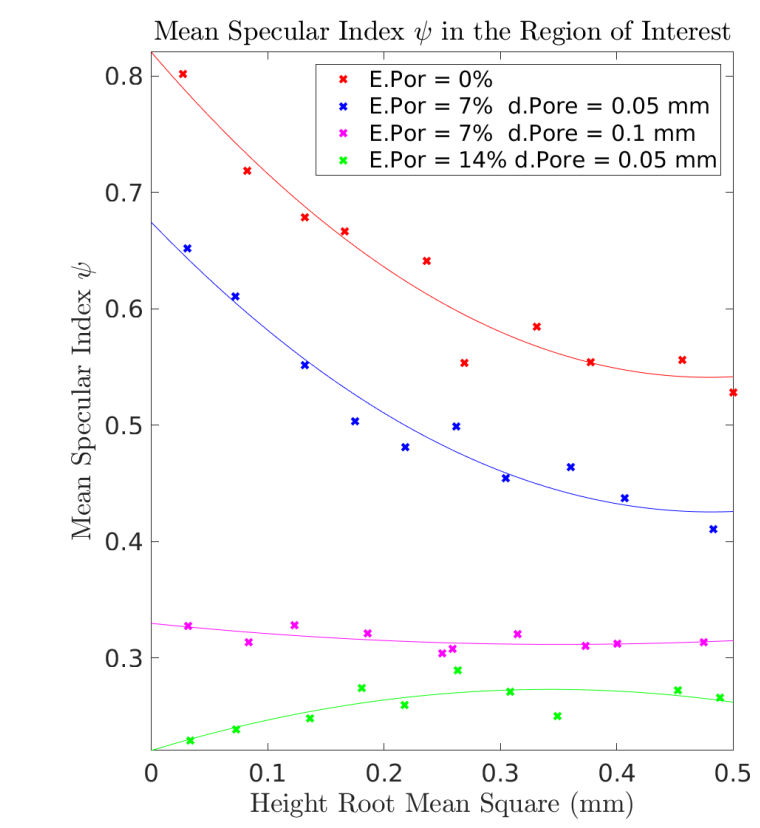}
    \caption{Evolution of the mean specular index (MSI) in the endosteal region of interest as a function of surface roughness metric (rms height), for different local microstructures: no porosity (E.Por = 0\%, red), E.Por = 7\% with a pore diameter of 0.05~mm (blue), E.Por = 7\% with a pore diameter of 0.1~mm (magenta), and E.Por = 14\% with a pore diameter of 0.05~mm (green).}
    \label{suppfig:specularity_vs_rugo_por}
\end{figure}

\hl{
Figure~\ref{suppfig:specularity_vs_rugo_por} illustrates the evolution of the mean specular index (MSI) as a function of surface rugosity for different microstructures near the endosteal region. For low-porosity or non-porous microstructures, the MSI decreases markedly with increasing rugosity. By contrast, in more porous microstructures, the MSI is lower overall and shows weaker sensitivity to rugosity. This suggests that the specular response in the endosteal region is governed by the combined effects of surface roughness and local porosity. For microstructures with high porosity or large pore diameters, porosity appears to be the dominant factor.
}

Tables~\ref{supptable:spearman-rho-onewl} and \ref{supptable:AdjustedR2-onewl} report the Spearman rank correlations and the performance of univariate linear regression models between endosteal microstructural parameters and specularity metrics when the region of interest extends over one wavelength (1.4~mm).

Tables~\ref{supptable:spearman-rho-halfwl} and \ref{supptable:AdjustedR2-halfwl} present the corresponding analyses for a smaller region of interest of half a wavelength (0.7~mm), allowing assessment of the spatial scale dependence of the observed relationships.

\begin{table}[ht!]
	\centering
	\begin{tabular}{lllllllllll}
		 & E.RMS & $\rho_{length}$ & E.Por & IQR.E.Por & E.Po.Dm & IQR.E.Po.Dm & $N_{30}$ &$N_{40}$&$N_{50}$& MSI \\
		$N_{30}$ & n.s. & n.s. & -0.86$^{**}$ & -0.71$^{*}$ & -0.89$^{**}$ & -0.75$^{*}$ & - & - &- & - \\
		$N_{40}$ & n.s. & n.s. & -0.93$^{***}$ & -0.70$^{*}$ & -0.92$^{***}$ & -0.80$^{*}$ & - & - &- & - \\
		$N_{50}$ & n.s. & n.s. & -0.62$^{*}$ & n.s. & -0.71$^{*}$ & -0.60$^{*}$ & - &- & -& -  \\
		MSI & n.s. & n.s. & -0.69$^{*}$ & n.s. & -0.79$^{*}$ & -0.67$^{*}$ & 0.89$^{**}$ &0.90$^{***}$&0.95$^{***}$& - \\
	\end{tabular}
    \caption{Spearman rank correlation coefficients ($\rho$) between endosteal microstructural variables within a ROI of 1.4~mm around the endosteal surface and specularity metrics computed using different thresholds for the number of specular pixels ($N_{30}$, $N_{40}$, $N_{50}$) and the MSI. Significance levels: $^*p~<$ 0.05; $^{**}p~<$ 0.001; $^{***}p~<$ 0.0001; n.s.: not significant.}
	\label{supptable:spearman-rho-onewl}
\end{table}
\begin{table}[ht!]
	\centering
	\begin{tabular}{l|l|l|l|l|l|l|l}
\hline
    &\multirow{2}{*}{Quantity} &  \multicolumn{2}{c|}{$N_{30}$}  &  \multicolumn{2}{c|}{$N_{40}$}  &  \multicolumn{2}{c}{$N_{50}$} \\
		 && Adj-R$^2$  & RMSE &Adj-R$^2$  & RMSE&Adj-R$^2$  & RMSE\\
\hline
		\multirow{2}{*}{Porosity } &E.Por & 71$^{**}$ & 5$^{}$& 84$^{***}$ & 5$^{}$& 32$^{*}$ & 6$^{}$ \\
		&IQR.E.Por & 45$^{*}$ & 6.4$^{}$& 44$^{*}$ & 9$^{}$& n.s. & $-$ \\
\hline
		\multirow{2}{*}{Pore diameter} &E.Po.Dm & 77$^{**}$ & 4$^{}$& 84$^{***}$ & 5$^{}$& 46$^{*}$ & 6$^{}$ \\
		&IQR.E.Po.Dm & 52$^{*}$ & 6$^{}$& 60$^{*}$ &7$^{}$& 29$^{*}$ & 7$^{}$ \\
\hline

	\end{tabular}
    \caption{Performance of univariate linear regression models relating endosteal microstructural variables (porosity and pore diameter) to specularity metrics with different thresholds ($N_{30}$, $N_{40}$, and $N_{50}$) within a ROI of 1.4~mm (corresponding to one wavelength). Model performance is summarized using the adjusted coefficient of determination (Adj-R$^2$, in \%) and the root mean square error (RMSE). Statistical significance refers to the regression model: $^*p~<$ 0.05; $^{**}p~<$ 0.001; $^{***}p~<$ 0.0001; n.s.: not significant.}
	\label{supptable:AdjustedR2-onewl}

\end{table}
\begin{table}[ht!]
	\centering
	\begin{tabular}{lllllllllll}
		 & E.RMS & $\rho_{length}$ & E.Por & IQR.E.Por & E.Po.Dm & IQR.E.Po.Dm & $N_{30}$ & $N_{40}$ &$N_{50}$ & MSI \\
		E.RMS & - & - & - & - & - & - & - & -  & - & -\\
		$\rho_{length}$ & n.s. & - & - & - & - & - & - & -  & - & -\\
		E.Por & n.s. & n.s. & - & - & - & - & - & -  & - & -\\
		IQR.E.Por & n.s. & n.s. & 0.88$^{**}$ & - & - & - & - & -  & - & -\\
		E.Po.Dm & n.s. & n.s. & 0.96$^{***}$ & 0.85$^{**}$ & - & - & - & - & - & - \\
		IQR.E.Po.Dm & n.s. & n.s. & 0.84$^{**}$ & 0.69$^{*}$ & 0.88$^{**}$ & - & - & -  & - & -\\
		$N_{30}$ & n.s. & n.s. & -0.88$^{**}$ & -0.83$^{**}$ & -0.92$^{***}$ & -0.73$^{*}$ & - & -  & - & -\\
		$N_{40}$ & n.s. & n.s. & -0.97$^{***}$ & -0.85$^{**}$ & -0.96$^{***}$ & -0.80$^{*}$ & - & -  & - & -\\
		$N_{50}$ & n.s. & n.s. & -0.85$^{**}$ & -0.71$^{*}$ & -0.87$^{**}$ & -0.71$^{*}$ & - & - & - & - \\
		MSI & n.s. & n.s. & -0.82$^{*}$ & -0.73$^{*}$ & -0.84$^{**}$ & -0.68$^{*}$ & 0.87$^{**}$ &0.89$^{**}$&0.96$^{**}$& - \\
	\end{tabular}
    \caption{Spearman rank correlation coefficients ($\rho$) between endosteal microstructural variables within a ROI of 0.7~mm around the endosteal surface and specularity metrics computed using different thresholds for the number of specular pixels ($N_{30}$, $N_{40}$, $N_{50}$) and the MSI. Significance levels: $^*p~<$ 0.05; $^{**}p~<$ 0.001; $^{***}p~<$ 0.0001; n.s.: not significant.}
	\label{supptable:spearman-rho-halfwl}
\end{table}

\begin{table}[ht!]
	\centering
	\begin{tabular}{l|l|l|l|l|l|l|l}
\hline
    &\multirow{2}{*}{Quantity} &  \multicolumn{2}{c|}{$N_{30}$}  &  \multicolumn{2}{c|}{$N_{40}$}  &  \multicolumn{2}{c}{$N_{50}$} \\
		 && Adj-R$^2$  & RMSE &Adj-R$^2$  & RMSE&Adj-R$^2$  & RMSE\\
\hline
		\multirow{2}{*}{Porosity } &E.Por & 76$^{**}$ & 4$^{}$ & 93$^{***}$ & 3$^{}$& 70$^{**}$ & 6$^{}$ \\
		&IQR.E.Por & 66$^{**}$ & 5$^{}$ & 70$^{**}$ & 7$^{}$& 46$^{*}$ & 7$^{}$ \\
\hline
		\multirow{2}{*}{Pore diameter} &E.Po.Dm & 84$^{***}$ & 4$^{}$& 91$^{***}$ & 4$^{}$& 73$^{**}$ & 5$^{}$ \\
		&IQR.E.Po.Dm & 48$^{*}$ & 6$^{}$&61$^{*}$ & 8$^{}$& 46$^{*}$ & 7$^{}$ \\
\hline
	\end{tabular}
    \caption{Performance of univariate linear regression models relating endosteal microstructural variables (porosity and pore diameter) to specularity metrics with different thresholds ($N_{30}$, $N_{40}$, and $N_{50}$) within a ROI of 0.7~mm (corresponding to half wavelength). Model performance is summarized using the adjusted coefficient of determination (Adj-R$^2$, in \%) and the root mean square error (RMSE). Statistical significance refers to the regression model: $^*p~<$ 0.05; $^{**}p~<$ 0.001; $^{***}p~<$ 0.0001; n.s.: not significant.}
	\label{supptable:AdjustedR2-halfwl}

\end{table}